\pdfoutput=1
\documentclass[a4paper,fleqn,usenatbib,useAMS]{mnras}

\usepackage{hyperref}

\usepackage{natbib}
\usepackage{color}
\usepackage{txfonts}

\usepackage{etoolbox}
\makeatletter
\newcount\c@additionalboxlevel
\newcount\c@maxboxlevel
\patchcmd\@combinedblfloats{\box\@outputbox}{%
  \stepcounter{additionalboxlevel}%
  \box\@outputbox
}{}{\errmessage{\noexpand\@combinedblfloats could not be patched}}

\AtBeginShipout{%
  \ifnum\value{additionalboxlevel}>\value{maxboxlevel}%
    \typeout{Warning: maxboxlevel might be too small, increase to %
      \the\value{additionalboxlevel}%
    }%
  \fi 
  \@whilenum\value{additionalboxlevel}<\value{maxboxlevel}\do{%
    \typeout{* Additional boxing of page `\thepage'}%
    \setbox\AtBeginShipoutBox=\hbox{\copy\AtBeginShipoutBox}%
    \stepcounter{additionalboxlevel}%
  }%
  \setcounter{additionalboxlevel}{0}%
}
\makeatother

\usepackage[T1]{fontenc}
\usepackage{aecompl}

\usepackage{enumitem}

\setlist[itemize]{itemindent=.1cm}

\usepackage{subfig}
\usepackage{verbatim}
\usepackage[table,xcdraw]{xcolor}

\usepackage{lastpage}
\usepackage{graphicx}	

\usepackage{savesym}
\savesymbol{iint}
\savesymbol{iiint}
\savesymbol{iiiint}
\savesymbol{idotsint}

\usepackage{morefloats}

\usepackage{amsmath}	
\usepackage{amssymb}	

\usepackage{multicol}        
\usepackage{multirow}
\usepackage{bm}		
\usepackage{pdflscape}	
\title[LITTLE THINGS in 3D]{LITTLE THINGS in 3D: robust determination of the circular velocity of dwarf irregular galaxies.}

\author[G. Iorio et al.]{
G. Iorio,$^{1,2}$\thanks{giuliano.iorio@unibo.it}
F. Fraternali,$^{1,3}$
C. Nipoti,$^{1}$
E. Di Teodoro,$^{4}$
J. I. Read$^{5}$
and G. Battaglia$^{6,7}$
\\
$^{1}$Dipartimento di Fisica e Astronomia, Universit\`a di Bologna, Viale Berti Pichat 6/2, I-40127, Bologna, Italy\\
$^{2}$INAF -  Osservatorio Astronomico di Bologna, via Ranzani 1, I-40127, Bologna, Italy\\
$^{3}$Kapteyn Astronomical Institute, University of Groningen, Landleven 12, 9747 AD Groningen, The Netherlands\\
$^{4}$Research School of Astronomy and Astrophysics - The Australian National University, Canberra, ACT, 2611, Australia  \\
$^{5}$Department of Physics, University of Surrey, Guildford, GU2 7XH, Surrey, UK\\
$^{6}$Instituto de Astrofisica de Canarias, calle Via Lactea s/n, E-38205 La Laguna, Tenerife, Spain\\
$^{7}$Universidad de La Laguna, Dpto. Astrofisica, E-38206 La Laguna, Tenerife, Spain
}

\date{Accepted XXX. Received YYY; in original form ZZZ}

\pubyear{2016}

\begin{document}
\label{firstpage}
\pagerange{\pageref{firstpage}--\pageref{lastpage}}
\maketitle

\begin{abstract}
{Dwarf Irregular galaxies (dIrrs) are the smallest stellar systems with extended HI discs. The study of the kinematics of such discs is a powerful tool to estimate the total matter distribution at these very small scales. In this work, we  study the HI kinematics of 17  galaxies extracted from   the `Local Irregulars That Trace Luminosity Extremes, The HI Nearby Galaxy Survey' (LITTLE THINGS). Our approach differs significantly from previous studies in that we directly fit 3D models (two spatial dimensions plus one spectral dimension) using the software $^\text{3D}$BAROLO, fully exploiting the information in the HI datacubes.
For each galaxy we derive the geometric parameters of the HI disc (inclination and position angle), the radial distribution of the surface density, the velocity-dispersion ($\sigma_v$) profile and the rotation curve. The circular velocity (V$_{\text{c}}$), which traces directly the galactic potential, is then obtained by correcting the rotation curve for the asymmetric drift. 
As an initial application, we show that these dIrrs lie on a baryonic Tully-Fisher relation in excellent agreement with that seen on larger scales.
The final products of this work are high-quality, ready-to-use kinematic data ($\textrm{V}_\textrm{c}$ and $\sigma_v$) that we make publicly available. These can be used to perform dynamical studies and improve our understanding of these low-mass galaxies.
}

\end{abstract}

\begin{keywords}
galaxies: dwarf -- galaxies: kinematics -- galaxies: ISM -- galaxies: structure
\end{keywords}


\section{Introduction}
The HI 21 cm emission line is a powerful tool for studying the dynamics of late-type galaxies since it is typically detected well beyond the optical disc and is not affected by dust extinction. 
Decades ago, the rise of the radio-interferometers  allowed  many authors  to obtain high-resolution rotation curves for large samples of spiral galaxies (e.g. \citealt{bosma1,bosma2,bosma3,van,begeman}). These studies found that the gas rotation remains nearly flat also in the outermost disc, where the visible matter fades and one would expect a nearly Keplerian fall-off of the rotation curve. The flat rotation curves of spirals are one of the most robust indications for the existence of dark matter (DM); gas-rich late-type galaxies are an ideal laboratory for studying the properties of DM halos. In this context, dwarf irregular galaxies (dIrrs) are also studied with great interest. Unlike large spirals, they appear to be dominated by DM down to their very central regions. Thus, the determination of the DM density distribution is nearly independent of the mass-to-light ratio of the stellar disc (e.g. \citealt{casy,cote}). 

The standard method to extract a galaxy rotation curve is based on the fitting of a tilted-ring model to a 2D velocity field (e.g. \citealt{begeman,scarparo}). The velocity field is extracted from the HI datacube by deriving the representative velocity of the line profile at each spatial pixel. There are different methods to define a representative velocity, for instance using the intensity-weighted mean (e.g. \citealt{rog2}),  or by fitting functional forms  (e.g. single Gaussian, \citealt{begeman} or multiple Gaussians, \citealt{RcLT}, hereafter O15).  
The 2D approach has been used in several numerical algorithms such as  ROTCUR \citep{begeman}, RESWRI \citep{scarparo}, KINEMETRY \citep{kinemetry} and DISKFIT \citep{diskfit}. All of these codes have been very useful to improve our understanding of the kinematics of late-type galaxies. However, working in 2D has a drawback: the results depend on the assumptions made when extracting the velocity field and are also affected by the instrumental resolution (the so-called beam-smearing; \citealt{bosma,begeman}).
These problems are especially relevant in the study of the kinematics of dIrrs since the HI line profiles can be heavily distorted by both non-circular motions and noise. As a consequence, the extraction of the velocity field can be challenging (see e.g. \citealt{2366oh}). Furthermore, dIrrs are often observed with a limited number of resolution elements because of their limited extension on the sky. The effect of beam smearing can also make the estimate of both the rotation curve and the inclination of the disc uncertain \citep{begeman}.

The pioneering work of \cite{swaters} showed that these problems can be solved with  an alternative approach: the properties of the HI disc can be retrieved with a direct modelling of the 3D datacube (two spatial axes plus one spectral axis) without explicitly extracting velocity fields. In practice, a 3D method consists of a data-model comparison of $n_\text{ch}$ maps (where $n_\text{ch}$ is the number of spectral channels)  instead of the single map represented by the velocity field. The best advantage of this approach is that the datacube models are convolved  with the instrumental response, so the final results are not affected by the beam smearing. \cite{swaters}, \cite{gentile}, \cite{lellimalin} and \cite{halol} used a 3D visual comparison between datacubes and model cubes  to correct and improve the results obtained with the classical 2D methods. However, a by-eye inspection of the data is time intensive and subjective. Modern software like TiRiFiC \citep{tirific}  and $^\text{3D}$BAROLO (hereafter 3DB; \citealt{barolo}) can perform a full 3D numerical minimisation  on the whole datacube.

In this paper, we exploit the power of the 3D approach by applying 3DB to a sample  of 17 dIrrs taken from the LITTLE THINGS (hereafter LT) sample \citep{LT}. 3DB
has been extensively tested on mock HI datacubes corresponding to low mass isolated dIrrs. \cite{mia} show that the rotation curve is well-recovered even for a star-bursting dwarf, provided that the inclination of the HI disc is higher than $40^\circ$.

The final products of our analysis are ready-to-use rotation curves\footnote{The final rotation curves can be downloaded from {\tt http://www.filippofraternali.com/styled-9/index.html}} corrected for the instrumental effects (see Sec. \ref{sec:3db}) and for the
asymmetric drift (see Sec. \ref{sec:asy}). The rotation curves can be used to study the dynamics of dIrrs (e.g. \citealt{intro2,mia2,mia}) or to extend the study of the scaling relations on small scales (e.g. \citealt{bt3,btfl}).
An unbiased estimate of the rotation curves of dIrrs is also essential to shed light on the well-known cosmological tensions on small scale: the long debated `cusp-core problem'  \citep{deb,ccoh,Jcore}, the `missing-satellites problem' \citep{miss,mia2} and the `too-big-to-fail problem' \citep{just_too,tbtf,mia2}. These ``problems" are discrepancies between the properties of DM halos predicted by cosmological simulations and those inferred from observations. In this context it is therefore crucial to obtain reliable estimates of the DM distribution in real galaxies.

In this work, we focus in particular on the asymmetric-drift correction (see Sec. \ref{sec:asy}). This correction term is usually applied without considering the errors made in its calculation, but these can be very large and should be taken into account. 
The propagation of the asymmetric-drift errors has a great impact on the determination of the circular velocity of galaxies in which the contribution of the gas velocity dispersion is important (see for example DDO 210 in Sec. \ref{sec:gls}). We developed a method to calculate and propagate the asymmetric-drift uncertainties: as a result, the quoted errors on the final corrected rotation curves are a robust description of the real uncertainties.

The paper is organised as follows. In Section \ref{sec:sample}, we illustrate the sample and the data used in this work. In Section \ref{sec:method}, we briefly introduce the tilted-ring model and 3DB.
In Section \ref{sec:datan}, we describe in detail the analysis applied to our data.
Section \ref{sec:gls} shows the results of our analysis for each galaxy in our sample. In  Section \ref{sec:scientific}, we look at whether the low mass dIrrs that we study here lie on the baryonic Tully-Fisher relation. In Section \ref{sec:disc},  we discuss the assumptions made to derive the rotation curves and we compare our results with previous work.
A summary is given in Section \ref{sec:sum}.

\section{Sample and Data} \label{sec:sample}

\subsection{Sample selection} \label{sec:sample_sel}

 \begin{figure*}
  \centering
    \subfloat{%
      \includegraphics[width=0.48\textwidth]{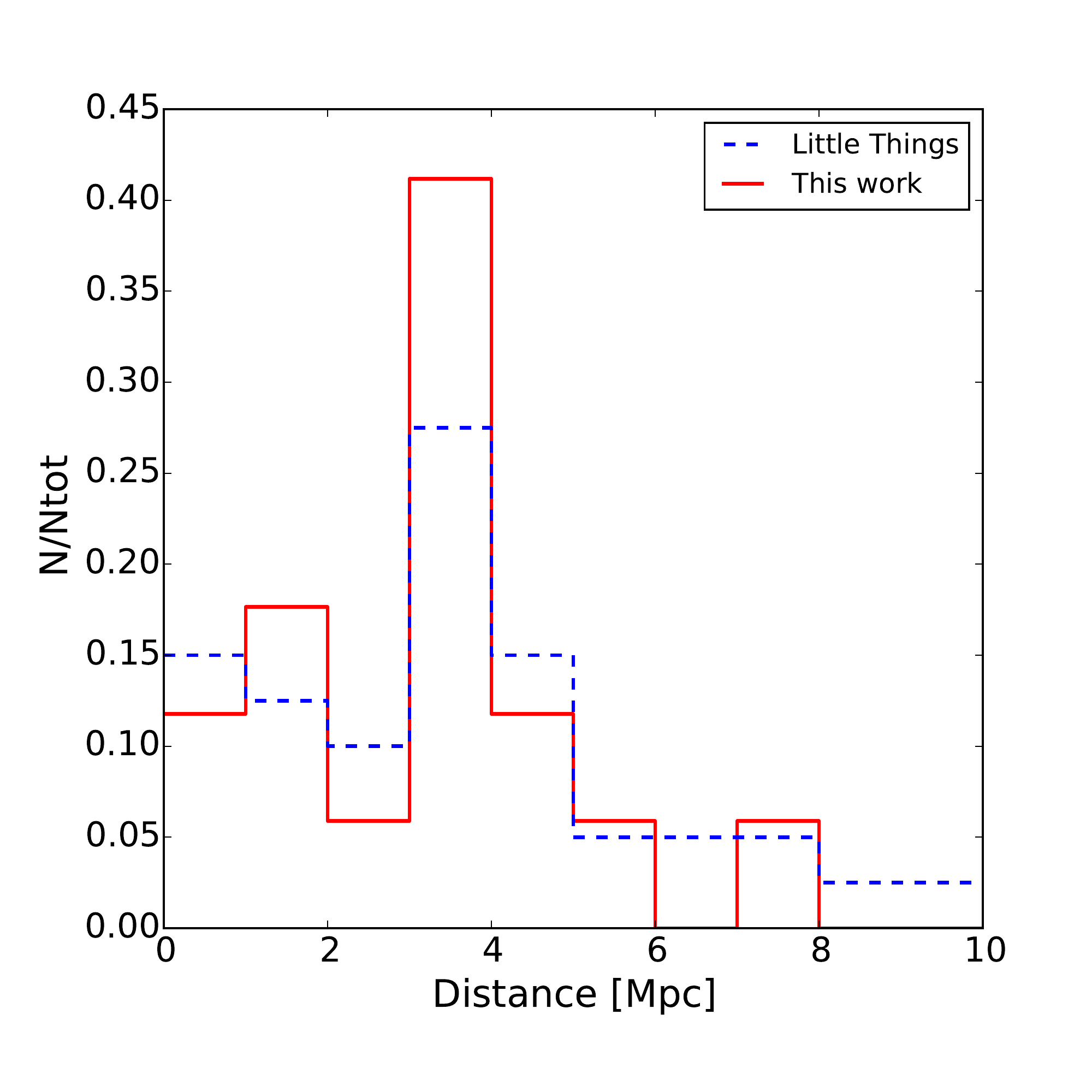}
    }
    \hfill
    \subfloat{%
      \includegraphics[width=0.48\textwidth]{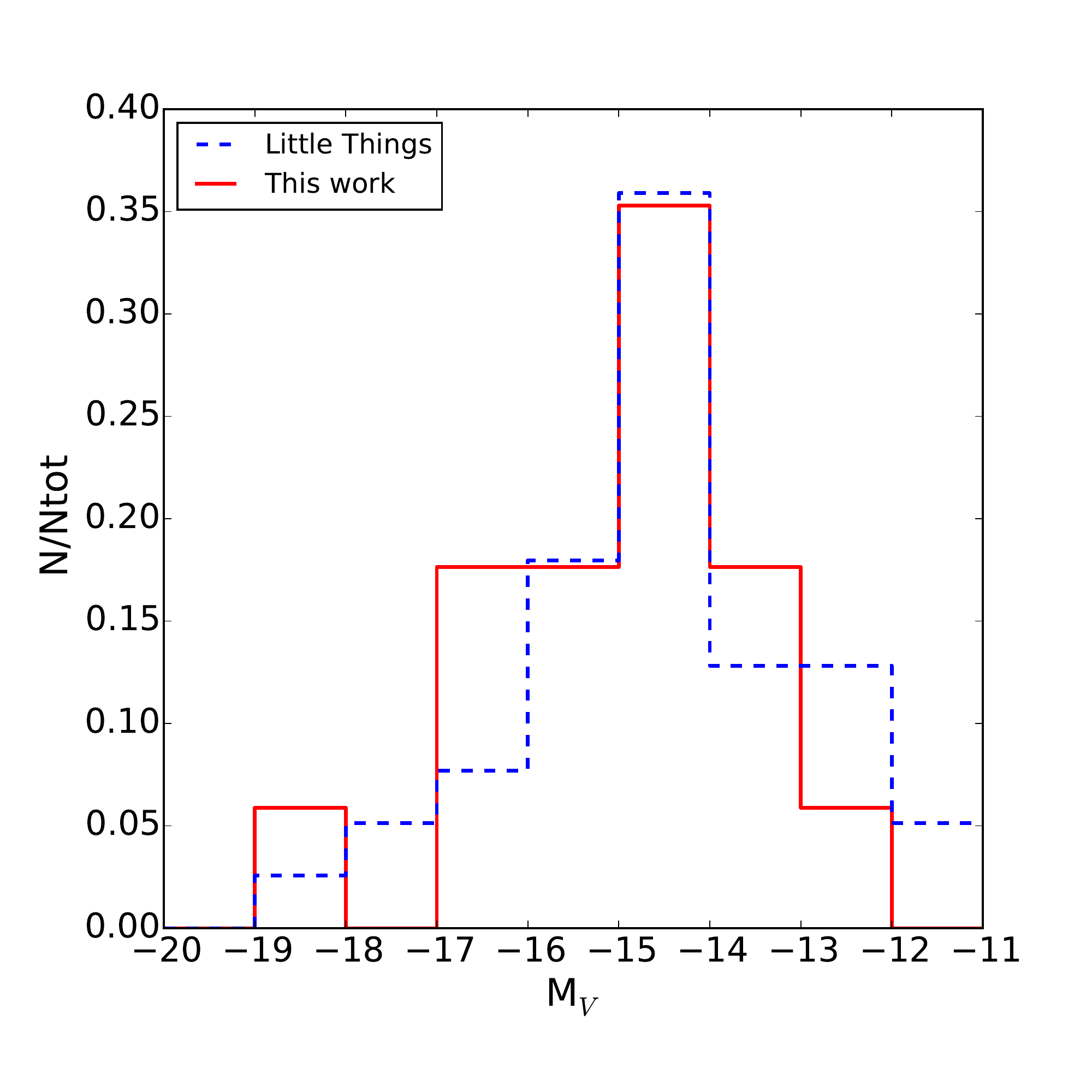}
    }
    \caption{Distribution of the Galactocentric distances (left-hand 
    panel) and  of the absolute magnitudes (right-hand panel) for the 
    sample of galaxies studied in this work (red histograms) and for the 
    whole LT sample (blue dashed histograms).
    Data from \protect\cite{LT} and references 
    therein.}
    \label{fig:sample}
  \end{figure*}

The galaxies used in this work  represent a sub-sample of the galaxies of the LT survey \protect\citep{LT} which is a sample of 37 dIrrs and 4 Blue Compact Dwarfs (BCDs) located in the Local Volume within 11 Mpc from the Milky Way.
LT combines archival and new HI observations taken with the Very Large Array (VLA)  to obtain a very high spatial and spectral  resolution data set (see Table \ref{tab:sample_obs}). The objects in the LT sample have been chosen to be both  isolated and  representative  of the full range of dIrr properties \protect\citep{LT}.

We built our sample selecting dIrrs that are well-suited to study the circular motion of the HI discs. To this end, we excluded from our selection all the objects seen at low  inclination angles and the 4 BCDs. The inclination angle $i$ is defined as the angle between the plane of the disc and the line of sight such that $i=90^\circ$  for an edge-on galaxy and $i=0^\circ$  for a face-on galaxy.   
The estimate of $i$ in nearly face-on galaxies ($i<40^\circ$) is extremely difficult both from the kinematic fit \protect\citep{begeman} and from the analysis of the HI map \protect\citep{mia}. Moreover, for $i<40^\circ$, relatively small errors on $i$ have a great impact on the  deprojection of the observed rotational velocity (see Eq. \ref{eq:vtitlt}). As a consequence, the final rotation curves of these nearly face-on galaxies could be biased and unreliable with both 2D \protect\citep{oman2} and 3D methods \protect\citep{mia}.
Among the objects with $i>40^\circ$, we selected 17 dIrrs, 
about half of the original LT sample.  

We checked, using a Kolmogorov-Smirnov (KS) test, that our sub-sample preserves the statistical distribution of the galactic  properties such as distance (KS p-value  0.96), absolute magnitude (KS p-value  0.99), star formation rate density (SFRD, KS p-value  0.92) and baryonic mass (KS p-value  0.99).
Because of our rejection criterion, the average $i$ in our sub-sample ($58^\circ$) is higher than the one measured in the LT sample ($50^\circ$).
Figure \ref{fig:sample} shows the distributions of the distances and of the absolute magnitudes
of our sample and the original LT sample.

It is worth pointing out that our rejection criterion is based on $i$ estimated  for the stellar disc from the V-band photometry \protect\citep{HVBand}. When we re-estimated $i$ from the HI datacubes (see Sec. \ref{sec:tstage}), we found that four galaxies (DDO 47, DDO 50, DDO 53, and DDO 133) have an average $i$ lower than  40$^\circ$ (see Table \ref{tab:sample_res}). We did not exclude these galaxies from our sample, but our results for these objects should be treated with caution. In particular the formal  errors on the velocities and on $i$ could underestimate the real uncertainties (see notes on the individual galaxies in Sec. \ref{sec:gls}).

In conclusion, our sample comprises 17 objects covering the stellar mass range $5\times10^5 \lesssim \text{M}_*/ \text{M}_\odot \lesssim  10^8$, with a mean SFRD of about 0.006 $\text{M}_\odot \text{yr}^{-1} \text{kpc}^{-2}$ and a mean specific SFRD (SSFRD) of about 
$2.26\times10^{-10}$ $\text{yr}^{-1} \text{kpc}^{-2}$ (data from \protect\citealt{LT} and references therein). For comparison the SFRD and the SSFRD of the complete LT sample are 0.007 $\text{M}_\odot \text{yr}^{-1} \text{kpc}^{-2}$ and $2.95\times10^{-10}$ $\text{yr}^{-1} \text{kpc}^{-2}$ respectively (\protect\citealt{LT} and references therein). 

\subsection{The Data} \label{sec:sample_data}

The HI datacubes were taken from the the publicly available archive of the LT survey\footnote{{\tt https://science.nrao.edu/science/surveys/littlethings}}. 
For each galaxy, LT provides two datacubes that differ in the weight scheme used to reduce the raw data: the natural datacube offers the best signal-to-noise ratio, the robust datacube gives the best angular resolution. 
We preferred the natural datacubes for two main reasons: (i) an high signal-to-noise ratio allows us to have enough sensitivity to extend the study of the kinematics 
at large radii; (ii) 3D methods are not dramatically influenced by the resolution of the data (see Sec. \ref{sec:method}).
In some cases (DDO 50, DDO 154, NGC 1569, NGC 2366, UGC 8508 and WLM) the galaxies are so extended in the sky that we further smoothed the datacube maintaining  a good spatial resolution. Instead, we have chosen the robust datacube when the natural datacube gives too few resolution elements to sample the HI disc (5 cases, see Tab. \ref{tab:sample_obs}).
The mean spatial resolution  is about 240 pc ranging from about 60 pc (DDO 210) to  about 460 pc (DDO 87).
These data are summarised in Cols. 4-6 of Table \ref{tab:sample_obs}. The rotation curves of all the galaxies in our sample have been already derived with the classical 2D approach in O15.  The masses and the surface-density profiles of the stellar discs for all the galaxies can be found in \protect\cite{ZLT}, except for DDO 47 for which they can be found in \protect\cite{d47starprof} and in \protect\cite{d47stellarmass}.

\begin{table*}
\centering
\tabcolsep=0.11cm
\begin{tabular}{l|ccc|cccc|ccc}
\hline
\multicolumn{1}{c|}{Galaxy} & \begin{tabular}[c]{@{}c@{}}D\\ (Mpc)\\ (1)\end{tabular} & \begin{tabular}[c]{@{}c@{}}M$_V$\\ (mag)\\ (2)\end{tabular} & \begin{tabular}[c]{@{}c@{}}$f_c$\\ (pc/arcsec)\\ (3)\end{tabular} &   \begin{tabular}[c]{@{}c@{}}cube\\ \\ (4)\end{tabular} & \begin{tabular}[c]{@{}c@{}}Beam\\ (arcsec$\times$arcsec)\\  (5)\end{tabular} & \begin{tabular}[c]{@{}c@{}}Ch. sep\\ (km/s)\\ (6)\end{tabular} & \begin{tabular}[c]{@{}c@{}}M$_\text{HI}$\\ ($10^7$ M$_\odot$)\\ (7)\end{tabular} & \begin{tabular}[c]{@{}c@{}}$\sigma_\text{ch}$\\ (mJy/Beam)\\ (8)\end{tabular} & \begin{tabular}[c]{@{}c@{}}$\sigma_\text{3T}$\\ (M$_\odot$/pc$^2$)\\ (9)\end{tabular} \\ \hline
\rowcolor[HTML]{CCCCCC} 
CVnIdwA & 3.6 & -12.4 & 17.4 & ro & 10.9 x 10.5   & 1.3 & $4.5\pm0.1$ & 0.63 & 0.96 \\
DDO 47 & 5.2 & -15.5 & 25.2  & na & 16.1 x 15.3   & 2.6 & $35.3\pm0.6$ & 0.61 & 0.41   \\
\rowcolor[HTML]{CCCCCC} 
DDO 50 & 3.4  & -16.6 & 16.5 & na$\dagger$ & 15.0 x 15.0 &  2.6 & $59.5\pm1.6$ & 0.82 & 0.65  \\
DDO 52 & 10.3  & -15.4 & 49.9 & na & 11.8 x 7.2  &  2.6 & $22.3\pm0.1$ & 0.46  & 1.20   \\
\rowcolor[HTML]{CCCCCC} 
DDO 53 & 3.6  & -13.8 & 17.4 & ro & 6.3 x 5.7  & 2.6 & $4.7\pm0.3$  & 0.57 & 3.80  \\
DDO 87 & 7.4  & -15.0 & 35.9 & na  & 12.8 x 11.9   & 2.6 & $20.7\pm1.0$ & 0.51 & 0.90  \\
\rowcolor[HTML]{CCCCCC} 
DDO 101 & 6.4  & -15.0 & 31.0 & ro  & 8.3 x 7.0   & 2.6 & $1.8\pm0.2$ & 0.50 & 0.90  \\
DDO 126 & 4.9  & -14.9 & 23.7 & na & 12.2 x 9.3    & 2.6 & $12.8\pm0.1$ & 0.41 & 1.27  \\
\rowcolor[HTML]{CCCCCC} 
DDO 133 & 3.5  & -14.8 & 17.0 & ro & 12.4 x 10.8  & 2.6 & $10.0\pm0.2$  &  0.35 & 0.89  \\
DDO 154 & 3.7  & -14.2 & 18.0 & na$\dagger$  & 15.0x15.0   & 2.6 & $25.1\pm0.1$ & 0.44 & 0.52  \\
\rowcolor[HTML]{CCCCCC} 
DDO 168 & 4.3  & -15.7 & 20.8 & na & 15.0 x 15.0   & 2.6 & $25.1\pm0.1$ & 0.47 & 0.54  \\
DDO 210 & 0.9 & -10.9 & 4.4  & na & 16.6 x 14.1   & 1.3 & $0.2\pm0.1$ & 0.75 & 0.59  \\
\rowcolor[HTML]{CCCCCC} 
DDO 216 & 1.1  & -13.7 & 5.3 & ro & 16.2 x 15.4   & 1.3 & $0.5\pm0.1$ & 0.91  & 0.40  \\
NGC 1569 & 3.4  & -18.2 & 16.5 & na$\dagger$& 15.0 x 15.0    & 2.6 & $17.3\pm0.1$ & 0.77  & 1.10\\
\rowcolor[HTML]{CCCCCC} 
NGC 2366 & 3.4  & -16.8 & 16.5 & na$\dagger$ & 15.0 x 15.0  & 2.6 & $64.0\pm1.2$ & 0.52 & 0.59  \\
UGC 8508 & 2.6  & -13.6 & 12.6 & na$\dagger$  & 15.0 x 15.0   & 1.3 & $1.7\pm0.1$ &  1.31 & 0.54\\
\rowcolor[HTML]{CCCCCC} 
WLM & 1.0 & -14.4 & 4.8  & na$\dagger$ & 25.0 x 25.0   & 2.6 & $5.6\pm0.1$ & 2.00 & 0.58  \\
\hline
\end{tabular}
\caption{ 
{\bf Properties of the sample of galaxies studied in this work.}  -  
{\bf (1)} distance in Mpc (\protect\citealt{LT} and references therein);  {\bf (2)}  V band absolute magnitude (\protect\citealt{LT} and references therein); {\bf (3)} conversion factor from arcsec to pc; {\bf (4)} weighted scheme used to produce the final datacube: "na" for natural "ro" for robust. A $\dagger$  means that we have further smoothed the original datacube to the beam indicated in Col. 5; {\bf (5)} beam major x minor axis in arcsec of the data used for the kinematic fits (see Sec. \ref{sec:sample_data}); {\bf (6)} channel separation in km/s (this value indicates also the velocity resolution of the datacube in terms of FWHM);  {\bf (7)} total HI mass (see Sec. \ref{sec:mapd}); {\bf (8)} rms noise per channel estimated with 3DB; {\bf (9)} 3-$\sigma$ pseudo noise in the total map (See Sec. \ref{sec:maprad}). Note that DDO 87 has a slightly different distance with respect to the one reported in \protect\cite{LT} because their value is a typo (the correct value is given in \protect\citealt{typo})
}
\label{tab:sample_obs}
\end{table*}

\section{The Method} \label{sec:method}
\subsection{Tilted-ring model} \label{sec:tring}
Since \cite{tring},  the most common approach  to study the kinematics of the HI discs is the so-called `tilted-ring model'. In practice the rotating disc is broken into a series of independent circular rings with radius R, each with its kinematic and geometric properties.
The projection of each ring on the sky is an elliptical ring with semi-major axis R and semi-minor axis R $\cos i$. Throughout the paper R indicates both the radius of the circular ring and the semi-major axis of its projection.
For a given projected ring, the projected velocity along the line of sight ($\text{V}_\text{los}$) is 
\begin{equation}
\label{eq:vtitlt}
\textrm{V}_\text{los}=\textrm{V}_\text{sys} + \textrm{V}_\text{rot}(\textrm{R})  \cos \theta \sin i,
\end{equation} where $\textrm{V}_\text{sys}$ is the systemic velocity of the galaxy, $\textrm{V}_\text{rot}$ is the rotation velocity of the gas at radius R, $\theta$ is the azimuthal angle of the rings in the plane of the galaxy (related to $i$, to the galaxy centre and to the position angle; Eq. 2a and 2b in \citealt{begeman}).  The positon angle (PA, hereafter)  is defined as the angle between the north direction on the sky and the projected major axis of the receding half of the rings. 

It is worth noting that the Eq. \ref{eq:vtitlt}
is strictly valid only assuming that the gas is settled in a razor-thin disc and it describes only the circular motion of the gas. Deviations from  pure circular orbits can include radial motions due to inflow-outflow, non-circular motions due to  deviation from symmetry of the galactic potential (spiral arms, mergers, misaligned DM halo and so on; see e.g. \citealt{scarparo} and \citealt{swatlop}) and small-scale perturbations due to the star formation activity (stellar winds, SNe, see e.g. \citealt{mia}). 
The tilted-ring model allows us to trace the radial variation of the HI disc geometry and model the so-called `warp' \citep{warp,gwarp}.
Moreover, using the ring decomposition it is possible also to obtain radial profiles from the integrated 2D maps (e.g. the HI surface density, see Sec. \ref{sec:map})

In this work we measure the kinematic and geometric properties of the galaxies in our sample with the publicly available software 3DB\footnote{{\tt http://editeodoro.github.io/Bbarolo/.}}
\citep{barolo} that is a 3D code based on the tilted-ring approach.

\subsection{$^\text{3D}$BAROLO} \label{sec:3db}

3DB is a 3D method that performs a tilted-ring analysis on the whole datacube. 
In practice, for each sampling radius it builds a ring model based on Eq. \ref{eq:vtitlt}, then it calculates the residuals 
between the emission of the model and of the data  pixel-by-pixel along the ring in the datacube. The final  parameters of each ring are found through the minimisation of these residuals. Before the comparison between the datacube and the model, the latter is smoothed to the spatial and spectral instrumental resolution. This ensures full control of the observational effects and in particular a proper account of beam smearing that can strongly affect the derivation of the rotation velocities in the inner regions of dwarf galaxies (see e.g. \citealt{swaters}). 
Moreover 3DB fits, at the same time, the rotation velocity and the velocity dispersion instead of treating them as separate components as done in the classical 2D approach (e.g. \citealt{tambvdisp}, O15).

In conclusion, 3DB fits up to 8 parameters for each ring in which the galaxy is decomposed: central coordinates,  $\textrm{V}_\text{sys}$, $i$, PA, HI surface density ($\Sigma$), HI thickness ($\text{z}_d$), $\text{V}_\text{rot}$ and the velocity dispersion ($\sigma_v$). 
3DB separates the genuine HI emission from the noise by building a mask: only the pixels containing a signal above a certain threshold from the noise are taken into account. The noise estimated with 3DB for each galaxy is reported in the Col. 8 of Table \ref{tab:sample_obs}.
Further details on 3DB can be found in \cite{barolo}.

3DB works well
both on high-resolution and low-resolution data  \citep{barolo}. This allow us to use the optimal compromise between spatial resolution and sensitivity, as already discussed in Sec. \ref{sec:sample_data}.
\cite{mia} showed  that 3DB is capable of obtaining a good estimate of the kinematic parameters of mock HI datacubes corresponding to low-mass isolated dIrrs, similar to those that we will study here.

\subsubsection{Assumptions} \label{sec:ass}
As the focus of this work is  the kinematics ($V_\text{rot}$ and $\sigma_v$) of the HI discs, we can consider the other 6 variables fitted by 3DB as ``nuisance parameters''. 
We decided to reduce the relatively high number of parameters making assumptions on the HI surface density, HI scale height, systemic velocity and the position of the galactic centre.
\begin{itemize}
\item{\textit{HI surface density:}} we remove the surface density from the list of the free parameters by normalising the HI flux. 3DB implements two different normalisation techniques: pixel-by-pixel  or azimuthally averaged. In the first case the model is normalised  locally to the value of the total HI map, while in the second case the model is normalised to the azimuthally-averaged flux in each ring.
For a full description of these normalisation techniques, see the 3DB reference paper (\citealt{barolo}). We decided to use the local pixel-by-pixel option, beacuse this approach is convenient to take into account the asymmetry of the HI distribution and to avoid that regions with peculiar emission  (e.g. clumps or holes) affect excessively the global fit (see \citealt{lellinorm} and \citealt{lellinorm2}).
However, we checked that the results obtained with the two normalisations are fully compatible within the errors.

\item{\textit{HI scale height:}} 3DB includes the possibility to fit the HI scale 
height. This is something exclusive to 3D methods since there is no information about
the scale height in the integrated 2D velocity field. However, the assumption of a thick
disc is somewhat inconsistent with the `tilted-ring model' (see Sec. \ref{sec:tring}):
in the presence of a thick disc  along the line of sight we are accumulating  emission
form different rings, hence a ring-by-ring analysis can not incorporate this effect.
To be fully consistent, we should set the value of the scale height $\text{z}_\text{d}$ to 0: however, it turns out that, assuming $\text{z}_\text{d}=0$ the galactic models made with 3DB have a too sharp cut-off at the border of rings. Therefore, we 
decided to set for all the galaxies a scale height of 100 pc, constant in radius. In Sec. \ref{sec:hscale} we discuss in detail the effect of this assumption.
\item{\textit{Systemic velocity:}} we fix the systemic velocity for all rings to the value calculated as
\begin{equation}
\text{V}_\text{sys}=0.5(\text{V20}_\text{app} + \text{V20}_\text{rec}),
\label{eq:vsys}
\end{equation}
where \text{V20} is the velocity where the flux of the global HI profile is 20\% of the flux peak, while `app' and `rec' indicate the approaching and the receding halves of the galaxy. 
\item{\textit{Centre of the galaxy:} We fix the position of the galactic centre for all rings to the value found with one of the methods described in Appendix \ref{sec:iguess}.}
\end{itemize}

\section{Data analysis} \label{sec:datan}
\subsection{HI total map} \label{sec:map}
Before starting the kinematic fit procedure, for each galaxy we produce HI total maps from the datacubes. This step is needed both to
have an initial rough estimate of the geometrical properties of the HI discs and to define the maximum radius to use in the kinematic fit. Moreover, we use it to extract the HI surface-density profiles using the best-fit parameters of the disc (Sec. \ref{sec:tstage}). The HI surface-density profile is crucial to correct the estimated rotation curve for the asymmetric drift (Sec. \ref{sec:asy}).

\subsubsection{HI spatial distribution} \label{sec:mapd}
Integrating the signal of each pixel along the spectral axis one obtains a 2D total map that represents the spatial distribution (on the sky) of the HI emission.
To build these maps, we masked the original datacube with the following procedure: first we smoothed the original 
data to obtain a low-resolution cube (between 2 and 3 times the original beam, depending on the galaxy); then, a pixel is included in the mask only if its emission is above a certain threshold
(2.5 of the noise  $\sigma_\text{ch}$ of the smoothed cube) in at least three adjacent channels.  Finally we sum, along the spectral axis,  all the flux of the pixel included in the mask obtaining a 2D map. This method produces a
very clean  final map with only a small contribution from the noise.

Following \cite{roberts}, we can directly relate the intensity of the radio emission ($S$) to the projected surface-density of the gas ($\Sigma_\text{obs}$) as
\begin{equation}
\frac{\Sigma_\text{obs}(x,y)}{\text{M}_\odot/\text{pc}^2}= 8794 \left( \frac{S(x,y)}{\text{Jy/Beam}} \right)   \left( \frac{\Delta
V}{\text{km/s}}  \right) \left( \frac{B_\text{maj}
B_\text{min}}{\text{arcsec}^2} \right)^{-1},
\end{equation} where $B_\text{maj}$ and $B_\text{min}$ are the FWHM of the major and minor axis of the beam and $\Delta V$ is the channel separation of the datacube.
We obtained the radial profile of $\Sigma_\text{obs}$  averaging the  total map along elliptical rings defined by the disc geometrical parameters (see Sec. \ref{sec:tstage}). The errors on $\Sigma_\text{obs}$ are calculated as the standard deviation of the averaged values\footnote{ Pixels in datacubes are correlated by the convolution  with the instrumental beam occurring during the observations. 
As a consequence, the estimated errors are an underestimate of the real uncertainties (see \protect\citealt{sick}). However, the estimated errors for the surface density are already very large and we did not correct them. Moreover, the elliptical rings are usually much larger than the beam, so the correlation of the data is not expected to affect significantly the final estimate of the average and the error of the surface density.}.
The intrinsic surface density ($\Sigma_\text{int}$) is defined as the density of the HI disc integrated along the vertical axis of the galaxy. Assuming a razor-thin HI disc, the relation between $\Sigma_\text{obs}$ and $\Sigma_\text{int}$ is simply
\begin{equation}
\Sigma_\textrm{int}(\text{R})=\Sigma_\text{obs}(\text{R}) \cos i,
\label{eq:cosinc}
\end{equation}  because the area of the projected ring is smaller than the one of the intrinsic ring by a factor $\cos i$.
Finally, we can have a measure of the total mass of the HI disc summing the observed surface densities of all pixels in the total map multiplied by the physical area of the pixels
\begin{equation}
\frac{\text{M}_\text{HI}}{\text{M}_\odot}=23.5  \left( \frac{\delta}{\text{arcsec}} \right)^2  \left( 
\frac{D}{\text{Mpc}} \right)^2  \sum_\text{pixels} \left( \frac{\Sigma_\text{obs} (x,y)}{\text{M}_\odot \text{pc}^{-2}} \right),
\label{eq:mass}
\end{equation} where $\delta$ is the size of the pixel and $D$ is the distance of the galaxy. To have an estimate of the uncertainties in this measure we have built two other maps: a low-resolution map using the smoothed cube that we used to build the mask and an extremely low-resolution map smoothing again the cube to a beam of 40-60 arcsec. We applied the primary beam correction to the total HI map using the task PBCORR of GIPSY \protect\citep{gipsy}. For each map the mass is estimated using Eq. \ref{eq:mass} and the method described above.
In particular, we take as measure of the disc mass the average and the associated error the standard deviation of the values obtained in the three maps.
The final estimates of $\text{M}_\text{HI}$ are listed in 
Table \ref{tab:sample_obs} (Col. 7). The intrinsic surface-density profile and the HI map contours are shown respectively  in the box B (bottom right panel) and in the box C (left panel) of the summary plot of each galaxy (Sec. \ref{sec:gls}). 

\begin{figure} 
 \centering 
 \includegraphics[width=1.0\columnwidth]{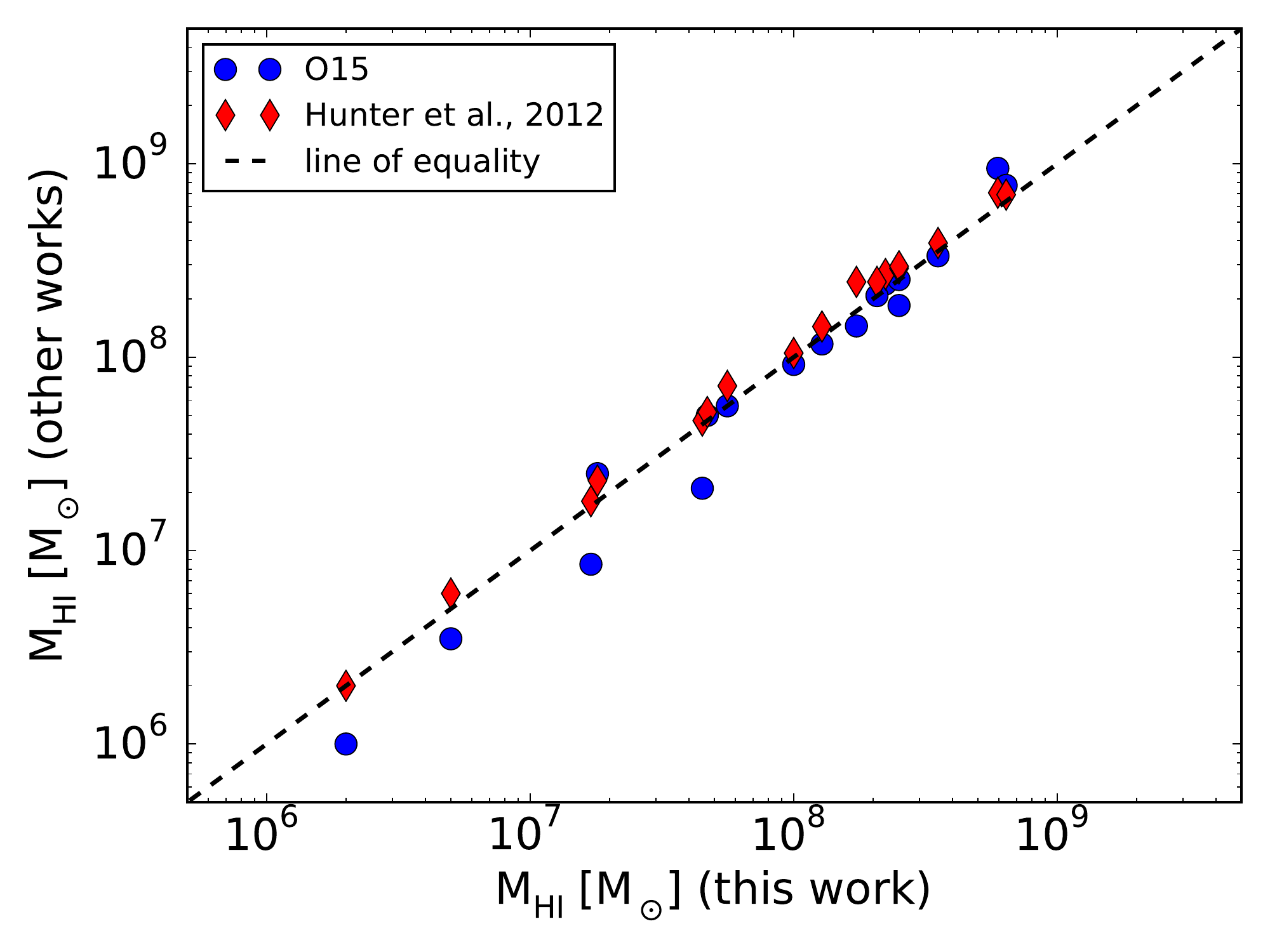}
 \caption{Comparison of the HI mass estimated in this work and in other works (O15, blu circles;  \protect\cite{LT}, red diamonds.)}
\label{fig:compmass}
\end{figure}

Figure \ref{fig:compmass} shows the comparison between the mass estimated with our method and those estimated by O15 and \protect\cite{LT}. The relative difference  between our measures and those of O15 is on average 20$\%$, with the largest discrepancies (up to 50$\%$) found at the low-mass end; these differences are mainly due to the different method used to estimate the total mass. The method used in \protect\cite{LT} is similar to the one used in this work (\citealt{LT} used only robust datacubes while we used both robust and natural datacubes; see Col. 4 in Tab. \ref{tab:sample_obs}): in this case the average relative difference is about 10$\%$.
Because of the data-reduction methods used in \protect\cite{LT}, the natural datacubes have systematically lower flux with respect to the robust ones (see \protect\citealt{LT} for the details), therefore our values are slightly lower.

\subsubsection{Noise in the total map and maximum radius} \label{sec:maprad}
In order to define a maximum radius for the HI disc, we need to calculate the noise level in the total map. If we sum N 
independent channels each with noise $\sigma_\text{ch}$ the final noise will be equal to $\sigma_\text{ch}\sqrt{N}$. However, using a mask, the number of summed 
channels is different from pixel to pixel, moreover the datacubes in our sample have been Hanning-smoothed and 
adjacent channels are not independent (see \protect\citealt{Vnoise}). To have a final estimate of the noise in the total map we follow the approach of 
\protect\cite{Vnoise} and \protect\cite{lnoise}: 
we constructed a signal-to-noise map and we defined as the 3-$\sigma$ pseudo level
($\sigma_\text{3T}$) the mean value of the pixels with a S/N between 2.75 and 3.25.
We fit our kinematic model only to the portion of the datacube within  the contour defined by $\sigma_\text{3T}$ and avoiding to use rings which do not intercept, around the major axis, HI emission coming from the disc. This ensures a robust estimate of the 
galactic kinematics avoiding regions of the galaxy  dominated by the noise and with poor information about  the gas rotation. The 3-$\sigma$ pseudo noise levels $\sigma_\text{3T}$ are listed in Table \ref{tab:sample_obs} (Col. 9), while the maximum radii $\textrm{R}_\text{max}$ used in the fit are reported in Table \ref{tab:sample_res} (Cols. 1-2).
\subsection{Datacube fit} \label{sec:tstage}

Given the assumptions we made (Sec. \ref{sec:ass}), we are left with two geometrical ($i$, PA) and two kinematic (V$_\text{rot}$ and $\sigma_v$) parameters.

Eq. \ref{eq:vtitlt} shows that the rotational velocity and the geometrical parameters are coupled: in particular they become degenerate for galaxies with rising rotation curves \citep{kamp1} as it is the case for most of the dIrrs. As a consequence, the fitting algorithm tends to be sensitive to the initial guesses, so it is important to initialise the fit with educated guess values. 
We estimated $i$ and PA using both the HI datacube and the optical data as explained in the Appendix \ref{sec:iguess}. The initial values for V$_\text{rot}$ and $\sigma_v$ have been set respectively to  30 km/s and 8 km/s for all the galaxies. 3DB allows to use an azimuthal weighting function $w(\theta)$ (see Eq. \ref{eq:vtitlt}) to ``weigh'' the residuals non-uniformly across the rings (see \protect\citealt{barolo}). We decided to use  $w(\theta)=\cos \theta$ to weigh most the regions around the major axis,  where most of the information on the galactic rotation lies.
If we fit at the same time the four parameters, we can obtain  rotation curves and velocity-dispersion profiles that show unphysical discontinuities due to the scatter noise of the geometrical parameters. For this reason, we run 3DB with the option {\it TWOSTAGE} (see Appendix \ref{sec:bsetup}) turned on: first a fit with four free parameters is obtained, then the geometrical parameters are regularised with a polynomial, and finally a new fit of only V$_\text{rot}$ and $\sigma_v$ is obtained.
We chose to use the lowest polynomial order allowed by the data: in practice, we set the geometrical parameters to a constant value, unless there is a clear evidence of radial trends of $i$ and/or of PA as for example in DDO 133  and in DDO 154. We assessed the existence of a radial trend and set the degree of the polynomial through the visual inspection of the radial profile of $i$ and/or PA, and also of the HI map,  of the velocity field and of the datacube channels.

\begin{figure*} 
 \centering 
 \includegraphics[width=0.94\textwidth]{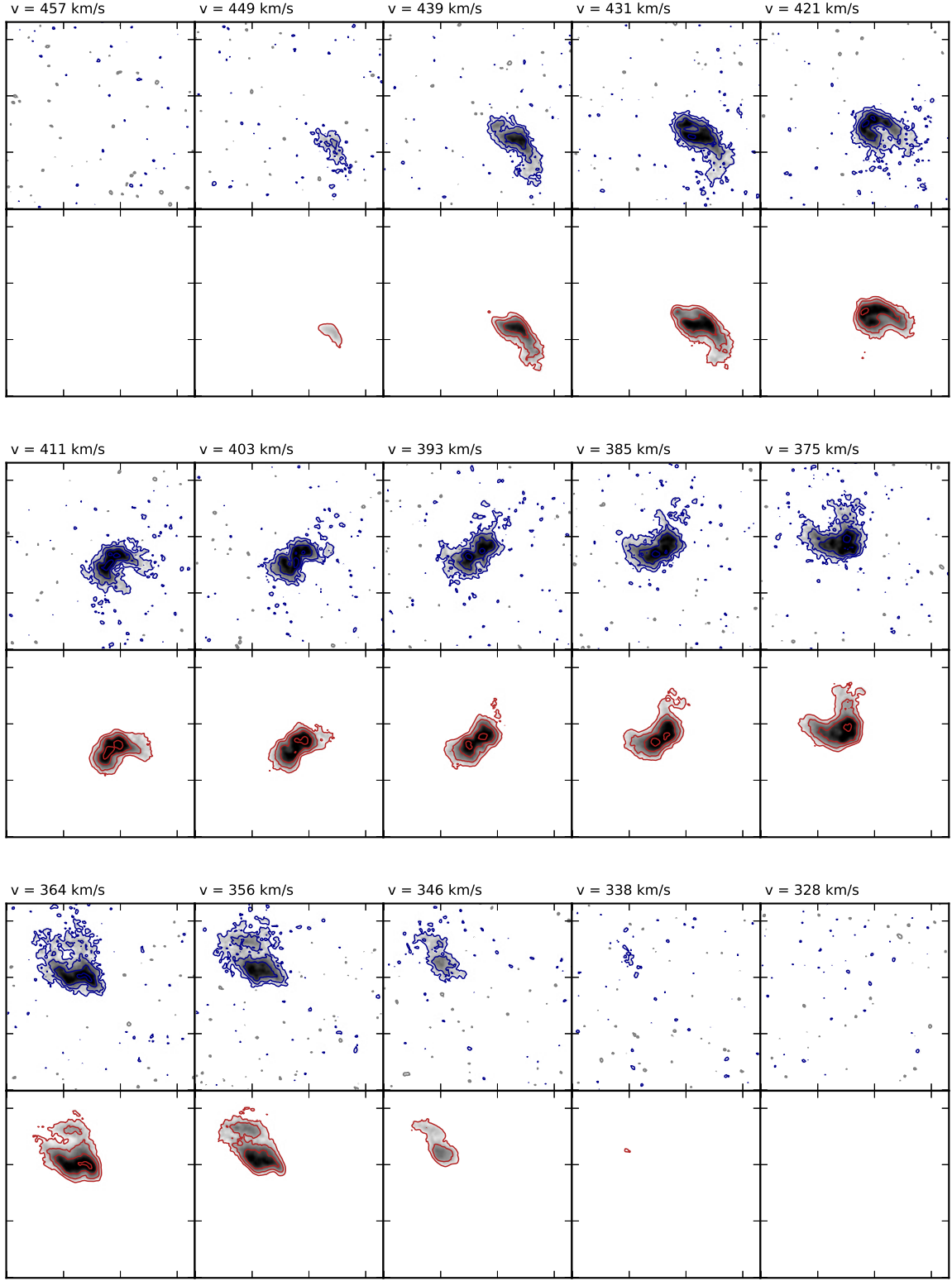}
 \caption{Channel maps showing the HI emission for DDO 154: data (top panels, blue contours) and the best-fit model found with 3DB (bottom panels, red contours). The velocity of each channel is reported above the panels. The minimum contour is 3 $\sigma_\text{ch}$ (Col. 8 in Tab. \ref{tab:sample_obs}) and  adjacent contours differ by a factor of 2 in emissivity; the grey contours in the the data channels indicate the emission at -3 $\sigma_\text{ch}$. The size of the field is about 14$^\prime$ x 14$^\prime$.}
\label{fig:chan154}
\end{figure*}

\begin{figure*} 
 \centering 
 \includegraphics[width=0.94\textwidth]{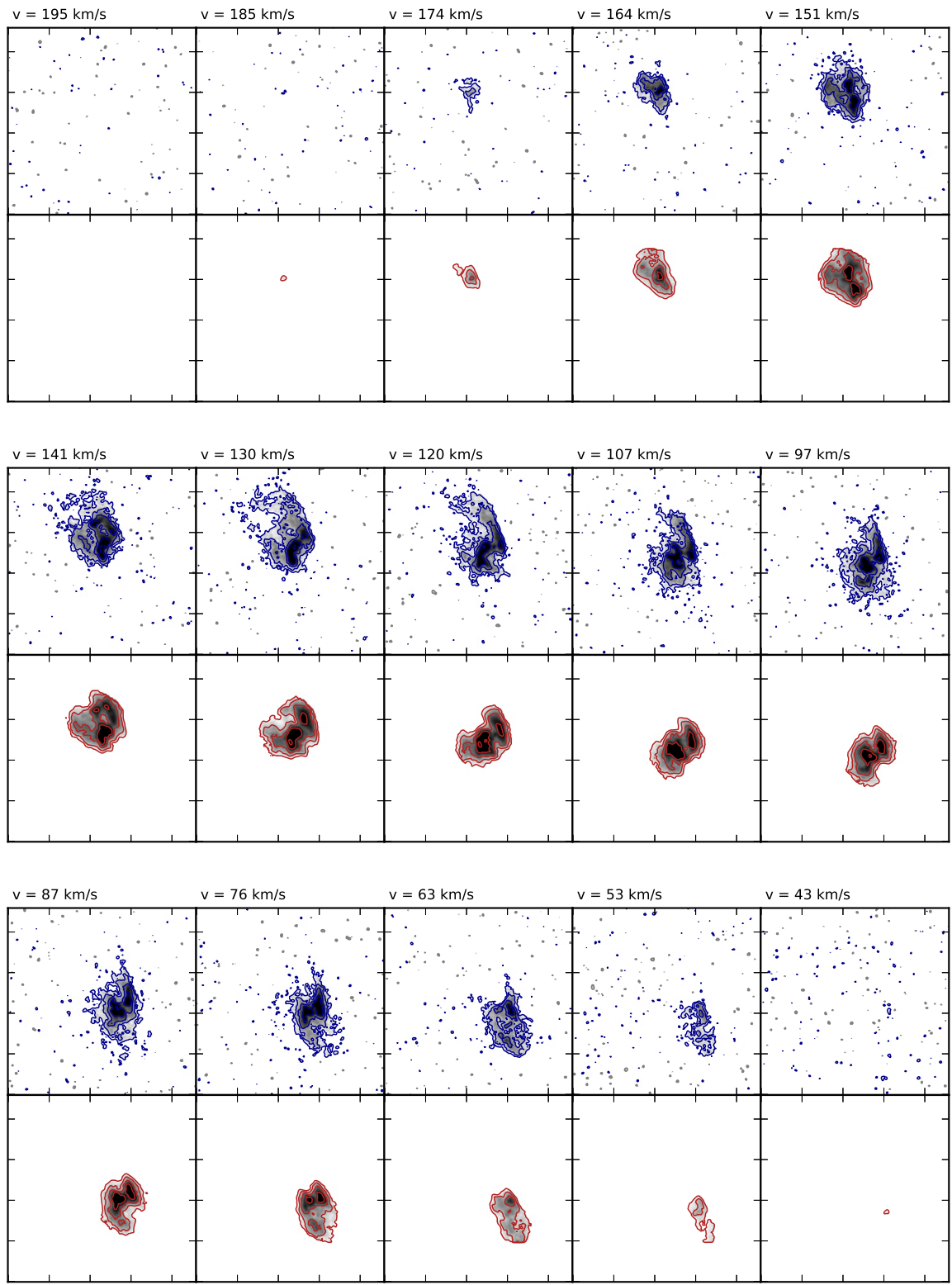}
 \caption{ Same as Fig. \ref{fig:chan154} but for NGC 2366. The size of the field is about 20' x 20'. }
\label{fig:chan2366}
\end{figure*}

The result of the fit is compared by-eye with the real datacube both analysing the position-velocity diagrams (PV, hereafter) along the major axis and the minor axis, and
channel by channel. Figures \ref{fig:chan154} and \ref{fig:chan2366} show the comparison in the channel maps between the observations and the best-fit model found with 3DB for DDO 154 and NGC 2366: the emission of DDO 154 is reproduced quite well; the best-fit model found for NGC 2366 traces the global kinematics but it is not able to reproduce the extended feature in the North-West of the galaxy (see e.g. channel at 120 km/s). The presence of such kind of peculiarities  makes the study of the dIrrs HI disc kinematics challenging and requires careful visual inspection. Whenever we notice that the model was not a good representation of the data,  we restart the fit changing the  initial guesses and/or varying the order of the
polynomial regularisation.

Concerning the errors on the estimated parameters, it is worth spending a few words about the differences between  the standard 2D approach and 3DB. In the first case the errors are a combination of (i) nominal errors coming from the tilted ring fit (typically using a Levenberg-Marquardt algorithm) and (ii) the differences between the fit of the approaching and receding sides. It is well know that errors (i) are very small and underestimate the real uncertainties: so the final errors are dominated by (ii) and  the definition of this `asymmetric-error' is somewhat arbitrary. 3DB uses a Monte-Carlo method to estimate the errors further exploring the space parameters around the best fit solutions (see \citealt{barolo} for further details), therefore our quoted errors should be a robust and statistically significant measure of the uncertainties of the kinematic parameters.

Given that we use the errors estimated with 3DB, we did not fit separately the approaching and receding halves of the galaxy as often done in the classical 2D methods. The only exception is DDO 126: in this case the best-fit model found with 3DB using the whole galaxy does not give a good representation of the datacube because of the strong kinematic asymmetries (see Sec. \ref{sec:gls}).

\subsection{The final circular velocity}
Assuming that the gas is in equilibrium,
the  rotation velocity of the gas (V$_\textrm{rot}$) can be related to the galactic gravitational potential  ($\Phi$)
through the radial component of the momentum equation  
\begin{equation}
\frac{1}{\rho}\frac{\partial \rho \sigma_v^2}{\partial \textrm{R}}=-\frac{\partial \Phi}{\partial \textrm{R}} + \frac{\textrm{V}^2_\textrm{rot}}{\textrm{R}},
\label{eq:asy_ps}
\end{equation} where $\rho$ is the volumetric density of the gas and $\sigma_v$ is the velocity dispersion.
Therefore, the observed V$_\textrm{rot}$  is not a direct tracer of the galactic potential when the pressure support term ($\rho \sigma^2_v$) due to the random gas motions is non-negligible.

\subsubsection{Asymmetric-drift correction} \label{sec:asy}
Eq. \ref{eq:asy_ps} can be re-written as
\begin{equation}
\textrm{V}^2_\textrm{c}-\textrm{V}^2_{\textrm{rot}}=-\frac{\textrm{R}}{\rho} \frac{\partial \rho \sigma_v^2}{\partial \textrm{R}}= - \textrm{R}\sigma_v^2 \frac{\partial \ln(\rho \sigma_v^2)}{\partial \textrm{R}}=\textrm{V}^2_\textrm{A},
\label{eq:asy_asy}
\end{equation} where $\textrm{V}_\textrm{c}=\sqrt{\textrm{R}\frac{\partial \Phi}{\partial \textrm{R}}}$ is the circular velocity and V$_\textrm{A}$ is the asymmetric drift, which becomes increasingly important at large radii and is heavily dependent on the value of the velocity dispersion. When the rotation velocity is much higher than the velocity dispersion the asymmetric-drift correction is negligible. This is the case in spiral galaxies where usually V$_\textrm{rot}/\sigma_v>10$  (e.g. \citealt{RcT}). Instead,  for several dIrrs the rotation velocity is comparable to the velocity dispersion, making the asymmetric-drift correction indispensable for an unbiased estimate of the galactic potential. 

The volumetric density in the equatorial plane (z=0) is proportional to  the ratio of the intrinsic surface density and the vertical scale height (z$_d$), therefore the asymmetric drift $\textrm{V}^2_\textrm{A}$ is given by
\begin{equation}
\textrm{V}^2_\textrm{A}=-\textrm{R}\sigma_v^2 \frac{\partial \ln \left( \sigma_v^2 \Sigma_\textrm{int} \textrm{z}_d^{-1} \right)}{\partial \textrm{R}}.
\label{eq:asy_asyqfin}
\end{equation}
We can ignore the term with the radial derivative of  $\textrm{z}_d$  assuming that the thickness of the gaseous layer is indipendent of radius (but see Sec.\ref{sec:hscale}). Furthermore, assuming that the HI disc is thin, the ratio between the intrinsic and the observed surface density is just the cosine of $i$ (Eq. \ref{eq:cosinc}).  Under these assumptions, from Eq. \ref{eq:asy_asyqfin}  we derive the classical formulation of the asymmetric-drift correction (see e.g. O15):
\begin{equation}
\textrm{V}^2_\textrm{A}=-\textrm{R}\sigma_v^2 \frac{\partial \ln \left( \sigma_v^2 \Sigma_\textrm{obs} \cos i \right) }{\partial \textrm{R}}.
\label{eq:asy_asyfin}
\end{equation}
Except for DDO 168, NGC 1569 and UGC 8508, all the analysed galaxies have a constant $i$ and the cosine in Eq. \ref{eq:asy_asyfin} can also be ignored.

\subsubsection{Application to real data} \label{sec:asy_data}
Fluctuations of the observed surface density and of the measured velocity dispersion at similar radii can have dramatic effects on the numerical calculation of the radial derivative in Eq. \ref{eq:asy_asyfin}. As a consequence, the final asymmetric-drift correction can be very scattered causing abrupt variations in the final estimate of the circular velocity. For this reason, we decided to use functional forms to describe both the 
velocity dispersion and the argument of the logarithm in Eq. \ref{eq:asy_asyfin}.
The velocity dispersion is regularised with a polynomial $\sigma_p(\textrm{R},n_p)$  with degree $n_p$ lower than 3. If there is not a clear radial trend, we consider a fixed velocity dispersion taking the median of $\sigma_v(R)$. The radial variation of the velocity dispersion is usually small, therefore the radial trend of $\Sigma_\textrm{int} \sigma_v^2$ is dominated by the behaviour of the surface density: this  falls off exponentially at large radii, while in the centre it is almost constant or  it shows an inner depression.
In analogy with \cite{Bureauasy} we chose to fit $\Sigma_\textrm{int} \sigma_v^2$ with the function
\begin{equation}
f(\textrm{R})=f_0\left( \frac{\text{R}_c}{\textrm{arcsec}}+1 \right) \left( \frac{\text{R}_c}{\textrm{arcsec}}+e^\frac{\textrm{R}}{\textrm{R}_d} \right)^{-1},
\label{eq:asy_fform}
\end{equation} where $f_0$ is a normalisation coefficient,
and $\textrm{R}_c$ and $\textrm{R}_d$ are characteristic radii. The function $f$ is characterised by an exponential decline at large radii and by an inner core almost equal to $f_0$. \cite{mia} showed that the Eq. \ref{eq:asy_fform} is a good compromise between a pure exponential that  overestimates the asymmetric-drift correction in the inner radii, and a functional form with an inner depression that can produce unphysical 
negative values of V$^2_\textrm{rot}$.
Combining Eq. \ref{eq:asy_fform} and Eq. \ref{eq:asy_asyfin} we can write the asymmetric-drift correction as
\begin{equation}
\textrm{V}^2_\textrm{A}= \textrm{R} \frac{\sigma^2_p(\textrm{R},n_p) e^\frac{\textrm{R}}{\textrm{R}_d}}{\textrm{R}_d} \left( \frac{\text{R}_c}{\textrm{arcsec}}+ e^\frac{\textrm{R}}{\textrm{R}_d}  \right)^{-1}.
\label{eq:asy_smooth}
\end{equation}

\subsubsection{Error estimates} \label{sec:asy_err}
We can calculate the final errors on  $\textrm{V}_\textrm{c}$ by applying the propagation of errors to Eq. \ref{eq:asy_asy} to obtain
\begin{equation}
\delta_\textrm{c}=\frac{\sqrt{\textrm{V}^2_\textrm{rot} \delta^2_\textrm{rot} +  \textrm{V}^2_\textrm{A} \delta^2_\textrm{A}} }{\textrm{V}_\textrm{c}},
\label{eq:asy_sigmac}
\end{equation} where $\delta_\textrm{rot}$ is the error found with 3DB and $\delta_\textrm{A}$ is the uncertainty associated with the estimated values of the asymmetric-drift term.
Usually $\delta_A$ is simply ignored (e.g. \citealt{Bureauasy}; O15), thus the uncertainties on the final corrected rotation curve are equal to $\delta_\textrm{rot}$. This is a reasonable assumption if the rotational terms
in Eq. \ref{eq:asy_sigmac} are dominant, but for some galaxies in our sample (e.g. DDO 210) the final circular rotation curve is heavily dependent on the asymmetric-drift correction. 
In these cases it is important that $\delta_\textrm{c}$ includes the uncertainties introduced by the operations described in Sec. \ref{sec:asy_data}.
We decided to estimate $\delta_\textrm{c}$ with a Monte-Carlo approach:
\begin{enumerate}
\item First we make $N$ realisations of the radial profile of both the velocity dispersion and of the surface density. For each sampling radius R the values of a single  realisation $\sigma^i_v(\textrm{R})$ or $\Sigma^i_\textrm{int}(\textrm{R})$ are extracted randomly from a normal distribution with the centre and the dispersion taken respectively from the values and errors of the parent populations.
\item For each of the $N$ realisations we apply the method described in Sec. \ref{sec:asy_data} to obtain the asymmetric-drift correction at each sampling radius $\textrm{V}^i_\textrm{A}(\textrm{R})$.
\item The final asymmetric-drift correction is calculated as $\textrm{V}_\textrm{A}(\textrm{R})=\textrm{median}_i(\textrm{V}^i_\textrm{A}(\textrm{R}))$, where each $\textrm{V}^i_\textrm{A}$ is
obtained with the Eq. \ref{eq:asy_smooth}. The associated errors as $\delta_\textrm{A}(\textrm{R})=K\times\textrm{MAD}(\textrm{V}^i_\textrm{A}(\textrm{R}))$ where the MAD is the median absolute deviation around the median. The factor $K$ links the MAD with the standard deviation of the sample ($K\approx1.48$ for a normal distribution). We chose to use the median and the MAD because they are less biased by the presence of outliers with respect to the mean and the standard deviation. 
\end{enumerate} We found that $N=1000$ is enough to obtain a good description of the error introduced by the asymmetric-drift correction.

\subsubsection{Final notes}
\label{sec:fnotes}

The method described in the above sections has been built to make the asymmetric-drift correction terms as smooth as possible. This ensures that our final rotation curves are not affected by non-physical noise related to the derivative term in Eq. \ref{eq:asy_asyfin}. The drawback of this approach is that intrinsic scatter on $\textrm{V}_\textrm{A}$ is hidden and the final errors $\delta_\textrm{A}$ could be several times larger than the  point-to-point scatter. However, we are confident that  the quoted values of $\delta_\textrm{A}$ truly trace the degree of uncertainties introduced by the asymmetric-drift correction at different radii.
In galaxies where the final circular velocity is totally dependent on the asymmetric-drift correction (e.g. DDO 210, see Sec. \ref{sec:gls} and Fig. \ref{fig:DDO210}) the scatter in $\textrm{V}_\textrm{rot}$ could be smoothed out, but  in these cases also the final errors are dominated by the errors on the asymmetric-drift and $\delta_\textrm{c}$ is still a good representation of the global uncertainties.

In some cases the velocity dispersion found with 3DB is not well constrained, so there are galaxies where at some radii the $\sigma_v$ is discrepant or peculiar (e.g. very small error) with respect to the global trend. In general, the presence of a single `rogue' $\sigma_v$  (e.g. DDO 87, Fig. \ref{fig:DDO87} or DDO 133, Fig. \ref{fig:DDO133}) does not have a significant influence on the estimate of the $\textrm{V}_\textrm{A}$ and on the median of $\sigma_v$ (see Tab. \ref{tab:sample_res}). However, if the
discrepant $\sigma_v$ is in a peculiar position (e.g. the last radius in DDO 210, Fig. \ref{fig:DDO210}) or the
region with `rogues' is extended by more than one ring (e.g. DDO 216,  empty circles in Fig. \ref{fig:DDO216}), the correction for the asymmetric drift  and the estimate of the circular velocities are biased. In these cases we exclude the radii with discrepant $\sigma_v$ both from the calculation of the asymmetric-drift correction and from the calculation of the median.


\section{Results} \label{sec:gls}

\begin{table*}
\centering
\tabcolsep=0.11cm
\begin{tabular}{l|ccc|ccccc|cccc}
\hline
\multicolumn{1}{c|}{\multirow{2}{*}{Galaxy}} & \multirow{2}{*}{\begin{tabular}[c]{@{}c@{}}R$_\text{max}$\\ (arcsec)\\ (1)\end{tabular}} & \multirow{2}{*}{\begin{tabular}[c]{@{}c@{}}R$_\text{max}$\\ (kpc)\\ (2)\end{tabular}} & \multirow{2}{*}{\begin{tabular}[c]{@{}c@{}}$\Delta$R\\ (pc)\\ (3)\end{tabular}} & \multicolumn{2}{c}{centre} & \multirow{2}{*}{\begin{tabular}[c]{@{}c@{}}V$_\text{sys}$\\ (km/s)\\ (5)\end{tabular}} & \multicolumn{1}{c}{\multirow{2}{*}{\begin{tabular}[c]{@{}c@{}}$i_\text{ini}$\\ ($^\circ$)\\ (6)\end{tabular}}} & \multicolumn{1}{c|}{\multirow{2}{*}{\begin{tabular}[c]{@{}c@{}}PA$_\text{ini}$\\ ($^\circ$)\\ (7)\end{tabular}}} & \multirow{2}{*}{\begin{tabular}[c]{@{}c@{}}V$_\text{o}$\\ (km/s)\\ (8)\end{tabular}} & \multirow{2}{*}{\begin{tabular}[c]{@{}c@{}}$<\sigma_v>$\\ (km/s)\\ (9)\end{tabular}} & \multirow{2}{*}{\begin{tabular}[c]{@{}c@{}}\textless$i$\textgreater\\ ($^\circ$)\\ (10)\end{tabular}} & \multirow{2}{*}{\begin{tabular}[c]{@{}c@{}}\textless PA \textgreater\\ ($^\circ$)\\ (11)\end{tabular}} \\
\multicolumn{1}{c|}{} &  &  &  & \begin{tabular}[c]{@{}c@{}}RA\\ (4a)\end{tabular} & \begin{tabular}[c]{@{}c@{}}DEC\\ (4b)\end{tabular} &  & \multicolumn{1}{c}{} & \multicolumn{1}{c|}{} &  &  &  &  \\ \hline
\rowcolor[HTML]{CCCCCC} 
CVnIdwA & 90 & 1.6  & 170  & 12 38 40.2$^B$ & 32 45 52$^B$ & 307.9$^B$ & $41^B$ & $55^B$ & $21.5\pm3.9$ & $7.7\pm0.5$  & $49.2\pm10.9$ & $60.6\pm18.5$  \\
DDO 47 & 210 & 5.3  & 380 & 7 41 54.6$^H$ & 16 48 10$^H$ & 272.8$^B$ & $35^H$ & $310^H$  & $62.6\pm5.2^\dagger$ & $7.9\pm0.2$ & $37.4\pm1.7^\dagger$ & $317.4\pm7.3$ \\
\rowcolor[HTML]{CCCCCC} 
DDO 50 & 390 & 6.4  & 190   & 8 19 08.7$^S$  & 70 43 25$^S$ & 156.7$^B$ & $30^T$ & $180^T$ &  $38.7\pm10.1^\dagger$ & $9.0\pm1.5$  & $33.1\pm4.6^\dagger$ & $174.9\pm5.9$ \\
DDO 52 & 108 & 5.4  & 450   & 8 28 28.5$^S$ & 41 51 21$^S$   & 396.2$^B$ & $51^S$  & $5^S$  & $51.1 \pm 6.3$  & $8.0\pm0.8$ & $55.1\pm2.9$  & 
$6.5\pm2.0$ \\
\rowcolor[HTML]{CCCCCC} 
DDO 53 & 60 & 1.1  & 110  & 8 34 08.0$^S$ & 66 10 37$^S$ &  20.4$^B$ &  $35^T$ & $120^T$ & $20.3 \pm 6.6^\dagger$ & $7.9\pm0.1$ & $37.0\pm2.0^\dagger$ & $123.4\pm3.8$  \\
DDO 87 & 144 & 5.2  & 430  & 10 49 34.7$^S$ & 65 31 46$^S$ & 338.7$^B$ & $40^T$ & $240^B$ & $50.3\pm9.1$ & $6.3\pm2.4$ & $42.7\pm7.3$  & $238.6\pm4.7$  \\
\rowcolor[HTML]{CCCCCC}
DDO 101 & 60 & 1.9  & 190  & 11 55 39.4$^S$ & 31 31 8$^S$ &  586.6$^B$ &  $49^S$ & $290^S$ & $59.2\pm3.6$    & $4.8\pm1.7$ & $52.4\pm1.7$ &  $287.4\pm3.6$ \\
DDO 126 & 140 & 3.3  & 240 & 12 27 06.3$^H$ & 37 08 23$^H$ & 214.3$^B$ &  $63^B$ & $138^B$ & $38.6\pm3.1$ & $9.1\pm1.6$ & $62.2\pm2.9$ & $140.7\pm3.5$  \\
\rowcolor[HTML]{CCCCCC}
DDO 133 & 165 & 2.8   & 190   & 12 32 55.4$^S$ & 31 32 14$^S$ & 331.3$^B$ & $38^B$ & $20^B$ & $47.2\pm5.1^\dagger$ & $8.1\pm0.7$  & $38.9\pm3.7^\dagger$  & $-3.8\pm8.9$  \\
DDO 154 & 390 & 7.0  & 270   & 12 54 06.2$^S$ & 27 09 02$^S$ & 375.2$^B$ &  $65^S$ & $224^S$  & $47.1\pm5.1$ & $8.5\pm0.9$  & $67.9\pm1.1$ &  $226.1\pm2.6$  \\
\rowcolor[HTML]{CCCCCC}
DDO 168 & 225 & 4.7  & 310  & 13 14 27.9$^B$ & 45 55 24$^B$ & 191.9$^B$ &  $60^T$ & $300-270^T$ & $56.2\pm6.9$ & $8.8\pm1.3$ & $62.0\pm2.6$ & $272.7\pm4.2$ \\
DDO 210  & 100 & 0.4 & 40  & 29 46 52.0$^S$ & -12 59 51$^S$ & -140.0$^B$  &  $60^T$ & $65^T$  & $16.4\pm9.5$ & $6.2\pm0.6$ &  $63.2\pm3.2$ & $77.3\pm15.2$ \\
\rowcolor[HTML]{CCCCCC}
DDO 216 & 195 & 1.0  & 80  & 23  28 32.1$^H$ & 14 44 50$^H$ & -188.0$^T$ &  $65^H$ & $130^H$ & $13.6\pm5.5$ & $5.6\pm0.5$  & $70.0\pm5.0$ & $130.4\pm9.0$ \\
NGC 1569 & 150 & 2.5  & 250   & 4 30 49.8$^S$  & 64 50 51$^S$ & -75.6$^B$ & $61^S$ & $122^S$ & $55.6\pm22.4$ &  $21.0\pm4.0$ & $67.0\pm5.6$ & $114.6\pm4.0$ \\
\rowcolor[HTML]{CCCCCC}
NGC 2366 & 384 & 6.3  & 260  & 7 28 48.8$^S$ & 69 12 22$^S$ & 100.8$^B$  & $65^B$ & $35^B$ & $57.7 \pm 5.4$ &  $12.6\pm1.8$ & $65.1\pm4.2$  & $39.8\pm2.8$  \\
UGC 8508 & 110 & 1.4  & 130  & 13 39 44.9$^S$ & 54 54 29$^S$ & 59.9$^B$ & $65^H$ & $120^H$ & $33.8\pm6.4$ &  $9.1\pm1.8$ & $67.6\pm5.3$ & $123.2\pm1.7$ \\
\rowcolor[HTML]{CCCCCC}
WLM & 600 & 2.9  & 120   & 0 01 59.2$^S$ & -15 27 41$^S$ & -124.0$^B$ & $70^S$ & $178^S$ & $38.7\pm3.4$ & $7.7\pm0.8$ & $74.0\pm2.3$ & $174.0\pm3.1$  \\ \hline
\end{tabular}
\caption{ { \bf Fit assumptions:} -  
{\bf (1)} outermost radius in arcsec (see Sec. \ref{sec:maprad}); {\bf (2)} outermost radius in kpc (see Sec. \ref{sec:maprad}); {\bf (3)}  sampling radius used in the datacube fitting; {\bf (4a)(4b)} assumed galactic centre in RA (hh mm ss.s) and DEC (dd mm ss) coordinate system (J2000); {\bf (5)} assumed systemic velocity (see Sec. \ref{sec:ass}); {\bf (6)} initial guesses for $i$; {\bf (7)} initial guesses for PA. $^S$ indicates that the values is estimated from the stellar disc \citep{HVBand}, $^B$ using 3DB, $^H$ using the fit of the contours of the HI total map (see Appendix \ref{sec:iguess} for details). $^T$ indicates  peculiar cases (see the individual galactic description for details).   {\bf Results}: {\bf (8)} mean circular velocity of the outer disc  (see Sec. \ref{sec:BT});  {\bf (9)} median of the best-fit velocity dispersion found with 3DB; {\bf (10)} median of the best-fit PA found with 3DB; {\bf (11)} median of the best-fit $i$ found with 3DB.$^\dagger$  indicates nearly face-on galaxies in which the quoted errors on $<i>$ and $\text{V}_\text{o}$ could underestimate the real uncertainties.}
\label{tab:sample_res}
\end{table*}

\begin{figure*} 
 \centering 
 \includegraphics[width=0.915\textwidth]{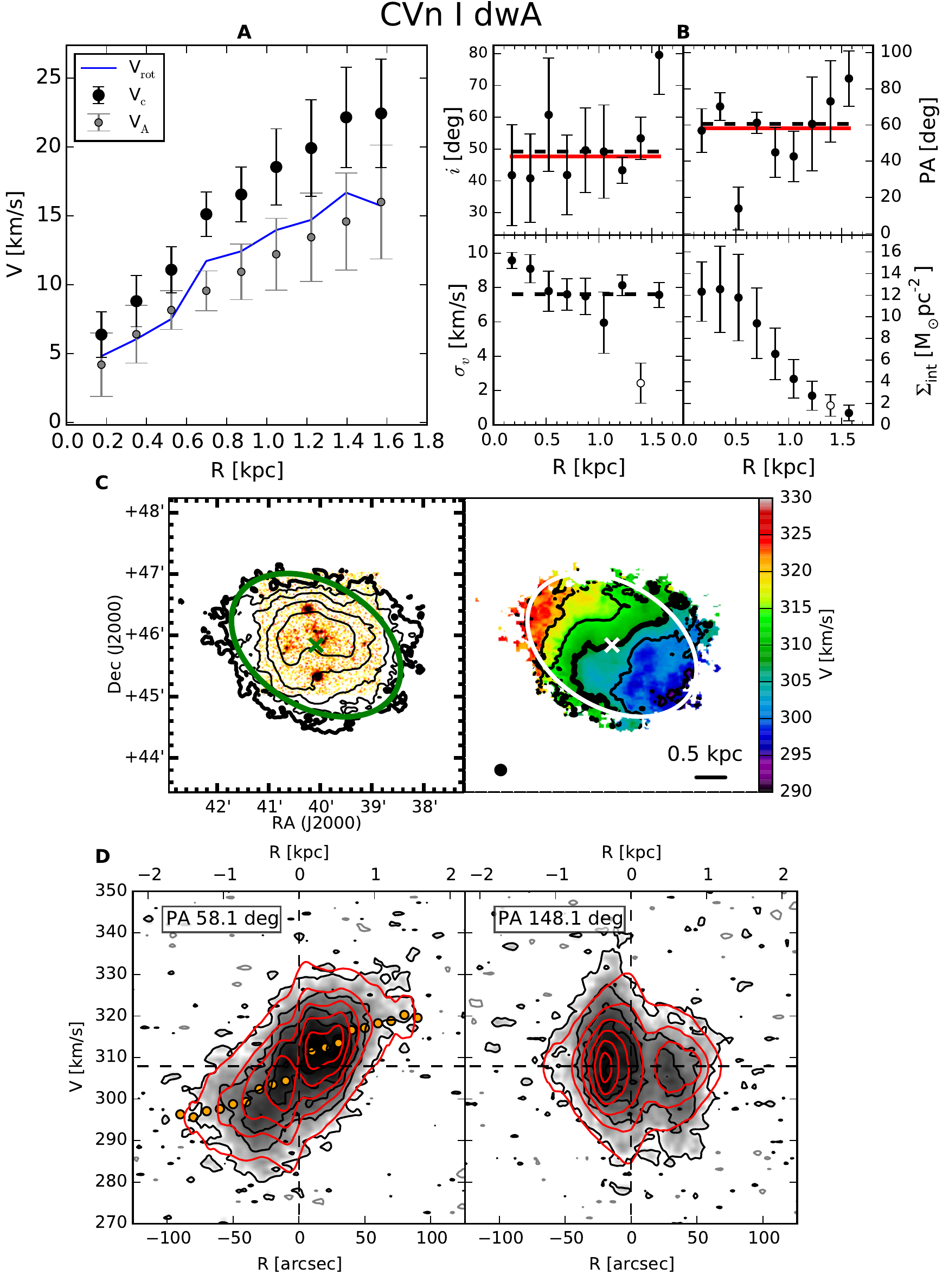}
 \caption{ See caption in Appendix \ref{sec:plot_layout}. Notes.   \textbf{C (left-hand panel):} Contours at $2^\text{n} \sigma_\text{3T}$ and $\sigma_\text{3T} = 0.96 \ \text{M}_\odot \text{pc}^{-2}$ (thick contour), stellar map in SDSS r band \protect\citep{stellarmap1}; \textbf{C (right-hand panel):} Contours at $\text{V}_{\textrm{sys}} \pm  \Delta \text{V}$ where  $\Delta \text{V}= 6.0$ km/s and $\text{V}_\text{sys}=307.9 $ km/s (thick contour). \textbf{D:} Contours at $(2+4n)\sigma_\textrm{ch}$, where $\sigma_\textrm{ch}=0.63$ mJy bm$^{-1}$, the grey contours are at -$2 \sigma_\textrm{ch}$.}
\label{fig:UA292}
\end{figure*}

\begin{figure*} 
 \centering 
 \includegraphics[width=0.915\textwidth]{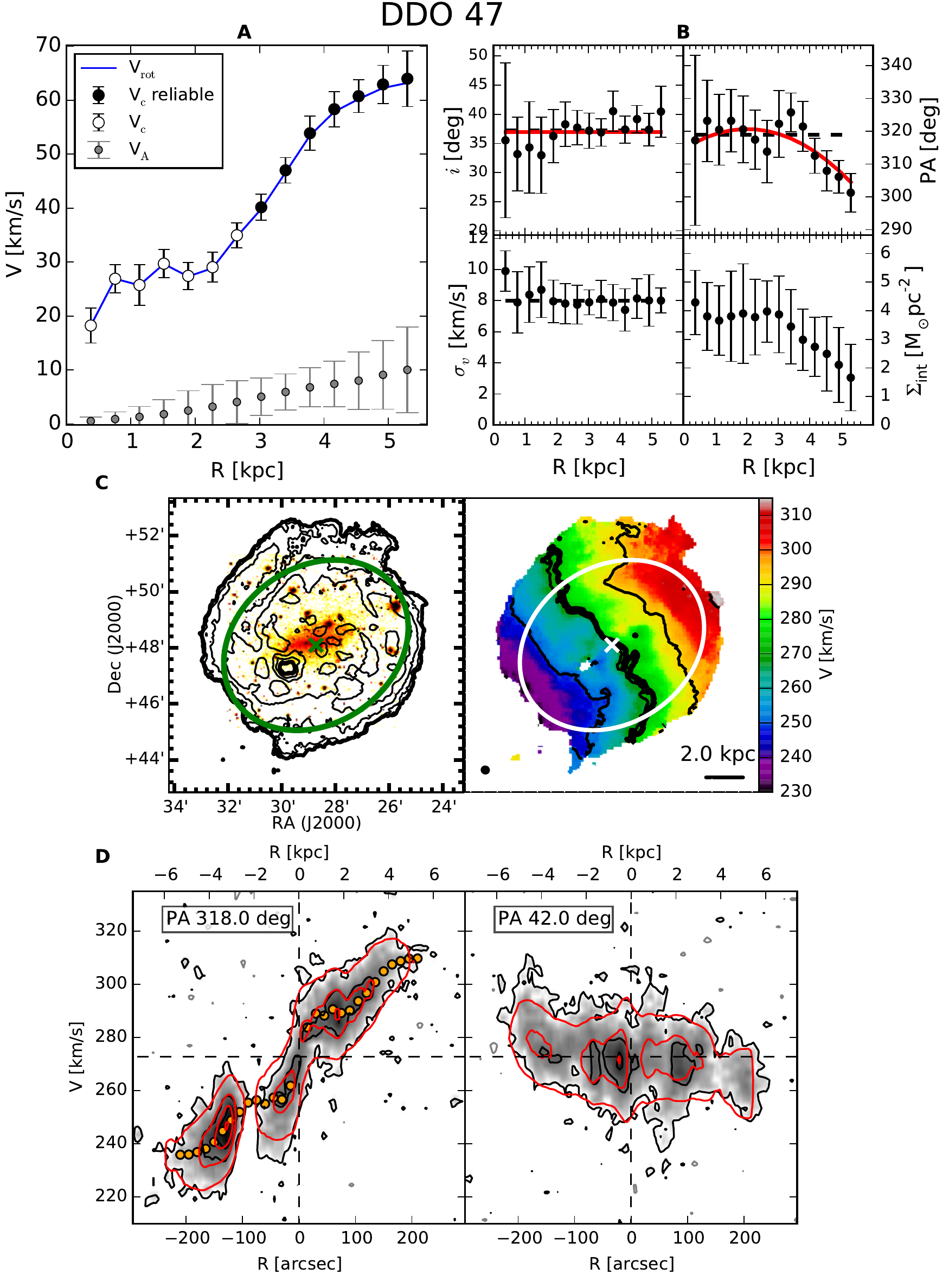}
 \caption{ See caption in Appendix \ref{sec:plot_layout}. Notes. \textbf{A:} The empty circles indicate the region in which the estimate of the rotation curve can be biased by small HI regions at anomalous velocity along the major axis (see the text for further details);  \textbf{C (left-hand panel):} Contours at $2^\text{n} \sigma_\text{3T}$ and $\sigma_\text{3T} = 0.41 \ \text{M}_\odot \text{pc}^{-2}$ (thick contour), stellar map in SDSS r band \protect\citep{stellarmap1}; \textbf{C (right-hand panel):} Contours at $\text{V}_{\textrm{sys}} \pm  \Delta \text{V}$ where  $\Delta \text{V}= 20.0$ km/s and $\text{V}_\text{sys}=272.8 $ km/s (thick contour). \textbf{D:} Contours at $(2+6n)\sigma_\textrm{ch}$, where $\sigma_\textrm{ch}=0.61$ mJy bm$^{-1}$, the grey contours are at -$2 \sigma_\textrm{ch}$. }
\label{fig:UGC3974}
\end{figure*}

\begin{figure*} 
 \centering 
 \includegraphics[width=0.915\textwidth]{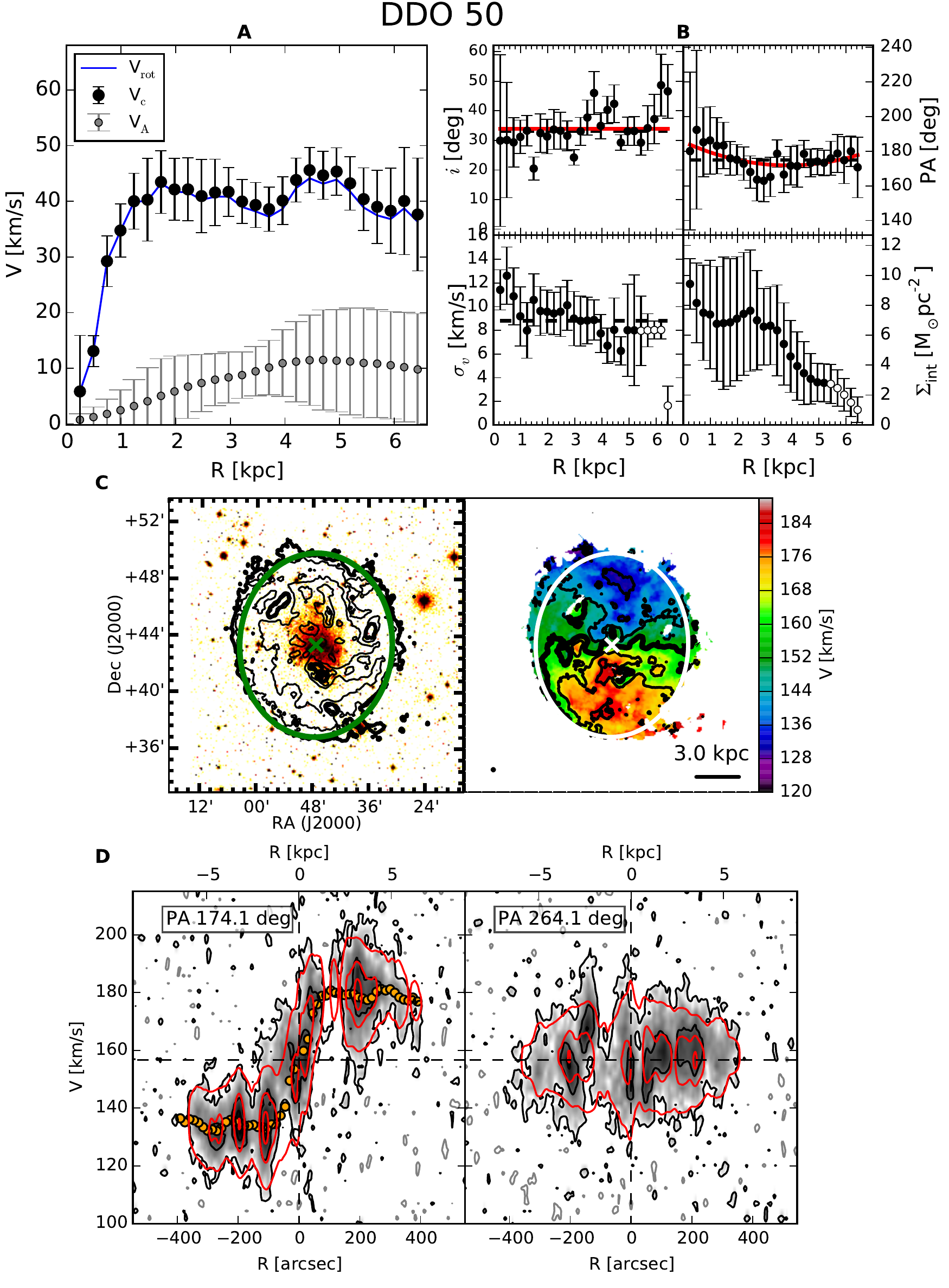}
 \caption{ See caption in Appendix \ref{sec:plot_layout}. Notes.   \textbf{C (left-hand panel):} Contours at $3^\text{n} \sigma_\text{3T}$ and $\sigma_\text{3T} = 0.65 \ \text{M}_\odot \text{pc}^{-2}$ (thick contour), stellar map in R band from \protect\cite{mapsr}; \textbf{C (right-hand panel):} Contours at $\text{V}_{\textrm{sys}} \pm  \Delta \text{V}$ where  $\Delta \text{V}= 12.0$ km/s and $\text{V}_\text{sys}=156.7 $ km/s (thick contour). \textbf{D:} Contours at $(2+7n)\sigma_\textrm{ch}$, where $\sigma_\textrm{ch}=0.82$ mJy bm$^{-1}$, the grey contours are at -$2 \sigma_\textrm{ch}$. }
\label{fig:DDO50}
\end{figure*}

\begin{figure*} 
 \centering 
 \includegraphics[width=0.915\textwidth]{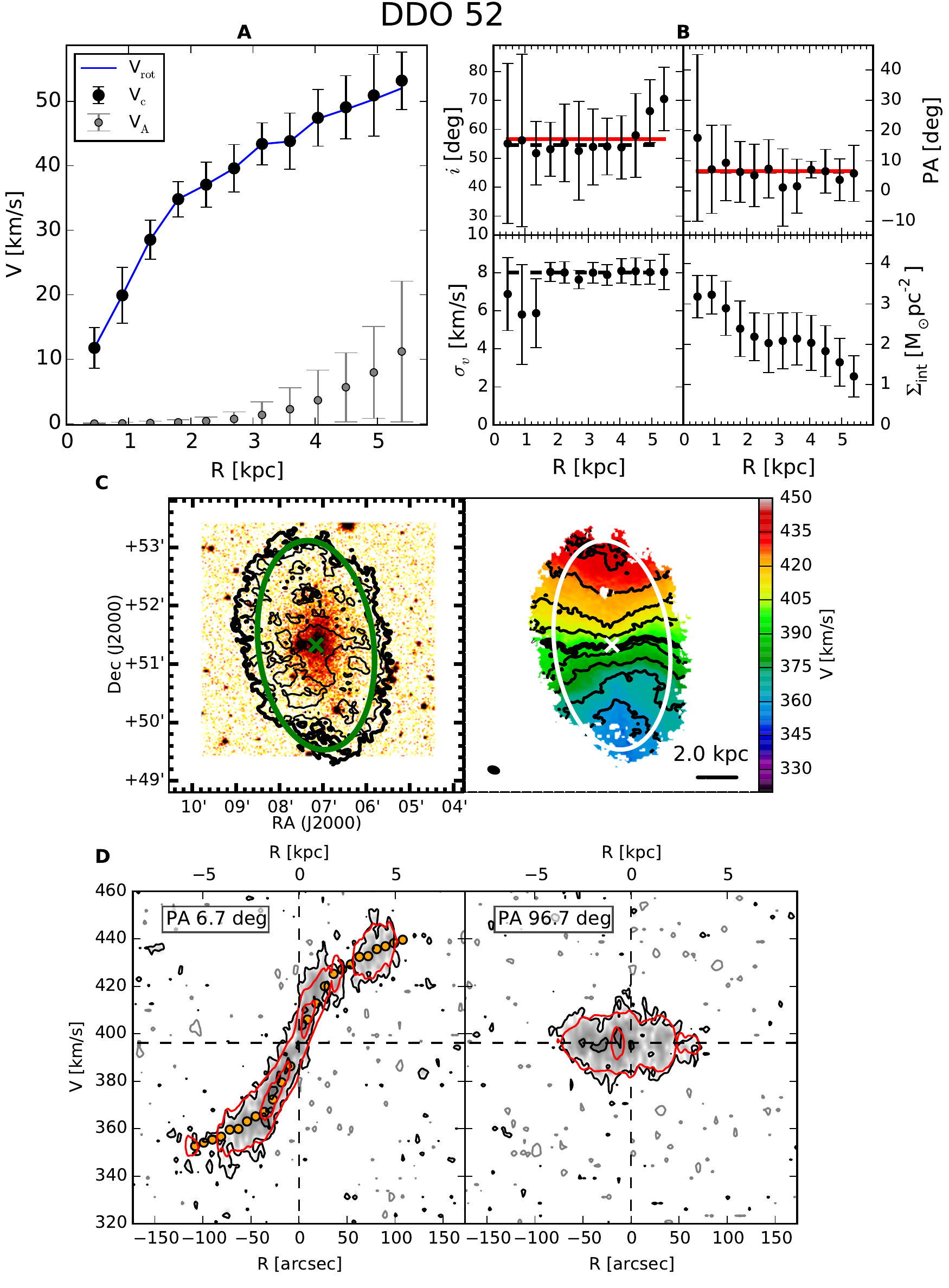}
 \caption{ See caption in Appendix \ref{sec:plot_layout}. Notes.   \textbf{C (left-hand panel):} Contours at $2^\text{n} \sigma_\text{3T}$ and $\sigma_\text{3T} = 1.20 \ \text{M}_\odot \text{pc}^{-2}$ (thick contour), stellar map in SDSS r band \protect\citep{stellarmap1}; \textbf{C (right-hand panel):} Contours at $\text{V}_{	\text{sys}} \pm  \Delta \text{V}$ where  $\Delta \text{V}= 10.0$ km/s and $\text{V}_\textrm{sys}=396.2 $ km/s (thick contour). \textbf{D:} The contours start from $(2+4n) \sigma_\textrm{ch}$, where $\sigma_\textrm{ch}=0.46$ mJy bm$^{-1}$, the grey contours are at -$2 \sigma_\textrm{ch}$. }
\label{fig:DDO52}
\end{figure*}

\begin{figure*} 
 \centering 
 \includegraphics[width=0.915\textwidth]{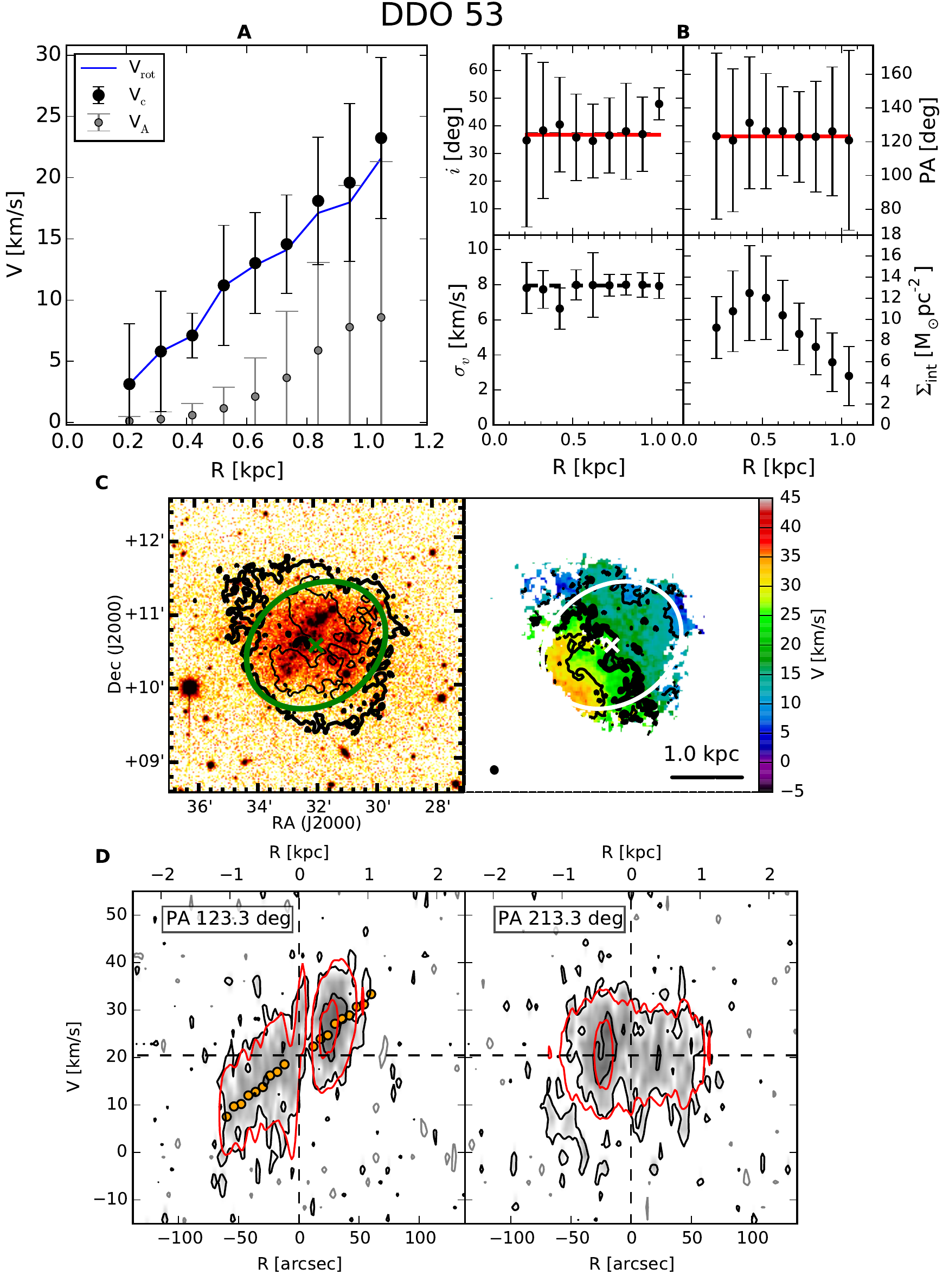}
 \caption{ See caption in Appendix \ref{sec:plot_layout}. Notes.   \textbf{C (left-hand panel):} Contours at $2^\text{n} \sigma_\text{3T}$ and $\sigma_\text{3T} = 3.80 \ \text{M}_\odot \text{pc}^{-2}$ (thick contour), stellar map in R band from \protect\cite{mapsr}; \textbf{C (right-hand panel):} Contours at $\text{V}_{\textrm{sys}} \pm  \Delta \text{V}$ where  $\Delta \text{V}= 8.0$ km/s and $\text{V}_\text{sys}=20.4 $ km/s (thick contour). \textbf{D:} Contours at $(2+4n)\sigma_\textrm{ch}$, where $\sigma_\textrm{ch}=0.57$ mJy bm$^{-1}$, the grey contours are at -$2 \sigma_\textrm{ch}$. }
\label{fig:DDO53}
\end{figure*}

\begin{figure*} 
 \centering 
 \includegraphics[width=0.915\textwidth]{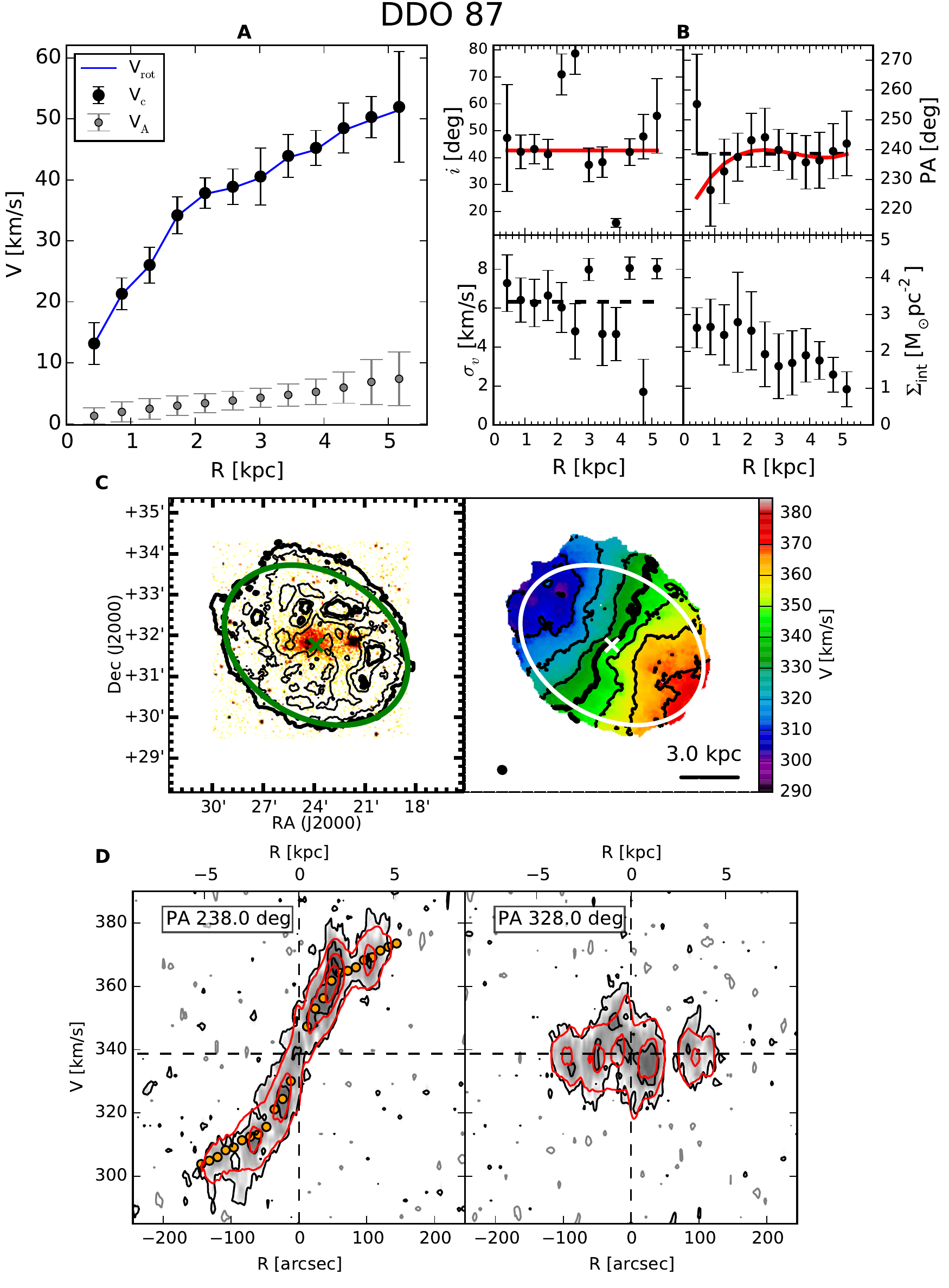}
 \caption{ See caption in Appendix \ref{sec:plot_layout}. Notes.   \textbf{C (left-hand panel):} Contours at $2^\text{n} \sigma_\text{3T}$ and $\sigma_\text{3T} = 0.90 \ \text{M}_\odot \text{pc}^{-2}$ (thick contour), stellar map in SDSS r band \protect\citep{stellarmap1}; \textbf{C (right-hand panel):} Contours at $\text{V}_{\textrm{sys}} \pm  \Delta \text{V}$ where  $\Delta \text{V}= 10.0$ km/s and $\text{V}_\text{sys}=338.7 $ km/s (thick contour). \textbf{D:} Contours at $(2+4n)\sigma_\textrm{ch}$, where $\sigma_\textrm{ch}=0.51$ mJy bm$^{-1}$, the grey contours are at -$2 \sigma_\textrm{ch}$. }
\label{fig:DDO87}
\end{figure*}

\begin{figure*} 
 \centering 
 \includegraphics[width=0.915\textwidth]{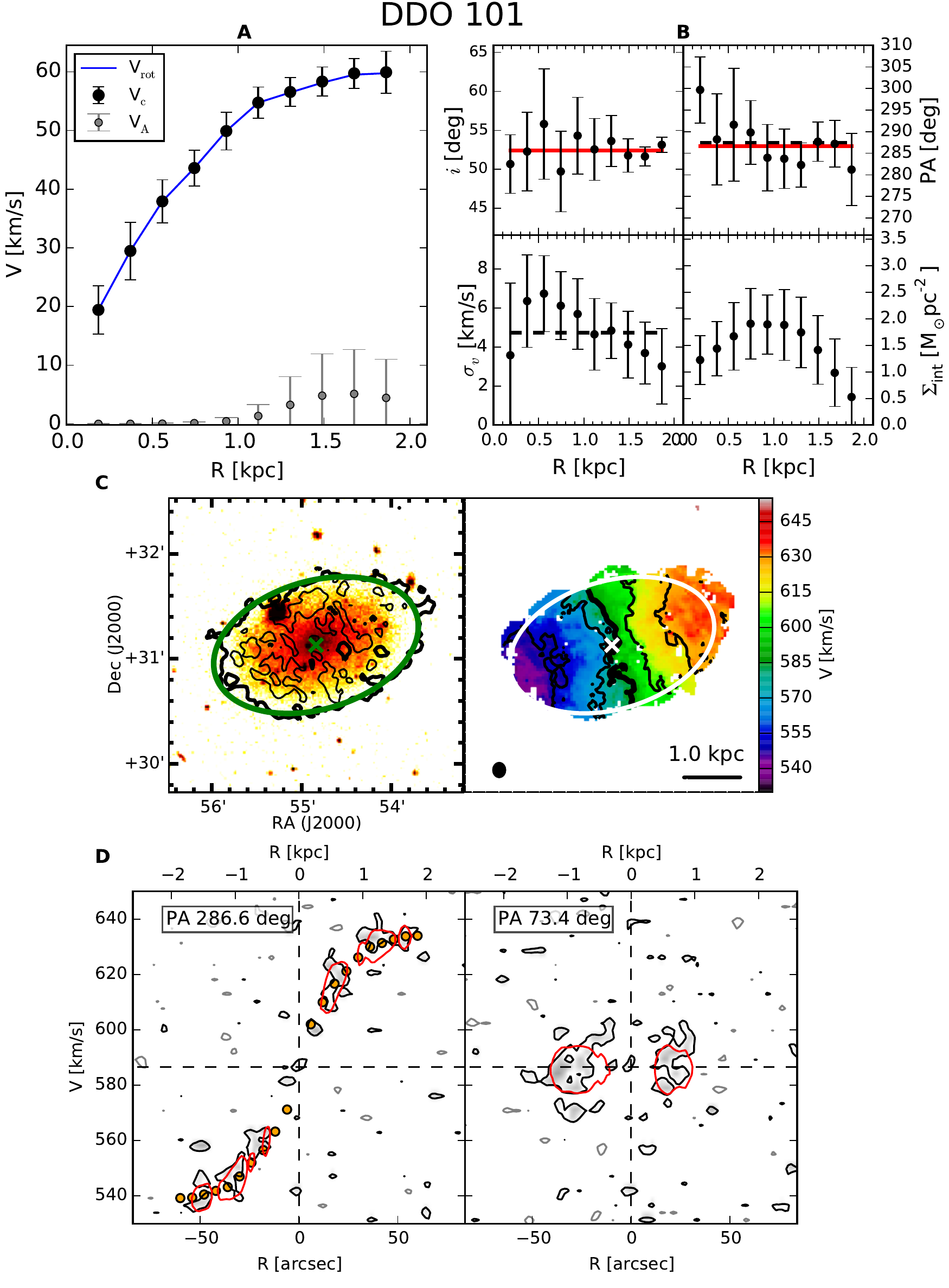}
 \caption{ See caption in Appendix \ref{sec:plot_layout}. Notes.   \textbf{C (left-hand panel):} Contours at $4^\text{n} \sigma_\text{3T}$ and $\sigma_\text{3T} = 0.90 \ \text{M}_\odot \text{pc}^{-2}$ (thick contour), stellar map in SDSS r band \protect\citep{stellarmap1}; \textbf{C (right-hand panel):} Contours at $\text{V}_{\textrm{sys}} \pm  \Delta \text{V}$ where  $\Delta \text{V}= 20.0$ km/s and $\text{V}_\text{sys}=586.6 $ km/s (thick contour). \textbf{D:} Contours at $(2+6n)\sigma_\textrm{ch}$, where $\sigma_\textrm{ch}=0.50$ mJy bm$^{-1}$, the grey contours are at -$2 \sigma_\textrm{ch}$. }
\label{fig:DDO101}
\end{figure*}

\begin{figure*} 
 \centering 
 \includegraphics[width=0.915\textwidth]{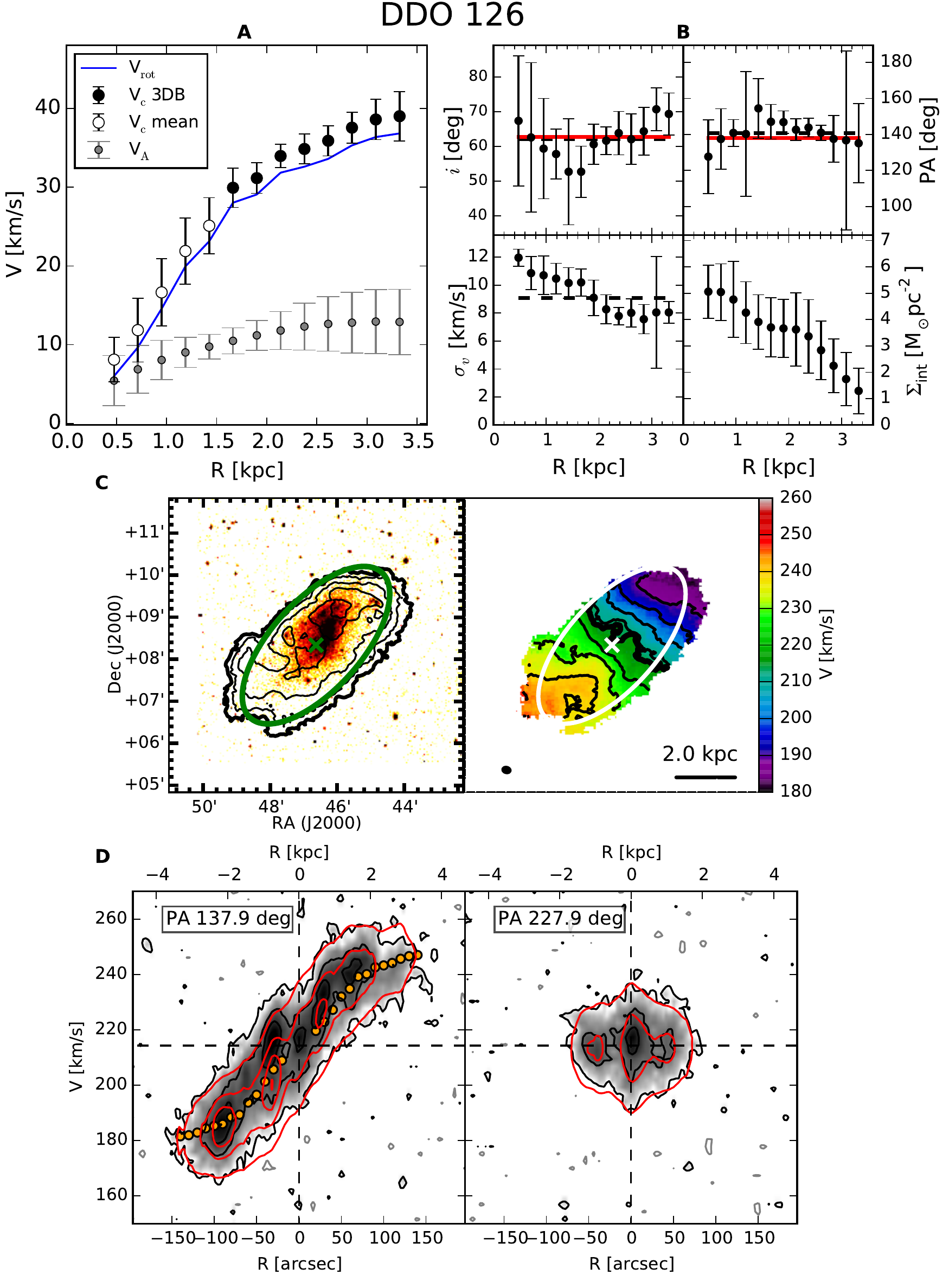}
 \caption{ See caption in Appendix \ref{sec:plot_layout}. Notes.  \textbf{A:} The empty circles indicates the region in which the circular velocity has been
 calculated as the mean between the approaching and receding rotation curves;  \textbf{C (left-hand panel):} Contours at $2^\text{n} \sigma_\text{3T}$ and $\sigma_\text{3T} = 1.27 \ \text{M}_\odot \text{pc}^{-2}$ (thick contour), stellar map in SDSS r band \protect\citep{stellarmap1}; \textbf{C (right-hand panel):} Contours at $\text{V}_{\textrm{sys}} \pm  \Delta \text{V}$ where  $\Delta \text{V}= 8.0$ km/s and $\text{V}_\text{sys}=214.3 $ km/s (thick contour). \textbf{D:} Contours at $(2+6n)\sigma_\textrm{ch}$, where $\sigma_\textrm{ch}=0.41$ mJy bm$^{-1}$, the grey contours are at -$2 \sigma_\textrm{ch}$. }
\label{fig:DDO126}
\end{figure*}

\begin{figure*} 
 \centering 
 \includegraphics[width=0.915\textwidth]{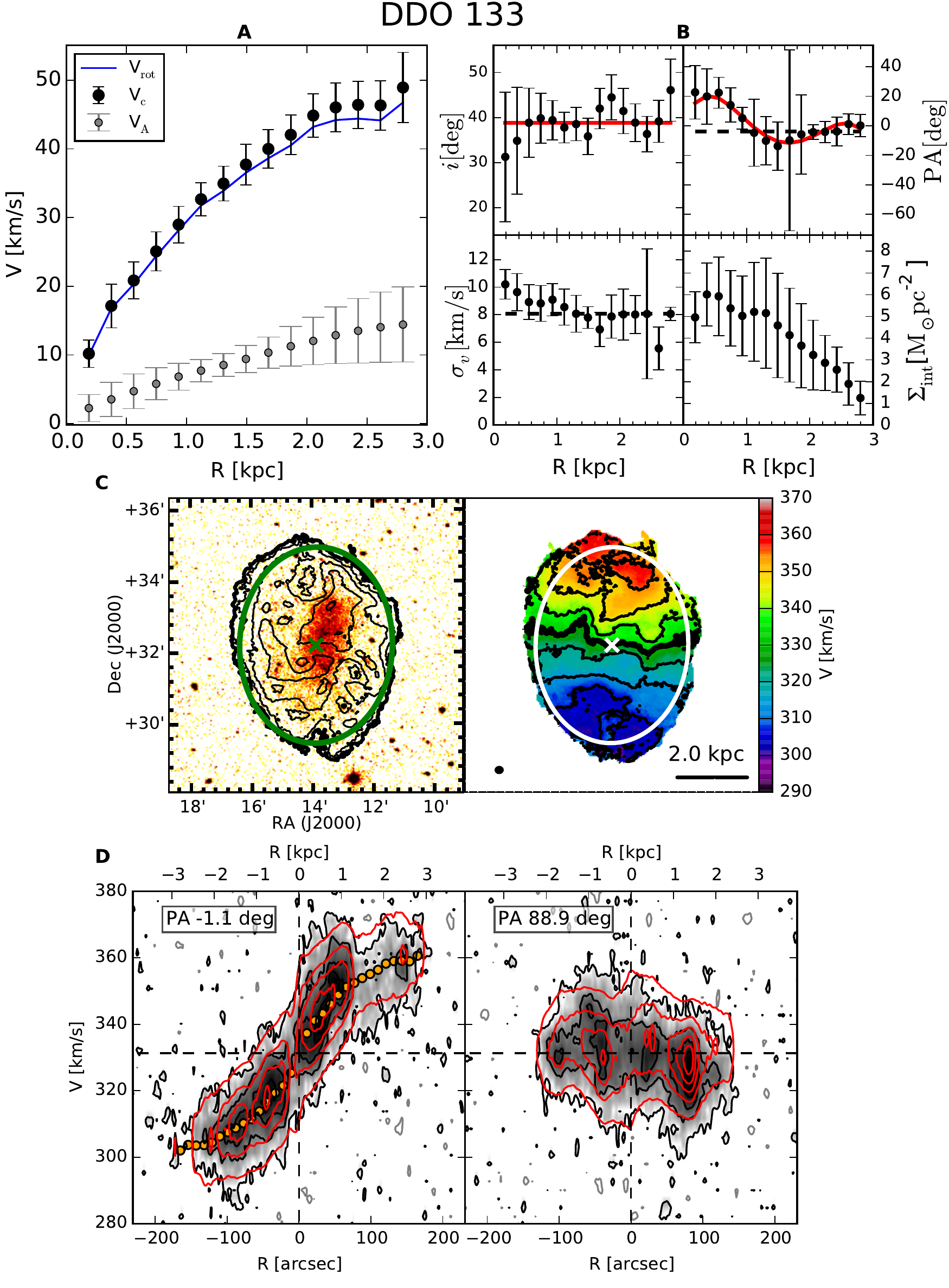}
 \caption{ See caption in Appendix \ref{sec:plot_layout}. Notes.   \textbf{C (left-hand panel):} Contours at $2^\text{n} \sigma_\text{3T}$ and $\sigma_\text{3T} = 0.89 \ \text{M}_\odot \text{pc}^{-2}$ (thick contour), stellar map in R band from \protect\cite{mapsr}; \textbf{C (right-hand panel):} Contours at $\text{V}_{\textrm{sys}} \pm  \Delta \text{V}$ where  $\Delta \text{V}= 8.0$ km/s and $\text{V}_\text{sys}=331.3 $ km/s (thick contour). \textbf{D:} Contours at $(2+5n)\sigma_\textrm{ch}$, where $\sigma_\textrm{ch}=0.35$ mJy bm$^{-1}$, the grey contours are at -$2 \sigma_\textrm{ch}$. }
\label{fig:DDO133}
\end{figure*}

\begin{figure*} 
 \centering 
 \includegraphics[width=0.915\textwidth]{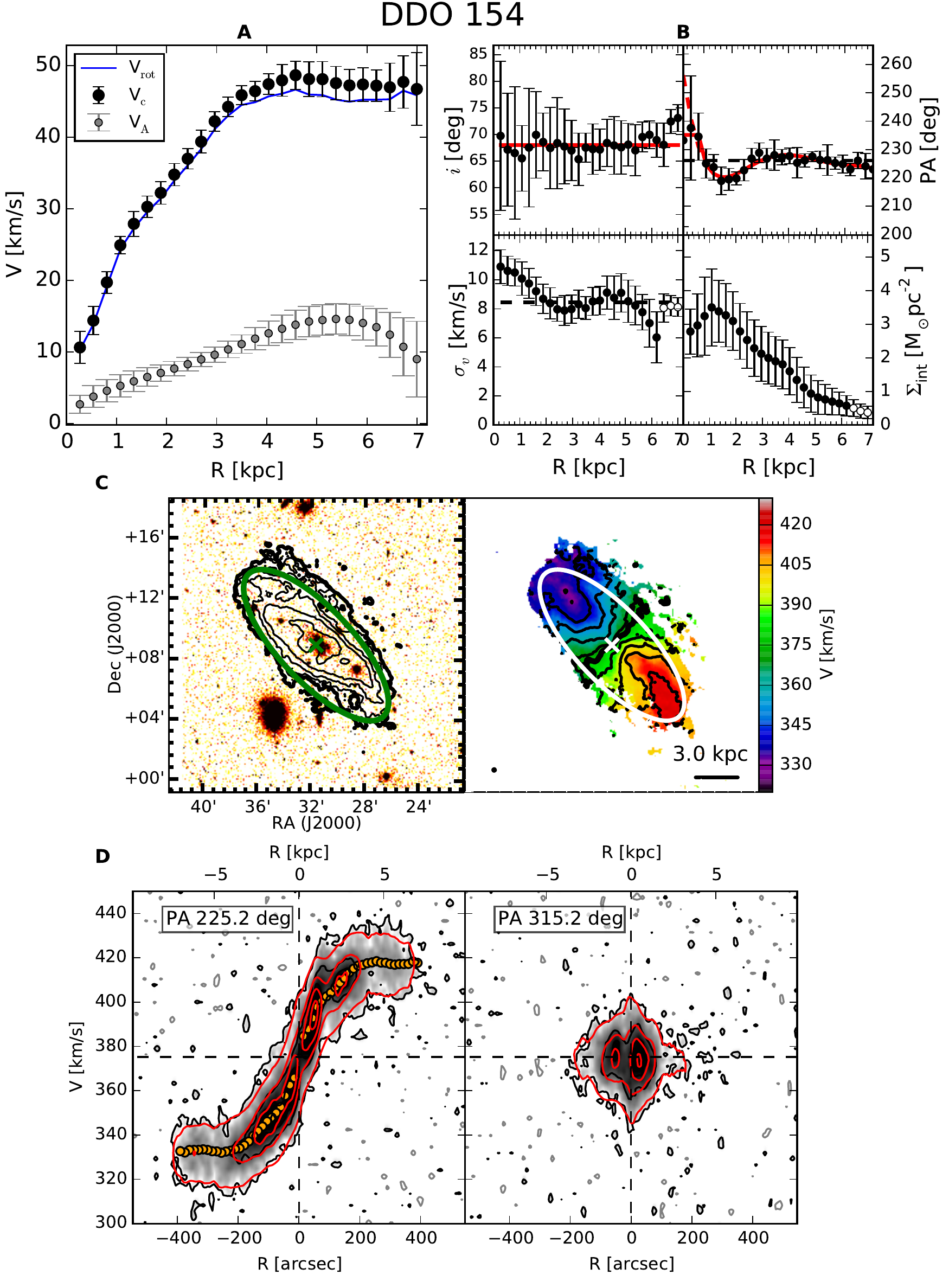}
 \caption{ See caption in Appendix \ref{sec:plot_layout}. Notes.\textbf{B (upper-left panel):} The red dashed line shows the best-fit polynomial to the data, the red continuous line shows the regularisation of the radial profile used for this galaxy (see text for details); \textbf{C (left-hand panel):} Contours at $2^\text{n} \sigma_\text{3T}$ and $\sigma_\text{3T} = 0.52 \ \text{M}_\odot \text{pc}^{-2}$ (thick contour), stellar map in R band from \protect\cite{mapsr}; \textbf{C (right-hand panel):} Contours at $\text{V}_{\textrm{sys}} \pm  \Delta \text{V}$ where  $\Delta \text{V}= 10.0$ km/s and $\text{V}_\text{sys}=375.2 $ km/s (thick contour). \textbf{D:} Contours at $(2+8n)\sigma_\textrm{ch}$, where $\sigma_\textrm{ch}=0.44$ mJy bm$^{-1}$, the grey contours are at -$2 \sigma_\textrm{ch}$. }
\label{fig:DDO154}
\end{figure*}

\begin{figure*} 
 \centering 
 \includegraphics[width=0.915\textwidth]{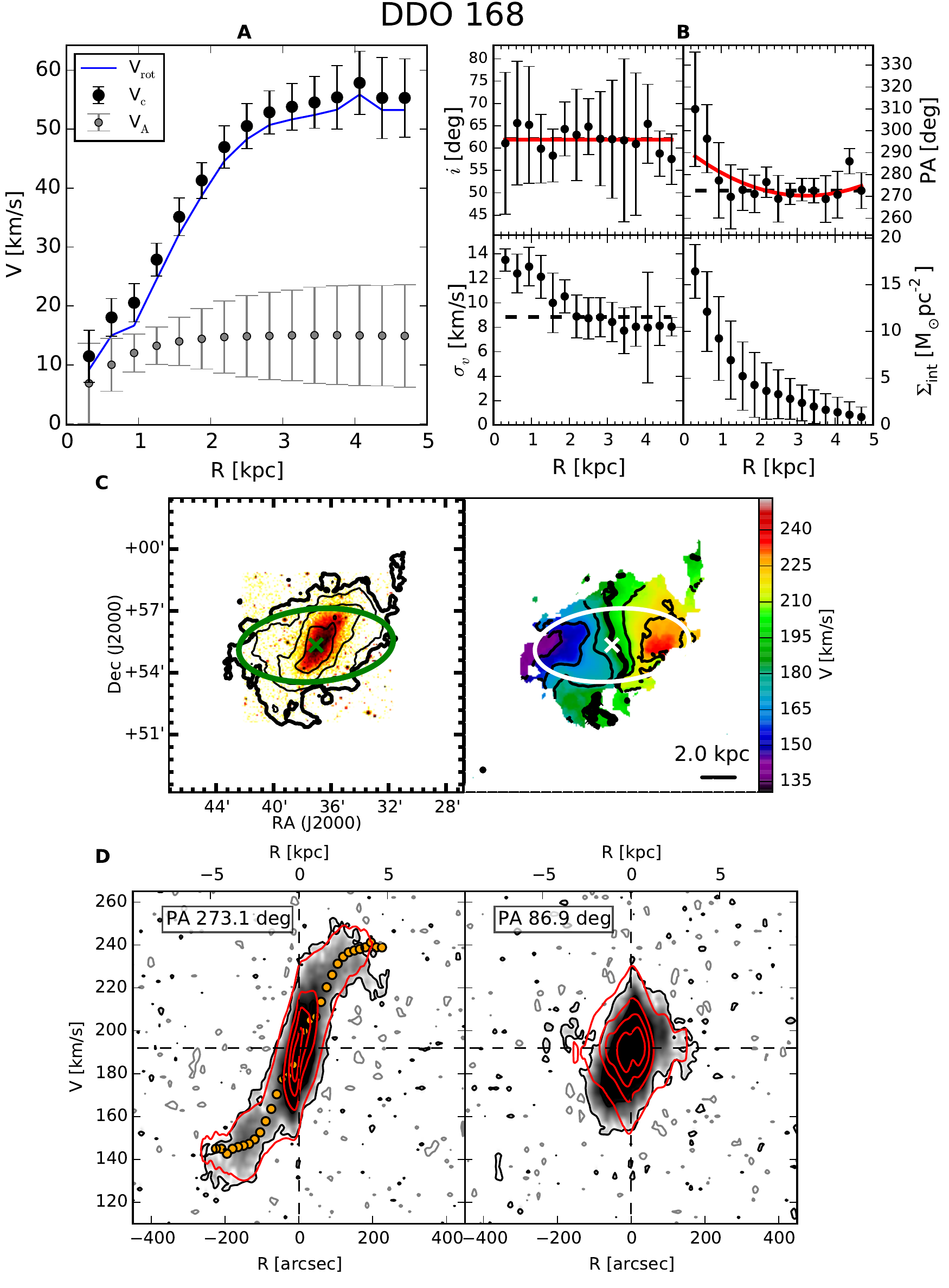}
 \caption{ See caption in Appendix \ref{sec:plot_layout}. Notes.   \textbf{C (left-hand panel):} Contours at $3^\text{n} \sigma_\text{3T}$ and $\sigma_\text{3T} = 0.54 \ \text{M}_\odot \text{pc}^{-2}$ (thick contour), stellar map in r band from \protect\citep{stellarmap1}; \textbf{C (right-hand panel):} Contours at $\text{V}_{\textrm{sys}} \pm  \Delta \text{V}$ where  $\Delta \text{V}= 15.0$ km/s and $\text{V}_\text{sys}=191.9 $ km/s (thick contour). \textbf{D:} Contours at $(2+15n)\sigma_\textrm{ch}$, where $\sigma_\textrm{ch}=0.47$ mJy bm$^{-1}$, the grey contours are at -$2 \sigma_\textrm{ch}$. }
\label{fig:DDO168}
\end{figure*}

\begin{figure*} 
 \centering 
 \includegraphics[width=0.915\textwidth]{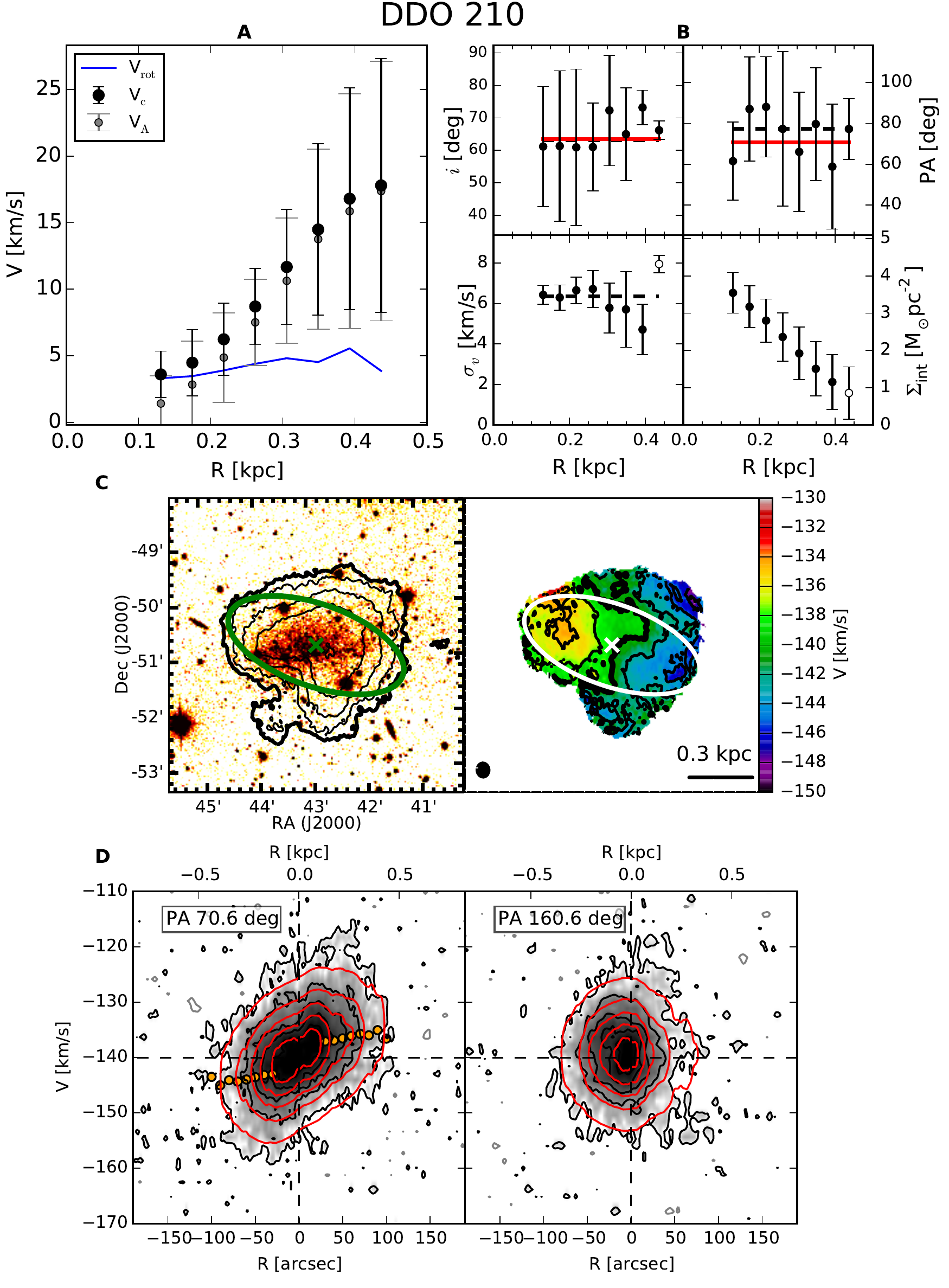}
 \caption{ See caption in Appendix \ref{sec:plot_layout}. Notes.   \textbf{C (left-hand panel):} Contours at $2^\text{n} \sigma_\text{3T}$ and $\sigma_\text{3T} = 0.59 \ \text{M}_\odot \text{pc}^{-2}$ (thick contour), stellar map in R band from \protect\cite{mapsr}; \textbf{C (right-hand panel):} Contours at $\text{V}_{\textrm{sys}} \pm  \Delta \text{V}$ where  $\Delta \text{V}= 2.5$ km/s and $\text{V}_\text{sys}=-140.0 $ km/s (thick contour). \textbf{D:} Contours at $(2+4n)\sigma_\textrm{ch}$, where $\sigma_\textrm{ch}=0.75$ mJy bm$^{-1}$, the grey contours are at -$2 \sigma_\textrm{ch}$. }
\label{fig:DDO210}
\end{figure*}

\begin{figure*} 
 \centering 
 \includegraphics[width=0.915\textwidth]{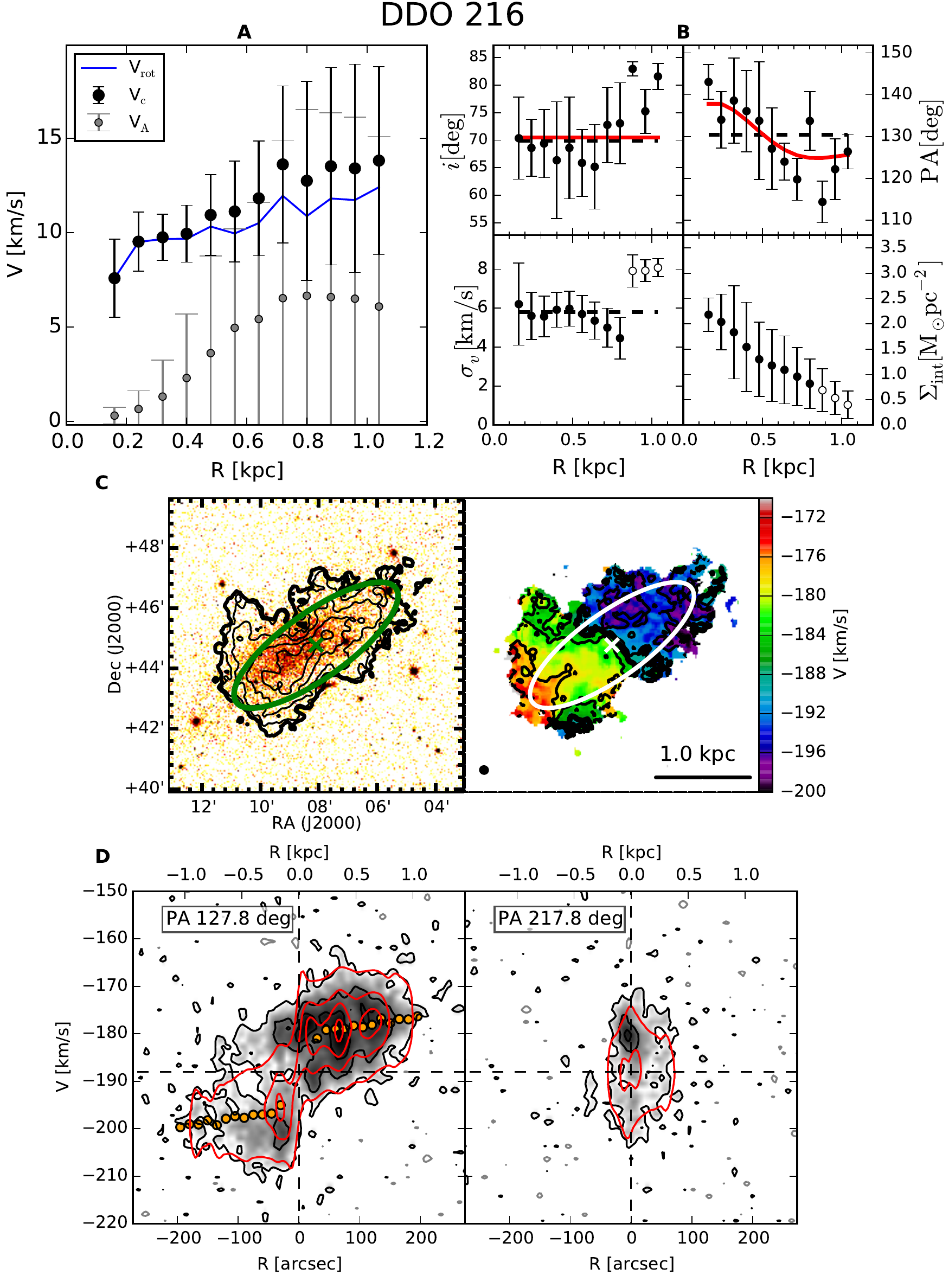}
 \caption{ See caption in Appendix \ref{sec:plot_layout}. Notes.   \textbf{C (left-hand panel):} Contours at $2^\text{n} \sigma_\text{3T}$ and $\sigma_\text{3T} = 0.40 \ \text{M}_\odot \text{pc}^{-2}$ (thick contour), stellar map in SDSS r band \protect\citep{stellarmap3}; \textbf{C (right-hand panel):} Contours at $\text{V}_{\textrm{sys}} \pm  \Delta \text{V}$ where  $\Delta \text{V}= 5.0$ km/s and $\text{V}_\text{sys}=-188.0 $ km/s (thick contour). \textbf{D:} Contours at $(2+6n)\sigma_\textrm{ch}$, where $\sigma_\textrm{ch}=0.91$ mJy bm$^{-1}$, the grey contours are at -$2 \sigma_\textrm{ch}$. }
\label{fig:DDO216}
\end{figure*}

\begin{figure*} 
 \centering 
 \includegraphics[width=0.915\textwidth]{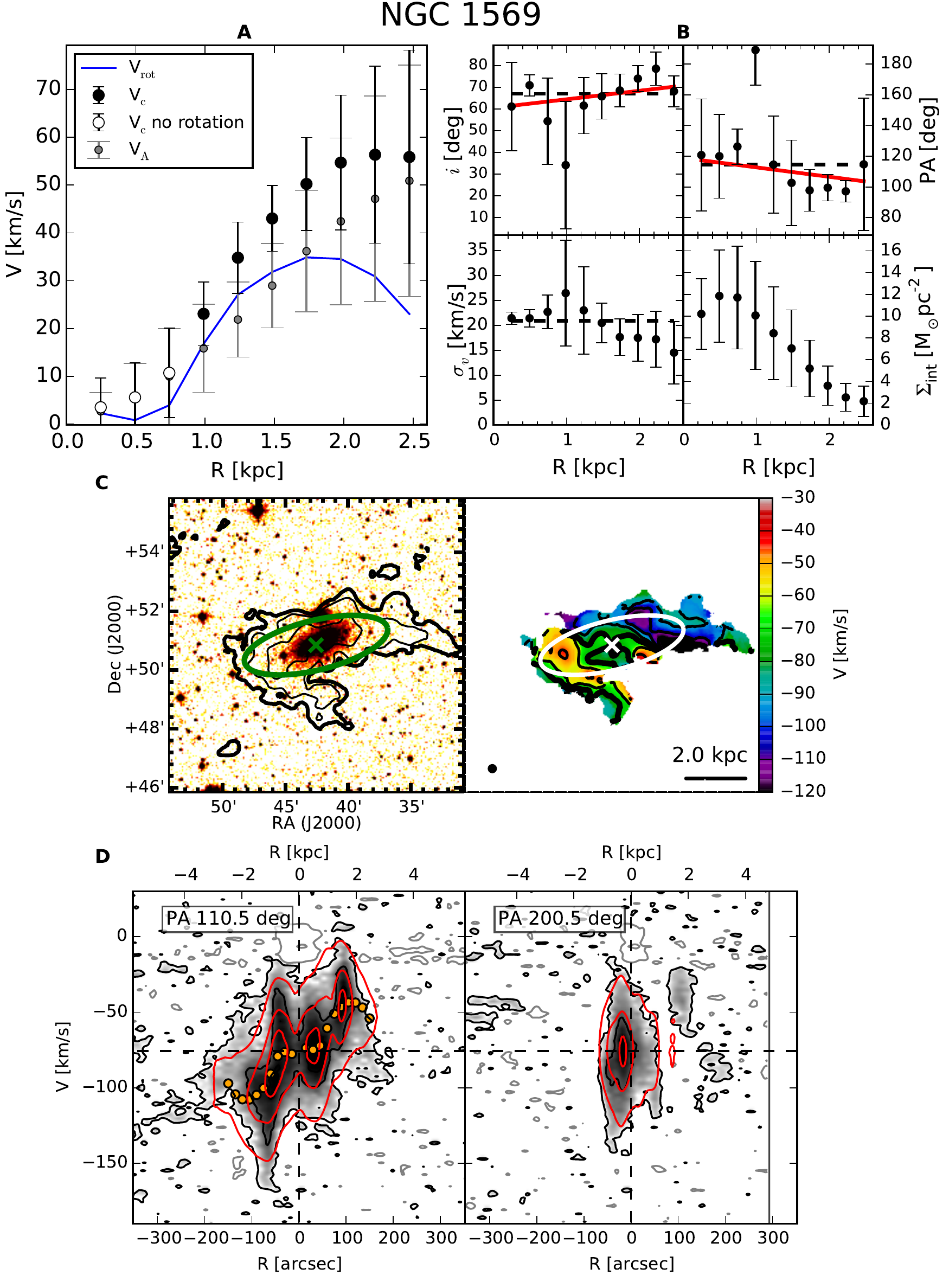}
 \caption{ See caption in Appendix \ref{sec:plot_layout}. Notes.  \textbf{A:} The empty circles indicate the region in which there are no significant signs of  gas rotation;  \textbf{C (left-hand panel):} Contours at $4^\text{n} \sigma_\text{3T}$ and $\sigma_\text{3T} = 1.10 \ \text{M}_\odot \text{pc}^{-2}$ (thick contour), stellar map in J band from \protect\citep{stellarmap3j}; \textbf{C (right-hand panel):} Contours at $\text{V}_{\textrm{sys}} \pm  \Delta \text{V}$ where  $\Delta \text{V}= 15.0$ km/s and $\text{V}_\text{sys}=-75.6 $ km/s (thick contour). \textbf{D:} Contours at $(2+8n)\sigma_\textrm{ch}$, where $\sigma_\textrm{ch}=0.77$ mJy bm$^{-1}$, the grey contours are at -$2 \sigma_\textrm{ch}$. }
\label{fig:NGC1569}
\end{figure*}

\begin{figure*} 
 \centering 
 \includegraphics[width=0.915\textwidth]{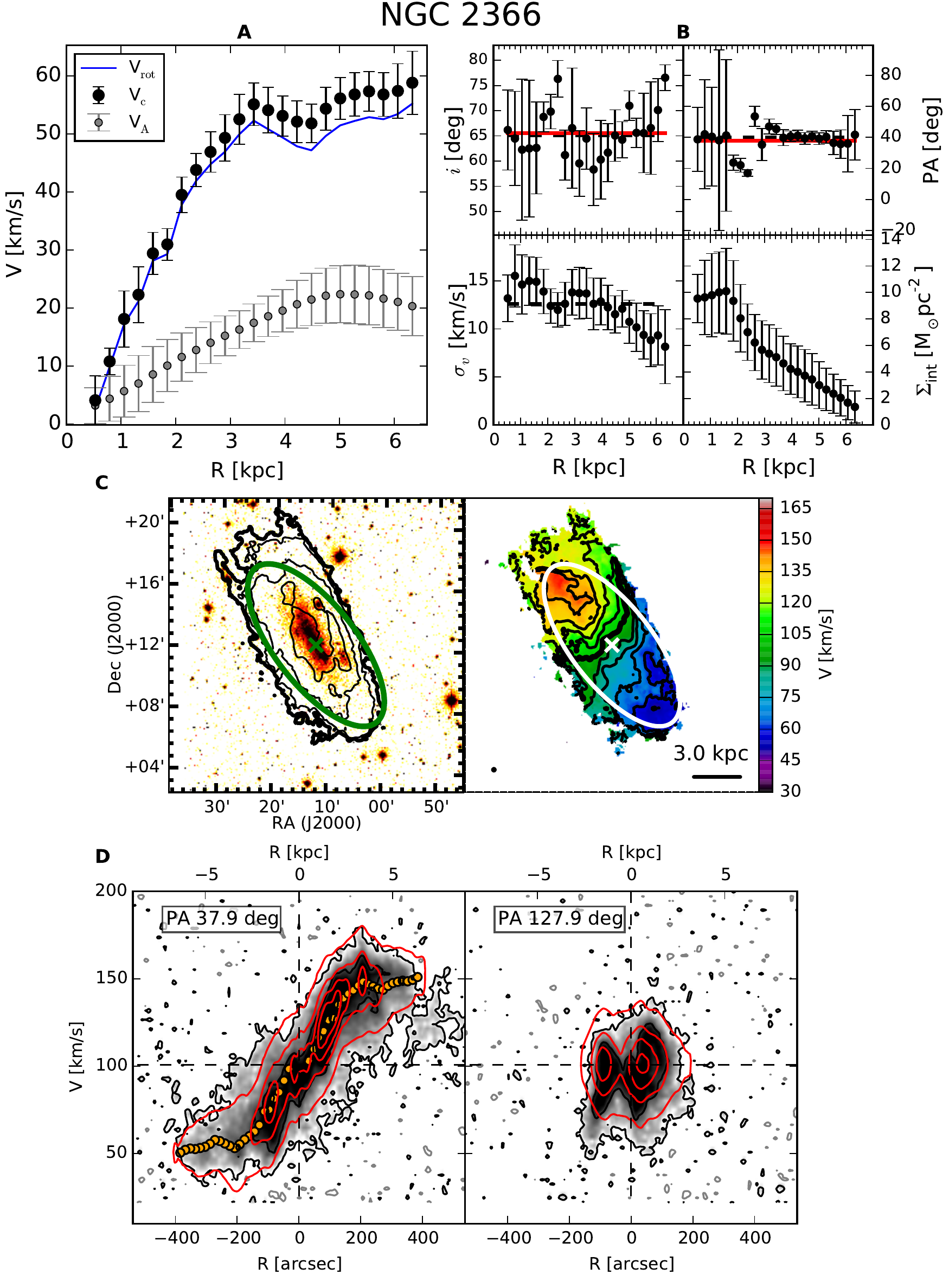}
 \caption{ See caption in Appendix \ref{sec:plot_layout}. Notes.   \textbf{C (left-hand panel):} Contours at $3^\text{n} \sigma_\text{3T}$ and $\sigma_\text{3T} = 0.59 \ \text{M}_\odot \text{pc}^{-2}$ (thick contour), stellar map in R band from \protect\cite{mapsr}; \textbf{C (right-hand panel):} Contours at $\text{V}_{\textrm{sys}} \pm  \Delta \text{V}$ where  $\Delta \text{V}= 10.0$ km/s and $\text{V}_\text{sys}=100.8 $ km/s (thick contour). \textbf{D:} Contours at $(2+10n)\sigma_\textrm{ch}$, where $\sigma_\textrm{ch}=0.52$ mJy bm$^{-1}$, the grey contours are at -$2 \sigma_\textrm{ch}$. }
\label{fig:NGC2366}
\end{figure*}

\begin{figure*} 
 \centering 
 \includegraphics[width=0.915\textwidth]{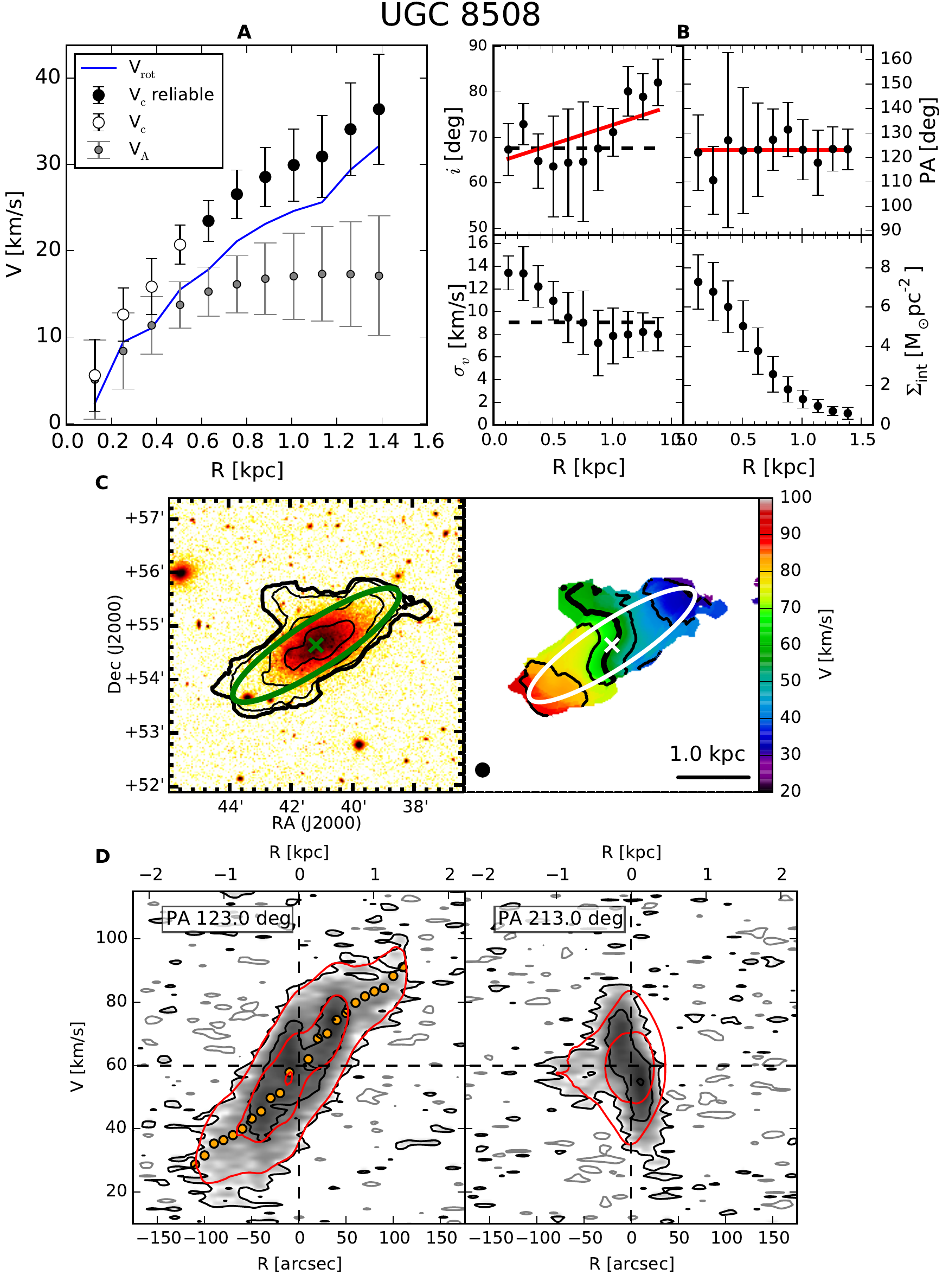}
 \caption{ See caption in Appendix \ref{sec:plot_layout}. Notes. \textbf{A:} The empty circles indicate the region of the galaxy with peculiar kinematics (see the text);  \textbf{C (left-hand panel):} Contours at $3^\text{n} \sigma_\text{3T}$ and $\sigma_\text{3T} = 0.54 \ \text{M}_\odot \text{pc}^{-2}$ (thick contour), stellar map in R band from \protect\cite{mapsr}; \textbf{C (right-hand panel):} Contours at $\text{V}_{\textrm{sys}} \pm  \Delta \text{V}$ where  $\Delta \text{V}= 10.0$ km/s and $\text{V}_\text{sys}=59.9 $ km/s (thick contour). \textbf{D:} Contours at $(2+6n)\sigma_\textrm{ch}$, where $\sigma_\textrm{ch}=1.31$ mJy bm$^{-1}$, the grey contours are at -$2 \sigma_\textrm{ch}$. }
\label{fig:UGC8508}
\end{figure*}

\begin{figure*} 
 \centering 
 \includegraphics[width=0.915\textwidth]{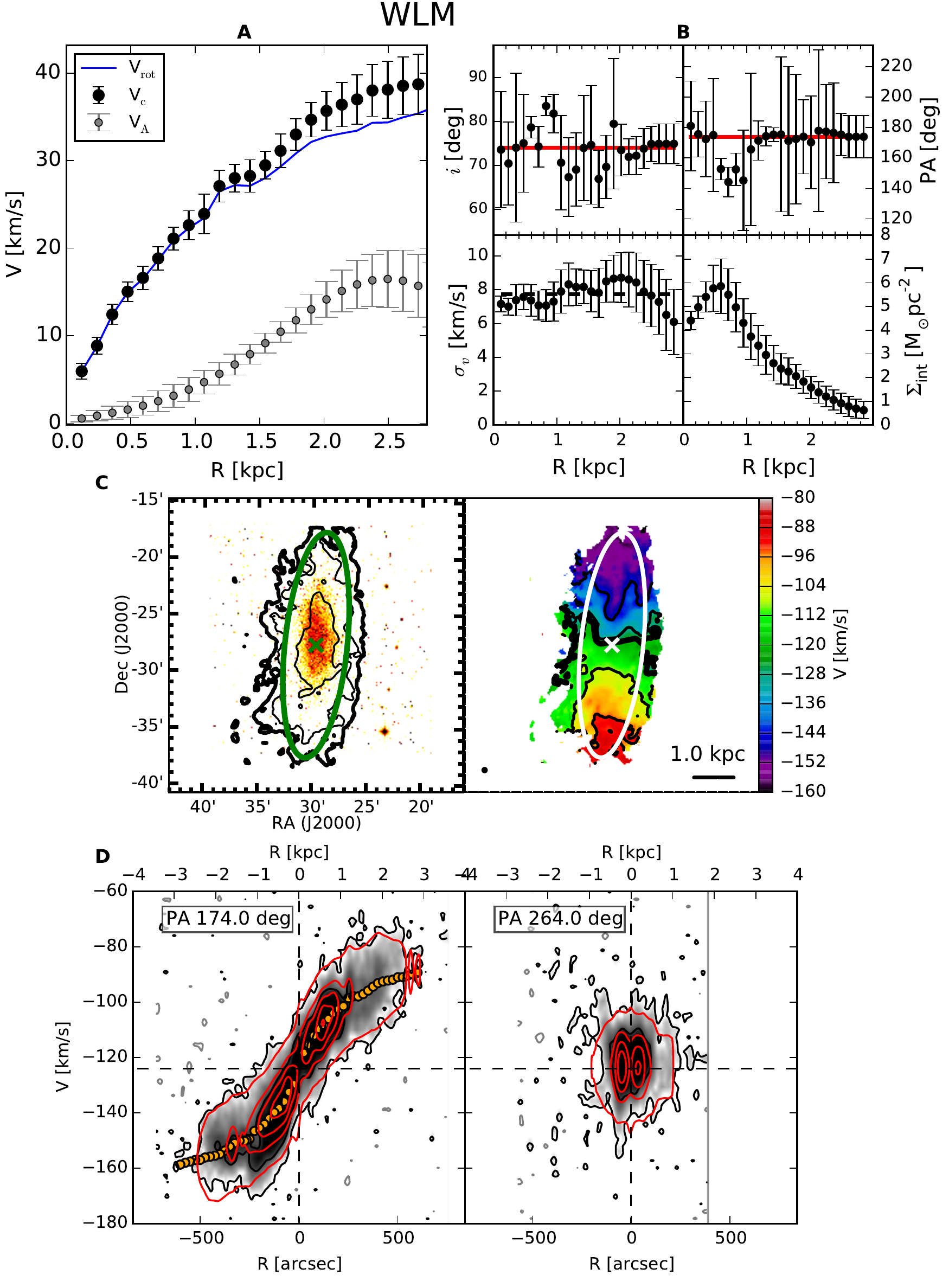}
 \caption{ See caption in Appendix \ref{sec:plot_layout}. Notes.   \textbf{C (left-hand panel):} Contours at $2^\text{n} \sigma_\text{3T}$ and $\sigma_\text{3T} = 0.58 \ \text{M}_\odot \text{pc}^{-2}$ (thick contour), stellar map in R band from \protect\cite{mapsr}; \textbf{C (right-hand panel):} Contours at $\text{V}_{\textrm{sys}} \pm  \Delta \text{V}$ where  $\Delta \text{V}= 20.0$ km/s and $\text{V}_\text{sys}=-124.0 $ km/s (thick contour). \textbf{D:} Contours at $(2+4n)\sigma_\textrm{ch}$, where $\sigma_\textrm{ch}=2.00$ mJy bm$^{-1}$, the grey contours are at -$2 \sigma_\textrm{ch}$. }
\label{fig:WLM}
\end{figure*}

For each analysed galaxy we produced a summary plot of its properties (Figs. \ref{fig:UA292}-\ref{fig:WLM}). 
Each plot is divided in four boxes: box A shows the rotation curve, the circular-velocity profile and the correction for the asymmetric 
drift; box B shows the velocity dispersion, the intrinsic surface density, $i$ and PA obtained 
with 3DB; box C shows the total HI map and the velocity field; box D shows the PV diagrams along 
the major and the minor axis for both model and data. The physical radii have been calculated assuming the distance in Tab. \ref{tab:sample_obs} (Col. 1).
A detailed description of the plot layout can be found in Appendix \ref{sec:plot_layout}.
For reference, Tab. \ref{tab:sample_res} lists the assumptions and the initial guesses used in 3DB (Cols. 1-7) and a summary of the results of the datacube fit (Cols. 8-11). Here below we report notes on individual galaxies. In the following notes $i_\text{ini}$ and PA$_\text{ini}$ indicate the initial guesses for $i$ and PA, respectively (see Cols. 6-7 in Tab. \ref{tab:sample_res}).

\begin{itemize}
\item[-] {\bf CVnIdwA.} The HI morphology of CVnIdwA (Fig. \ref{fig:UA292}) looks regular, but the orientation of the HI iso-density contours is quite different with respect to the direction of the velocity gradient. The emission of the stellar disc is very patchy, so we estimated the geometrical parameters of the HI disc using 3DB (see Appendix \ref{sec:iguess}).
\item[-] {\bf DDO 47.} The HI disc of DDO 47 (Fig. \ref{fig:UGC3974}) is nearly face-on, while the stellar disc looks highly inclined at 64$^\circ$ \citep{HVBand}. An inclination of 64$^\circ$ is not compatible with the HI contours of the total map and the stellar disc probably hosts a bar-like structure \citep{d47bar1}. Therefore we decided to estimate  $i_\text{ini}$ and PA$_\text{ini}$ by fitting ellipses on the contours of the total HI map (see Appendix \ref{sec:iguess}).  The discrepancy between the galactic centre estimated with the HI contours and the optical centre is very small (0.7 arcsec in RA and 2 arcsec in DEC), so we decided to use the HI centre. The best-fit $i$ (see Tab. \ref{tab:sample_res}) is consistent with previous works (35$^\circ$ in \citealt{d47inc}, 30$^\circ$ in \citealt{d47inc2} and  \citealt{1569still}) and seems a reasonable upper limit for this galaxy (see ellipses in Box C in Fig. \ref{fig:UGC3974}).
The final rotation curve shows a flattening at $R<120^{\prime \prime}$ that could be the sign of the presence of strong non-circular motion. Moreover in the approaching half of the galaxy there is a large hole that can be clearly seen both on the total map and on the PV along the major axis:  the presence of such hole could further bias the estimate of the gas rotation. For these reasons, we retain the velocities for $R<120^{\prime \prime}$ not completely reliable (empty circles in Box A in Fig. \ref{fig:UGC3974}).
\item[-] {\bf DDO 50.} The gaseous disc of DDO 50 (Fig. \ref{fig:DDO50}) is quite peculiar: it is 'drilled' by medium-large HI holes ranging from 100 pc to 1.7 kpc \citep{d50hole3} and it also shows clumpy regions with high density. Therefore, the HI surface density evaluated  across the rings could be very discontinuous with very high peaks and/or regions without emission. As a consequence, the  quoted errors on the intrinsic surface density profile (see Sec. \ref{sec:mapd}) are large and they are probably over-estimating the real statistical uncertainties, especially in the inner disc ($R<3$ kpc). Despite this peculiarity, the large scale kinematics is quite regular and typical of a flat rotation curve. The estimates of the $i$ and PA are difficult given that this galaxy is nearly face-on. The value of $i$ estimated with the isophotal fitting of the optical disc (47$^\circ$) is too high to be compatible with the flattening of the HI contours which favour an $i<40^\circ$. The best agreement between the datacube and the model has been found using initial values of 30$^\circ$ for $i_\text{ini}$ and  180$^\circ$  for PA$_\text{ini}$. The centre was set using the coordinates of the centre of the stellar disc \citep{HVBand}. 

\item[-] {\bf DDO 52.} The best-fit $i$  of DDO 52 (Fig. \ref{fig:DDO52}) tends to be more edge-on  at the end of the disc. However, the analysis of the channels and of the PVs indicates that the model with a constant $i$ gives a better representation of the data. 
\item[-] {\bf DDO 53.}  
The stellar disc and the HI disc of DDO 53 (Fig. \ref{fig:DDO53}) are misaligned and the galaxy is nearly face-on. As a consequence, $i$ and PA are very difficult to constrain: we tried different values in the range 30$^\circ$-50$^\circ$ for $i_\text{ini}$ and 100$^\circ$-140$^\circ$ for PA$_\text{ini}$.  The best match with the data has been found using  $i_\text{ini}=35^\circ$ and PA$=120^\circ$, but the rotational and the circular velocities should be taken with caution.
In the northern part of the galaxy there is some extra emission possibly connected with an inflow/outflow: this region is clearly visible in the PV along the minor axis around -50 arcsec at a velocity of about 5 km/s. 
\item[-] {\bf DDO 87.} The morphology of DDO 87 (Fig. \ref{fig:DDO87}) is clearly irregular in the inner part, but the outer disc looks more regular. We decided to set the initial guesses for the PA  using 3DB (see Appendix \ref{sec:iguess}). We tried different initial guesses for $i$: the best representation of the data has been obtained with $i_\text{ini}=40^\circ$ and PA$_\text{ini} = 240^\circ$, approximately 15$^\circ$  lower than the orientation of the optical disc \citep{HVBand}. 
\item[-] {\bf DDO 101.} The HI disc of DDO 101 (Fig. \ref{fig:DDO101}) is extended only  slightly beyond the optical disc and the HI emission is almost constant with some high density structures around 1.0 kpc.
Notice that the estimates of the distance for this galaxy are very uncertain (ranging from 5 to 16 Mpc) since they all rely on a poor distance estimator (Tully-Fisher relation, e.g. \citealt{d101tf}).
\item[-] {\bf DDO 126.}  DDO 126 (Fig. \ref{fig:DDO126}) shows kinematic asymmetries, so  we separately run 3DB also on the approaching and the receding halves of the galaxy. Beyond 1.5 kpc the fit on the whole galaxy gives a good representation of the datacube and the errors found with 3DB are larger than (or comparable to) the differences due to the kinematic asymmetries (black circles in Box A in Fig. \ref{fig:DDO126}). The inner regions are less regular and the best model has been found taking the mean between the approaching and receding rotation curves (empty circles in Box A in Fig. \ref{fig:DDO126}), while the errors have been calculated as half the  difference between the two values following the recipe of \cite{swaters}.
\item[-] {\bf DDO 133.} The HI disc of DDO 133 (Fig. \ref{fig:DDO133}) has a regular kinematics, although there is evidence of non-circular motions, especially in the region of the stellar bar \citep{d133bar}.  
The final PA found with 3DB is about 20$^\circ$ in the inner part of the disc (R$<$0.5 kpc) and it becomes almost zero in the outer disc. We decided to  take the radial trend of the PA into account with a fourth order polynomial.

\item[-] {\bf DDO 154.} The HI morphology and the kinematics of DDO 154 (Fig. \ref{fig:DDO154}) is quite regular. 
However, the contours of the HI map  clearly show  a radial trend of the PA, as confirmed by the values found with 3DB. In order to obtain a good description of the radial trend of the PA we used a fourth order polynomial. However, the best-fit polynomial shows a very steep gradient (about 25$^\circ$) in a very small region  (less than 1 kpc) that is not justified by the data (red dashed line in Box B in Fig. \ref{fig:DDO154}). The points located in this region are compatible with a constant PA, so we decided to fix it to the mean values of the first three points (see Box B in Fig. \ref{fig:DDO154}). As in O15, we do not confirm the almost Keplerian fall-off of the rotation curve claimed by \cite{d154h} beyond 5 kpc. The channel maps are shown in Fig. \ref{fig:chan154}

\item[-] {\bf DDO 168.} The velocity gradient of DDO 168 (Fig. \ref{fig:DDO168}) is misaligned with respect to both the HI and the stellar disc. Moreover, the presence of a prominent bar visible both in the stellar  disc \citep{HVBand}  and in the inner part of the HI disc makes the initial estimate of the geometrical parameters very uncertain. 
We  obtained a first estimate of the PA using the 2D tilted-ring fitting of the velocity field shown in Fig. \ref{fig:DDO168}: the resultant PA decreases from 300$^\circ$ to about  270$^\circ$. We used these values as initial guesses for all sampling radii in 3DB (see Appendix \ref{sec:iguess}). 
The centre was set to the value found with 3DB which roughly corresponds to the optical centre.
We tried different values for $i_\text{ini}$ between $40^\circ-70^\circ$: the best reproduction of the datacube has been obtained with $i_\text{ini}=60^\circ$.
The final results still show a variation of PA that we regularised with a second order polynomial. 
The outer disc of DDO 168 is quite irregular. There is extra emission at velocities close to V$_\text{sys}$ possibly related to inflow/outflow,  while the distortions of the iso-velocity contours (Box C. in Fig. \ref{fig:DDO168}) could be due to the presence of  an outer warp.

\item[-] {\bf DDO 210.} DDO 210 (also know as Aquarius dIrr, Fig. \ref{fig:DDO210}) is the least massive galaxy in our sample and it is classified as a transitional  dwarf galaxy (\citealt{mc} and reference therein). The HI map is quite peculiar with isodensity contours that are not elliptical. As a consequence the estimate of the galactic centre using the HI data is very uncertain and we decided to set it to the optical value \citep{HVBand}. The kinematics is dominated by the velocity dispersion, however a weak velocity gradient is visible. The velocity gradient looks misaligned with both the stellar and the HI disc, so we set the initial values of $i$ and PA (60$^\circ$ and 65$^\circ$, respectively) using a by-eye inspection of the velocity field.
This procedure is arbitrary, but it is important to note that the final circular-velocity curve and the related errors are independent of our procedure since they are totally dominated by the asymmetric-drift correction (see Sec. \ref{sec:asy} and Sec. \ref{sec:fnotes}). The V$_\text{rot}$ found with 3DB in the first two rings (10 and 20 arcsec) is not well-constrained, so we preferred to exclude them. Notice that along the minor axis there is an extended region with HI emission apparently not connected with the rotating disc. As in the case of DDO 53 this emission could trace an inflow/outflow.
\item[-] {\bf DDO 216.} DDO 216  (also know as Pegasus dIrr, Fig. \ref{fig:DDO216}) is defined as a transitional  dwarf galaxy \citep{d216def}. The galaxy shows a velocity gradient aligned with the HI and the optical disc, but it could be entirely due
to a single `cloud' at a discrepant velocity in the approaching side of the galaxy \citep{d216gradient}.
For the purpose of our work, we assumed the gradient genuine and caused entirely by the gas rotation. The analysis of the alternative scenario can be found in Appendix \ref{sec:d216b}.
Given the kinematic peculiarities (see Appendix \ref{sec:d216b}), we did not use 3DB to estimate the V$_\text{sys}$ (Sec. \ref{sec:ass}). We tried different values between $-186$ and $-190$ km/s and we decided to use $\textrm{V}_\text{sys}=-188$ km/s because this value minimises the kinematic asymmetries between the receding and the approaching halves of the galaxy. We set the centre, $i_\text{ini}$ (65$^\circ$)  and PA$_\text{ini}$ (130$^\circ$) using the values obtained from the elliptical fit of the outermost HI contours (R$>$600 pc). The best-fit PA  decreases from about 140$^\circ$ to about $125^\circ$.
\item[-] {\bf NGC 1569} NGC 1569 (Fig. \ref{fig:NGC1569}) is a starburst galaxy with a very disturbed HI kinematics and morphology \citep{1569still,1569j,halol}. We found that the best-fit model shows a slight increase of $i$ and a slight decrease of PA as a function of radius. The best-fit model is a good representation of the large scale structure and kinematics of the HI disc, but it fails to reproduce the small scale local features. The ISM of this galaxy is highly turbulent: the velocity dispersion found with 3DB is about 20 km/s and the asymmetric-drift correction dominates at all radii. For this reason the kinematic data reported here should be used with caution especially at the inner radii (empty circles in Box A in Fig. \ref{fig:NGC1569}) where no significant rotation of the gas is observed (see also \citealt{halol}).
\item[-] {\bf NGC 2366.} The HI disc of NGC 2366 (Fig. \ref{fig:NGC2366}) is quite regular, although it shows some peculiar features.
The HI emission on the channel maps (Fig. \ref{fig:chan2366}) indicates the presence of  two ridges located in the North-West and South-East (less prominent) of the disc running parallel to the major axis. The ridges show different kinematics with respect to the disc (see \citealt{2366oh}) and their origin is not clear (see \citealt{2366hunter} for a detailed discussion). We checked that 3DB was not affected by the presence of this feature.
From the PV along the major axis (Panel D in Fig. \ref{fig:NGC2366}) it is clear that the 3DB model does not reproduce some emission close to the systemic velocity, especially in the receding side where the gas is seen  also at `forbidden' velocities below V$_\text{sys}$. As already stated by \cite{halol} this is probably due to the presence of some extraplanar gas that is rotating at lower velocity with respect to the gas in the disc (see e.g. \citealt{halof}). As in \cite{halol} and in O15 we do not find that the rotation curve declines beyond 5 kpc, as instead claimed by both \cite{2366hunter} and  \cite{2366v}.
\item[-] {\bf UGC 8508.} The HI and the stellar disc of UGC 8508 (Fig. \ref{fig:UGC8508}) are aligned but the analysis of the HI map favours an $i$ slightly higher than the value obtained from the stellar disc \citep{HVBand}. We found that the datacube is better reproduced with a linearly 
increasing $i$.  Notice that in the inner part of the galaxy (R$<$0.6 kpc) the kinematics  is very peculiar as it is visible by the S-shaped iso-velocity contours (right-panel C in Fig. \ref{fig:UGC8508}). This kind of distortions can be related to an abrupt variation of the PA  and/or to 
the presence of radial motions \citep{frat1} as well as to a deviation from axisymmetry of the galactic potential \citep{swatlop}. We tested the  hypothesis of a radially varying PA  and the presence of non-zero radial velocities (V$_\text{rad}$) performing a 2D analysis of the velocity field with ROTCUR \citep{begeman}. We found that the combination of the two effects can partially explain the distortions of the velocity field, but their magnitude is too large to be physically plausible.
Fortunately,  the final rotation curves obtained including radial motions and/or the varying PA are compatibles with the results we found with 3DB, though the inner points (empty circles in Box A in Fig. \ref{fig:UGC8508} ) should be treated with caution.
\item[-] {\bf WLM.} The HI and the optical discs of WLM (Fig. \ref{fig:WLM}) are well aligned, but the best-fit $i$ looks slightly too edge-on with respect to the HI contours (see ellipses in Fig. \ref{fig:WLM}). The excess of the emission around the minor axis could be partially due to the thickness of the gaseous layer (Sec. \ref{sec:hscale}, see also \citealt{leam}).
Further details  on the analysis of WLM can be found in \cite{mia}.
\end{itemize}

\section{Application: test of the Baryonic Tully-Fisher relation} \label{sec:scientific}
\label{sec:BT}

\begin{figure} 
 \centering 
 \includegraphics[width=1.0\columnwidth]{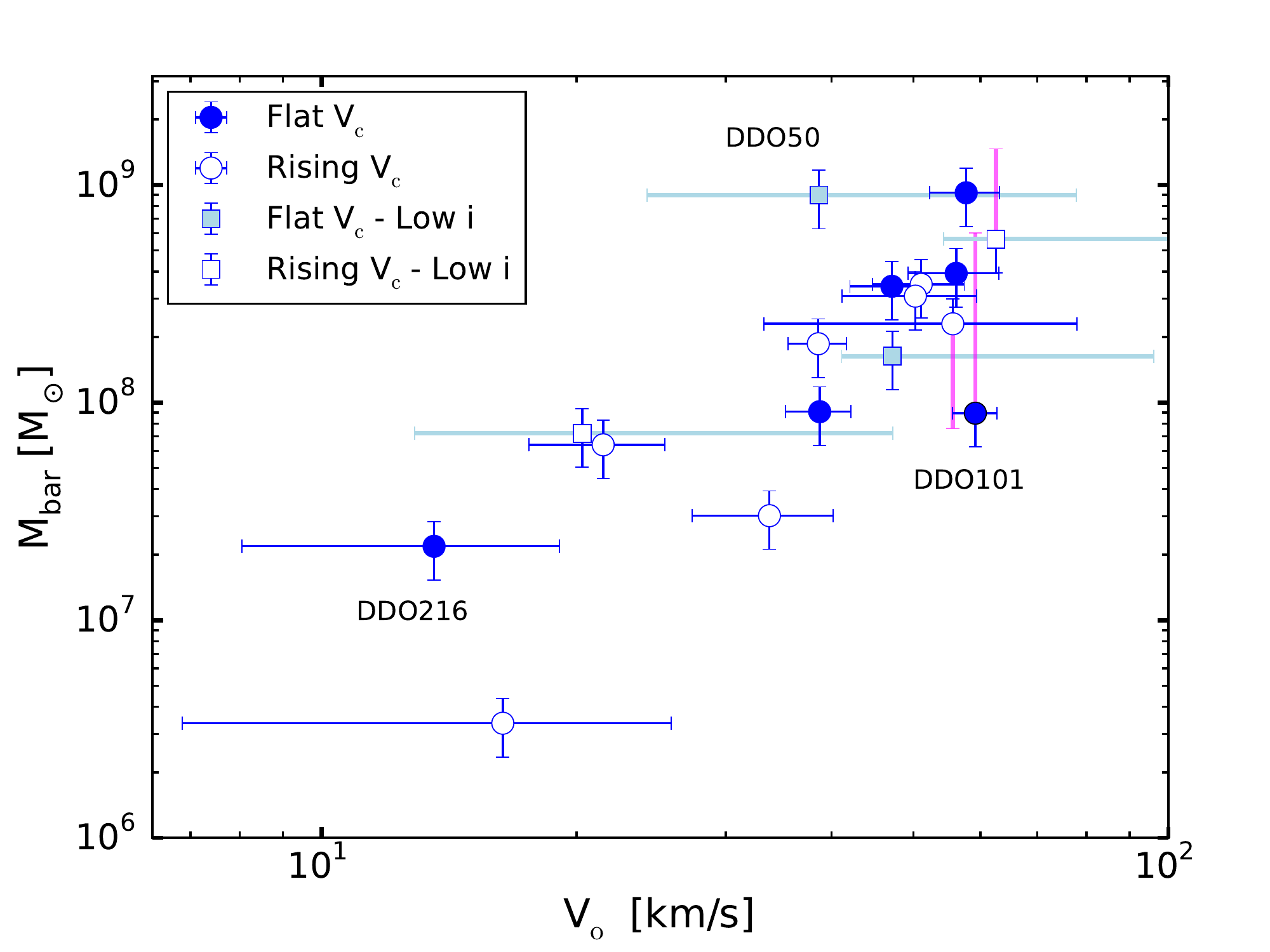}
 \caption{BTFR for the galaxies in our sample. The empty and the solid symbols indicate galaxies with rising or flat rotation curve, respectively. The squares indicate nearly face-on galaxies. The blue bars indicate 1-$\sigma$ errors. The light-blue and the magenta bars indicate intervals of allowed values (see text).}
\label{fig:btfr}
\end{figure}

\begin{figure}
 \centering 
 \includegraphics[width=1.0\columnwidth]{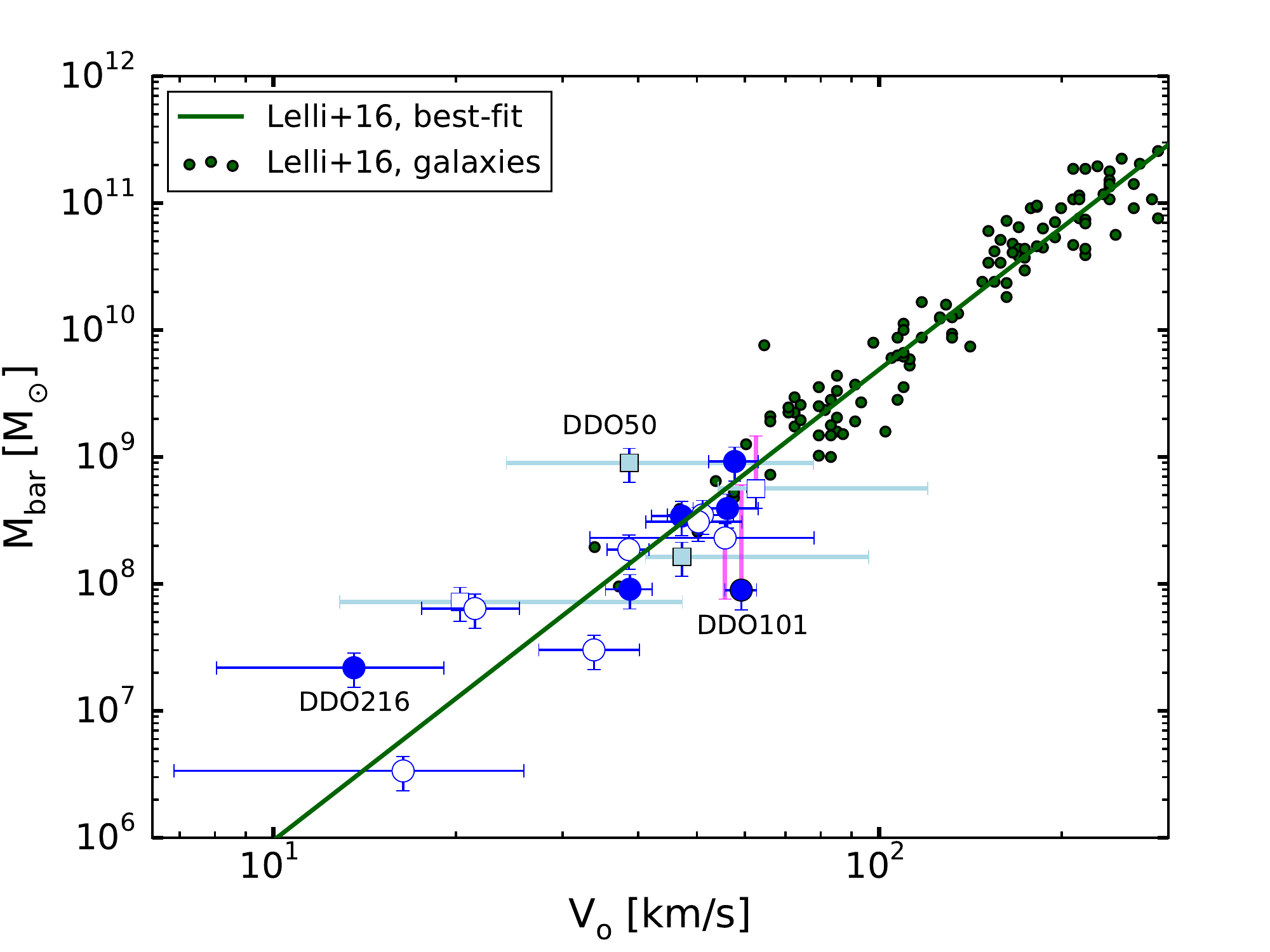}
 \caption{Same data as in \ref{fig:btfr} compared with the galaxies and the BTFR from \protect\cite{lellibtff}.}
\label{fig:btfrc}
\end{figure}

The baryonic Tully-Fisher relation (BTFR) links a characteristic circular  velocity (V) of a galaxy with its total baryonic mass (M$_\textrm{bar}$).  The relation, in the logarithmic form 
\begin{equation}
\log \left( \frac{\textrm{M}_\textrm{bar}}{\textrm{M}_\odot} \right)=\textrm{s}\log \left( \frac{\textrm{V}}{\textrm{km s}^{-1}} \right) + \textrm{A}
\label{eq:btfr}
\end{equation} is very tight and extends over 6 decades in M$_\textrm{bar}$ \citep{mgtf,lellibtff}.
The existence of this relation represents a fundamental benchmark for cosmological models and for galaxy formation theories \citep{mgtf2,dicintio,brook}. In this context it is very important to extend the study of the BTFR down to extremely low-mass dwarf galaxies (e.g. \citealt{begumtf}). In this section we present the BTFR  for the galaxies in our sample (Fig. \ref{fig:btfr}) and then we compare it with the results of \cite{lellibtff} (Fig. \ref{fig:btfrc}). 
The derivation of the baryonic mass and of the characteristic velocity is described below.

\begin{itemize}

\item{\bf Rotation velocity.}
The rotation velocities used in the Tully-Fisher relation (TFR) are W$_\textrm{20}$ and  V$_\textrm{flat}$: W$_\textrm{20}$ is the velocity width of the global HI line profiles, while V$_\textrm{flat}$ is the value of the flat portion of the rotation curve. \protect\cite{verTF}  found that V$_\textrm{flat}$ minimises the scatter in the TFR, so it is the best way to study the BTFR using high resolution HI data (see also \protect\citealt{brook}). However, several rotation curves of dIrrs do not reach the flat part (e.g DDO 53, Fig. \ref{fig:DDO53}) or the flattening is entirely due to the asymmetric-drift correction (e.g. NGC 1569, Fig. \ref{fig:NGC1569}).
As indicator of the rotation velocity, we therefore used the velocity of the outer disc (V$_\textrm{o}$) defined as the mean circular velocity of the last three fitted rings (see Sec. \ref{sec:tstage}).
V$_\textrm{o}$ is a measure of V$_\textrm{flat}$ for galaxies in which the flat part of the rotation curve is observed. For galaxies with a rising rotation curve V$_\textrm{o}$ is an estimate of the maximum circular velocity within the considered radial range.
Galaxies with rising rotation curves are indicated with empty markers in Figs. \ref{fig:btfr} and \ref{fig:btfrc}. 
The error on V$_\textrm{o}$ is conservatively assumed as the maximum error between the three values used to calculate V$_\textrm{o}$. In the five galaxies with a low $i$ (squares in Figs. \ref{fig:btfr} and \ref{fig:btfrc}) the uncertainties on $i$ and on the velocities can be largely underestimated (see Sec. \ref{sec:gls} and Tab. \ref{tab:sample_res}), so we present these galaxies
with a bar that indicates the  interval between a minimum (assuming $i=40^\circ$) and a maximum (assuming $i=20^\circ$)  value of V$_\textrm{o}$ (see Eq. \ref{eq:vtitlt}).
\newline

\item{\bf Baryonic mass.}
The total baryonic mass of the galaxies has been calculated as
\begin{equation}
\textrm{M}_\textrm{bar}=\textrm{M}_* + 1.33 \textrm{M}_\textrm{HI},
\end{equation} where M$_*$ is the stellar mass, M$_\textrm{HI}$ is the mass of the atomic hydrogen and the factor 1.33 takes into account the the presence of Helium \protect\citep{begumtf,btfl}. The molecular gas is likely irrelevant  in the mass budget of dwarf galaxies \protect\citep{moldwarf}. The mass of atomic gas is measured from the HI datacubes (Sec. \ref{sec:mapd} and Tab. \ref{tab:sample_obs}), while the stellar masses are from \protect\cite{d47stellarmass} for DDO 47 and from \protect\cite{ZLT} for all the other galaxies. 
The estimate of the mass is proportional to the square of the galactic distance, therefore errors on the distance add further uncertainties on the final estimate of the baryonic mass. 
Unfortunately, the works we used to take the stellar masses \protect\citep{d47stellarmass,ZLT} and the distances \protect\citep{LT} do not report the errors on their measures, so we assumed a conservative error of the 30$\%$ for $\text{M}_\textrm{bar}$.
We also performed a deeper analysis for the distance uncertainties.
The relative difference between the masses estimated assuming two different distances is 
\begin{equation}
\delta_\text{D}=\frac{\text{M}(\text{D}_1) - \text{M}(\text{D}_2)}{\text{M}(\text{D}_1)}=\frac{\text{D}^2_1-\text{D}^2_2}{\text{D}^2_1}.
\label{eq:mass-distance}
\end{equation} For each galaxy in our sample we choose the best distance estimator\footnote{The scale of distance estimators is, from the best to the worst: Cepheids, RGB-Tip, CMD, Brightest-Stars, Tully-Fisher relation.} 
available on NED (NASA/IPAC Extragalactic Database) and we considered the minimum ($\text{D}^{\min}_\textrm{NED}$) and the maximum ($\text{D}^{\max}_\textrm{NED}$) estimate of the distance, then using Eq. \ref{eq:mass-distance} we calculated
\begin{equation}
\delta^{\min/\max}_{\text{D}}= 1 - \left( \frac{\min(\text{D},\text{D}^{\min/\max}_\textrm{NED})}{\max(\text{D},\text{D}^{\min/\max}_\textrm{NED})} \right)^2,
\end{equation} where D is the distance assumed in this work (see Tab. \ref{tab:sample_obs}). When $\delta_{\text{D}}$ is large the error on the total mass is dominated by the uncertainty on the distance. Three galaxies have $\delta_{\text{D}}$ larger than the $60\%$: DDO 47 ($\delta^{\max}_{\text{D}}=62\%$), DDO 101 ($\delta^{\max}_{\text{D}}=85\%$) and NGC 1569 ($\delta^{\min}_{\text{D}}=68\%$). For these galaxies we do not show the 1-$\sigma$ error on the baryonic mass, but a magenta bar indicating the interval of mass found  assuming the distance $\text{D}$ or the distance $\text{D}^{\min/\max}_\textrm{NED}$.

\end{itemize}

In Fig. \ref{fig:btfrc} we compare our data with a recent fit to the BTFR \protect\citep{lellibtff}. Interestingly our data overlap between $10^8$ and $10^9$ M$_\odot$. 
In this range our data are perfectly compatible with the parameters of the BTFR estimated in \protect\cite{lellibtff} (s$=3.95$ and $\text{A}=1.86$ in Eq. \ref{eq:btfr}); we also confirm the relatively small scatter around the relation, with little increase towards lower mass galaxies, in contrast to \protect\cite{begumtf}. This remains the case, even when including galaxies with rising rotation curves: the only two outliers are DDO 50 (a nearly face-on galaxy) and DDO 101 (for which the uncertainity on the distance is large; Sec. \ref{sec:gls}, see also \protect\citealt{mia}).
Below $10^8$ M$_\odot$ the distribution of the galaxies looks again compatible with the relation of \protect\cite{lellibtff}. It appears more scattered, but this could owe entirely to the large error bars for these low-mass systems (that results from the increasingly important asymmetric-drift correction).
Furthermore, there are few galaxies in this mass range and all of them have rising rotation curves (with the exception of DDO 216 that has rather peculiar kinematics; see Fig. \ref{fig:DDO216}, Sec. \ref{sec:gls} and Appendix \ref{sec:d216b}).

\section{Discussion} \label{sec:disc}
\subsection{HI scale height} \label{sec:hscale}

All the galaxies in our sample have been analysed assuming a very thin HI disc with scale height 100 pc, independent of radius (see Sec. \ref{sec:ass}). This assumption would be fully justified in the case of spirals (e.g. \citealt{brinkbb,ol4244}), but in the shallow potential of the dIrrs the gaseous layer can be quite thick especially in the outer regions of the disc (see e.g. \citealt{hscaledw,tdisc}). In this Section we discuss how the presence of a thick disc could bias the results of our analysis. 

In the presence of a thin disc the observed HI emission can be easily related to the intrinsic properties of the galaxy: the observed velocity is a measure of gas rotation in the equatorial plane,  the observed velocity dispersion is an unbiased measure of the chaotic motion of the gas and the intrinsic profile of the HI surface density can be obtained by simply correcting the observed profile for the $i$ of the galaxy (Eq. \ref{eq:cosinc}).
In the presence of thick gaseous layers, the line of sight intercepts the emission coming from rings at different radial and vertical positions. As a consequence, the parameters obtained by assuming zero thickness may not be a precise measure of the  kinematics of the galaxy.

2D methods as in O15 work on integrated maps (velocity fields) and they can not  take into account the presence of a thick disc.
3DB is more promising since the scale height is one of the parameters needed in the datacube fitting.
We performed several tests with 3DB, but we found that the fit is essentially insensitive of the HI thickness for small-medium values of the scale height ($\text{z}_d\lesssim 0.7$ kpc) and returns unacceptable results for very thick HI discs. The reason for this is that 3DB fits one single ring at the time and it can not include the emission coming from the extended vertical layers of the other rings. 

We quantified the magnitude and the type of errors introduced by the assumption of thin disc as follows. 
We performed several tests on mock datacubes made with the task GALMOD \citep{sick} of the software package GIPSY \citep{gipsy}. We found that the assumption of a thin disc biases the results as follows:
\begin{enumerate}[label=(\roman*), itemindent=.4cm]
    \item the surface-density profile tends to be shallower than the real profile;
    \item the measured broadening of the HI line profiles is larger than the intrinsic velocity dispersion of the gas ($\sigma_v$ is overestimated);
    \item the representative velocity estimated from  the HI line profiles may not trace the gas rotation velocity at certain locations.
\end{enumerate} 
Obviously, these effects influence the estimate of the final rotation curve: due to (iii) the observed velocities are no longer tracing the rotation on the equatorial plane, while (i) and (ii) bias the calculation of the asymmetric-drift correction (see Sec. \ref{sec:asy}). We found that the combination of these effects causes the circular velocities, obtained with both 2D and 3D methods assuming a thin disc, to be lower than the intrinsic velocities at smaller radii (similar to the beam smearing) and  higher in the outermost disc. The magnitude of these differences depends mainly on the thickness of the HI discs, but the $i$ of the disc and the shape of the rotation curve are also important parameters. The biases described above are negligible for nearly face-on galaxies ($i<50^\circ$), while a rising rotation curve, typical of dIrrs, amplifies the errors.

The thickness of the HI disc has never been incorporated in the derivation of the rotational velocity.
In Iorio et al. (in prep) we will present an original method to estimate in a self-consistent way the intrinsic kinematic properties of the galaxies taking into account the thickness of the HI disc under the assumption of vertical hydrostatic equilibrium. Applying this method to WLM (Fig. \ref{fig:WLM}), we derived a scale height that is about 150 pc in the centre and flares linearly with radius up to 600 pc, well above the 100 pc assumed in our analysis. We compared the results obtained taking the HI thickness into account with the results obtained in the current work as described in Sec. \ref{sec:datan}: in the thick-disc model the peak of the profile of intrinsic surface density is higher by about 2 $\text{M}_\odot \text{pc}^{-2}$, the velocity dispersion is lower by an average of 0.5 km/s and the rotation velocities have a maximum difference of about 4 km/s. All these differences are compatible within the errors and we can conclude that our results for WLM are not seriously biased by the assumption of a thin disc.

The scale height of a HI disc can be related to the velocity dispersion of the gas and the total volumetric density in the plane of the galaxy as $\text{z}_d  \propto \sigma_v \rho^{-0.5}$ \citep{hf2,hf1}, we can define an average density in terms of circular velocity so that $\text{z}_d \text{R}^{-1} \propto \sigma_v \text{V}^{-1}_\text{c} $. 
Therefore,  WLM is expected to be representative of our sample of galaxies, in term of disc thickness (see Tab. \ref{tab:sample_res}): the bias due to the HI thickness is likely to be
significantly larger only for DDO 210, DDO 216 and NGC 1569. However, in  these galaxies the rotation curve is already very uncertain and the quoted errors are so large that should be still inclusive of the errors due to the thin-disc assumption. In conclusion, we are confident that all the rotation curves and velocity dispersions found in this work are not seriously  biased by the assumption that the HI disc is thin.

\subsection{Comparison to the standard 2D approach}

\begin{figure*} 
 \centering 
 \includegraphics[width=0.94\textwidth]{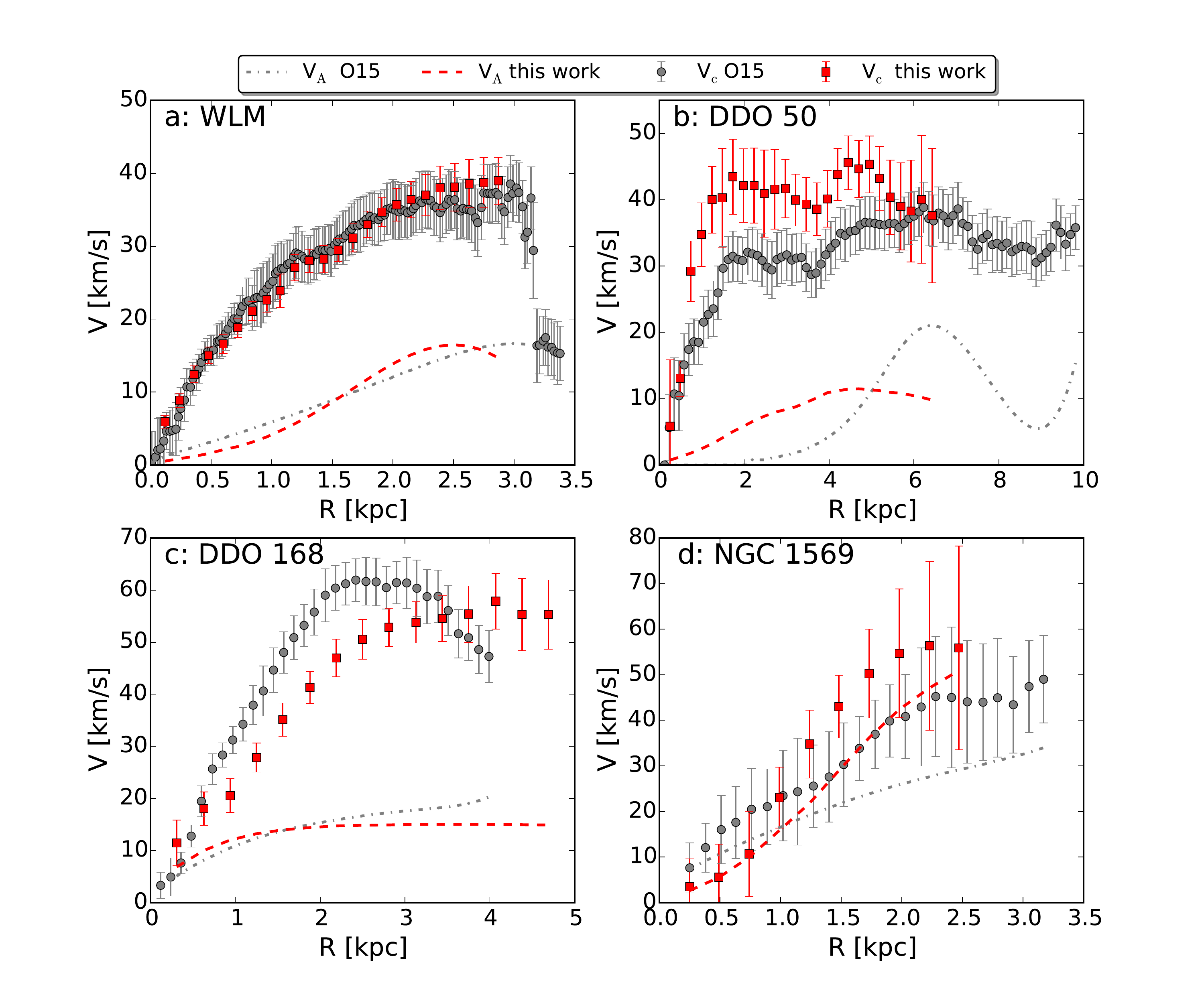}
 \caption{ Circular velocity (V$_\text{c}$) and asymmetric-drift term (V$_\text{A}$) for four representative galaxies in our sample (a: WLM, b: DDO50, c: DDO 168, d: NGC 1569). The red squares and the red curves show the results found in this work, while the grey circles and the grey curves show the results obtained in  O15.}
\label{fig:compare}
\end{figure*}

In this Section we compare the rotation curves obtained with our method with the ones obtained in O15 for the same dataset with the classical 2D approach.
As a general feature, the rotation curves of O15 reach larger radii: the median difference on the radial extension is about 700 pc and the  galaxies with the largest discrepancies are: DDO 50 (3.4 kpc), DDO 87 (2.2 kpc) and NGC 2366 (1.8 kpc). 
DDO 216 has rotation curves of similar extensions, while our rotation curves for DDO 210 and DDO 168 are more extended by about 200 pc and 700 pc respectively. These differences are due to the different approach in the choices on the outermost radii considered in the kinematic fit. We decided to include only the part of the galaxy with clear information on the rotation motion and not dominated by the noise (see Sec. \ref{sec:maprad}), as shown by the elliptical rings over plotted to the HI maps and HI velocity fields (Box C in Figs.  \ref{fig:UA292}-\ref{fig:WLM}). In contrast, O15 extended the fit as much as possible, but in some cases the outer part of the rotation curves might be extracted mainly from emission around the minor axis  which have poor information on the rotational motion. The peril of this approach is that the rotation curves at large radii can be quite unreliable or  show artifacts as for example in WLM (Fig. \ref{fig:compare}\textcolor{blue}{a}) where the rotation curve of O15 shows an abrupt drop-off after the outermost radii used in our analysis. 

Concerning the shapes of rotation curves, about half of our results are compatible with O15 within the errors. When differences are present they can be due to a combination of three different reasons: (i) an intrinsic difference in the best-fit projected rotational velocities (V$_\textrm{rot} \sin i$), (ii) a difference in the best-fit $i$, PA or centre and (iii) a difference in the asymmetric-drift correction terms (V$_A$).  
Most of the cases are due to (ii) and (iii); only one galaxy (DDO 168) seems to have a significant discrepancy caused by the two different fitting methods.
This galaxy is shown in Fig. \ref{fig:compare}\textcolor{blue}{c}, while examples of (ii) and (iii) are shown in Figs.  \ref{fig:compare}\textcolor{blue}{b} and \ref{fig:compare}\textcolor{blue}{d}.
The galaxies showing the largest discrepancies are listed below.
\begin{itemize}[itemindent=.1cm]
\item[-]{\bf CVnidwa.} 
Our final rotation curve is systematically higher than that of O15 by about 5 km/s. This difference is totally dominated by the discrepancy in the asymmetric-drift correction term. Our V$_A$ is already significant at the inner radii, while the V$_A$ of O15 is negligible out to 2 kpc. These differences are ascribed to the different radial trend of the intrinsic surface density profile caused by discrepancies in the best-fit PA . 
\item[-]{\bf DDO 47.} The circular velocity shows a discrepancy of about 7 km/s in the inner part of the disc ($R<2$ kpc) due to a discrepancy on the best-fit $i$: O15 used $i\approx 55^\circ$ while
we used $i=37^\circ$ independent of radius.
\item[-]{\bf DDO 50.} Discrepancies of about 10 km/s (see Fig. \ref{fig:compare}\textcolor{blue}{b}) are due to the different assumption on the value of $i$ (about 50$^\circ$ in O15 and 33$^\circ$  in this work).  Our curve is also less extended (see Fig. \ref{fig:DDO50}).
\item[-]{\bf DDO 52.} The final rotation curve of O15 is systemically higher due to the different values of $i$ (about 37$^\circ$ in O15 and 
55$^\circ$  in this work).
\item[-]{\bf DDO 87.} Our final rotation curve shows a steeper rising due to a discrepancy in the assumed $i$: we used $i=40^\circ$ independent of radius, while in O15 $i$  varies between 60$^\circ$ and 40$^\circ$.
\item[-]{\bf DDO 168.} The final rotation curves are quite different (see Fig. \ref{fig:compare}\textcolor{blue}{c}). In O15 the curve flattens at about 2 kpc and starts to decrease at 3 kpc, while our rotation curve shows a less steep rising in the inner part and a flattening at about 3.5 kpc. These discrepancies are due mainly to an intrinsic difference in the best-fit projected rotation curve found with the two methods. 
The differences are further increased  by the different assumptions on $i$ ($i\approx60^\circ$ in this work, $i\approx40^\circ$ in O15).
\item[-]{\bf NGC 1569.} The circular rotation curve found in this work rises steeper with respect to the rotation curve reported in O15 (see Fig. \ref{fig:compare}\textcolor{blue}{d}). The difference is caused by the asymmetric-drift correction given that with our 3D method we find a different radial trend of both the velocity dispersion and the intrinsic surface density.
\item[-]{\bf NGC 2366.} Our final rotation curve is systematically lower in the inner disc (R$<2$ kpc). The cause is a combination of a different position of the galactic centre and  different PA. We used a constant PA of about 40$^\circ$, while in O15 the PA grows from 20$^\circ$  to 40$^\circ$ in the first two kpc.
\end{itemize}

\section{Summary} \label{sec:sum}

We presented a study of the kinematics of the HI discs for 17 dwarf irregular galaxies taken from the public survey  LITTLE THINGS \citep{LT}. The main goal of this work is to make available to to the community a sample of high-quality rotation curves of dIrrs ready-to-use to perform dynamical studies. The tabulated quantities ($\text{V}_\text{rot}$, $\text{V}_\text{c}$, $\text{V}_\text{A}$, $\sigma_v$)  and HI surface-density profile ($\Sigma_\text{int}$) are available in the online version of the paper.
The key points of this work are listed below.

\begin{itemize}[itemindent=.2cm]
\item[1.] We derived the rotation curves from the HI datacubes with a state-of-art technique: the publicly available software $^\text{3D}$BAROLO \citep{barolo}.  It fits 3D models to the datacubes  without explicitly extracting velocity fields.  The fit in the 3D space of the datacubes ensures full control of the observational effects and in particular a proper account of beam smearing. 
\item[2.] The maximum radii used in the fits have been chosen with great care from the HI total map. This ensures a robust estimate of the kinematics avoiding regions of the galaxy  dominated by the noise and with scant information about the gas rotation.
\item[3.]  We developed a method to take into account the uncertainties of the asymmetric-drift correction. As a consequence, the quoted errors on the rotation curves are representative of the real uncertainties. The inclusion of these errors are fundamental in galaxies for which the calculation of the circular velocity is  highly dominated by the asymmetric-drift correction (e.g. DDO 210).
\item[4.] We estimated how our results can be biased by the assumption of a thin HI disc. We analysed in detail the galaxy WLM and we found that the HI layer flares linearly from about 150 pc to 600 pc, well above the 100 pc assumed in our analysis. The presence of such thick disc introduces systematic errors on the estimate of the velocity dispersion, intrisic HI surface density and  circular velocity. However, the differences are compatible within the quoted uncertainties. The thickness of WLM should be representative of the galaxies of our sample, so we can conclude that our results are not seriously biased by the assumption of thin disc.
\item[5.]  The rotation curves obtained in this work have been used to test the baryonic Tully-Fisher relation in the low-mass regime ($10^6 \lesssim {\rm M}_{\rm bar}/{\rm M}_\odot \lesssim 10^9$). We found that our results are compatible in slope, normalisation and scatter with the work of \cite{lellibtff} at higher mass (${\rm M}_{\rm bar} > 10^8    {\rm M}_\odot$).

\end{itemize}

\hfill \break

{\it Acknowledgments:} This research made use of the LITTLE THINGS data sample. We would like to thank  the anonymous referee and Antonino Marasco for useful comments that improved this manuscript and Se-Heon Oh and Federico Lelli for kindly making their data available.  G. Battaglia acknowledges financial support by the Spanish Ministry of Economy and Competitiveness (MINECO) under the Ramon y Cajal Programme (RYC-2012-11537). J.I. Read would like to acknowledge support from STFC consolidated grant ST/M000990/1 and the MERAC foundation.
The research has made use of the NASA/IPAC Extragalactic Database (NED) which is operated by the Jet Propulsion Laboratory, California Institute of Technology, under contract with the National Aeronautics and Space Administration.


\bibliographystyle{mnras}
\bibliography{references}


\appendix
\section{Estimate of the geometrical parameters} \label{sec:iguess}
It is important to initialise the fit of the HI datacubes with good estimates of the geometrical parameters: centre, $i$ and PA of the HI disc (see Sec. \ref{sec:tstage}).
We used three different methods:
\begin{itemize}
\item{\textit{3DB:}} 3DB can automatically estimate the geometrical parameters analysing the 0th moment (HI total map) and the 1th moment (velocity field) of the datacube.
The $i$ is estimated by fitting HI model maps to the observed HI total map, while the PA is calculated as the orientation of the line that maximises the velocity gradient in the velocity field. Finally, the centre is estimated as the flux-weighted average position of the source. 3DB builds the total map and the velocity field using a source-finding algorithm (DUCHAMP,  \citealt{sourceF}), separating the source pixels from the noise pixels on the basis of a S/N threshold. As a consequence, the final maps and the estimate of the disc parameters depend on the choice of this threshold. For this reason we repeat this analysis using S/N threshold  ranging from 2.5 to 5. Finally, we average the values on the range of S/N where 
the estimated parameters are almost constant.

\item{\textit{RingFit:}} We  developed a simple script, called RingFit, to fit ellipses to the iso-density contours at different radii of our total
HI maps (Sec. \ref{sec:mapd}). From the properties of the fitted ellipses we can 
estimate the geometrical parameters of the HI disc at different radii.
\item{\textit{Isophotal fitting of the optical maps:}} The geometrical parameters are derived from the V-Band photometry by \cite{HVBand}.
\end{itemize}

\begin{figure}
 \centering 
 \includegraphics[width=1.1\columnwidth]{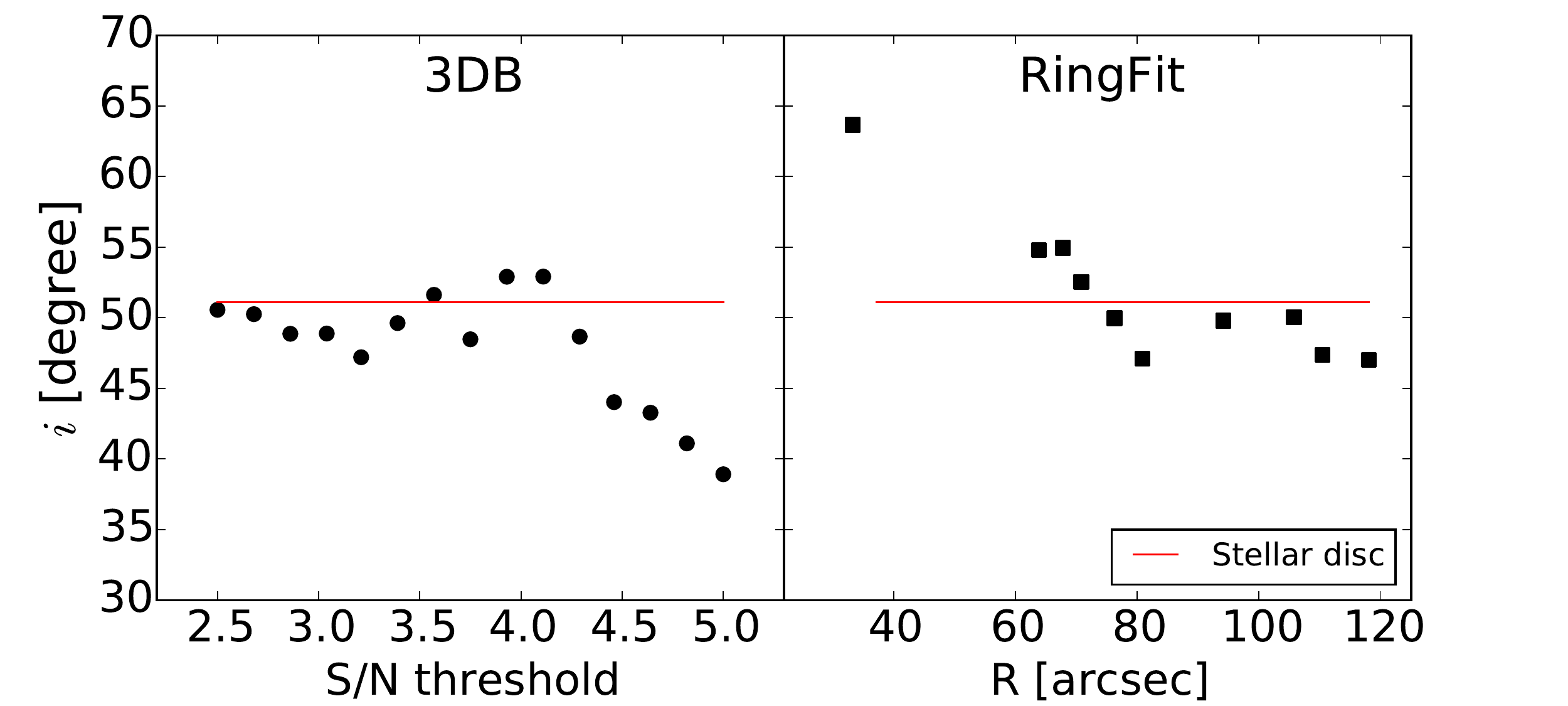}
 \caption{Initial estimate of the inclination angle for DDO 52 with 3DB (left-hand panel) and with  the fit of the iso-density contours of the HI total map (right-hand panel). The red lines show the inclination of the optical disc. See the text for details.}
\label{fig:d52iguess}
\end{figure}

\begin{figure} 
 \centering 
 \includegraphics[width=1.1\columnwidth]{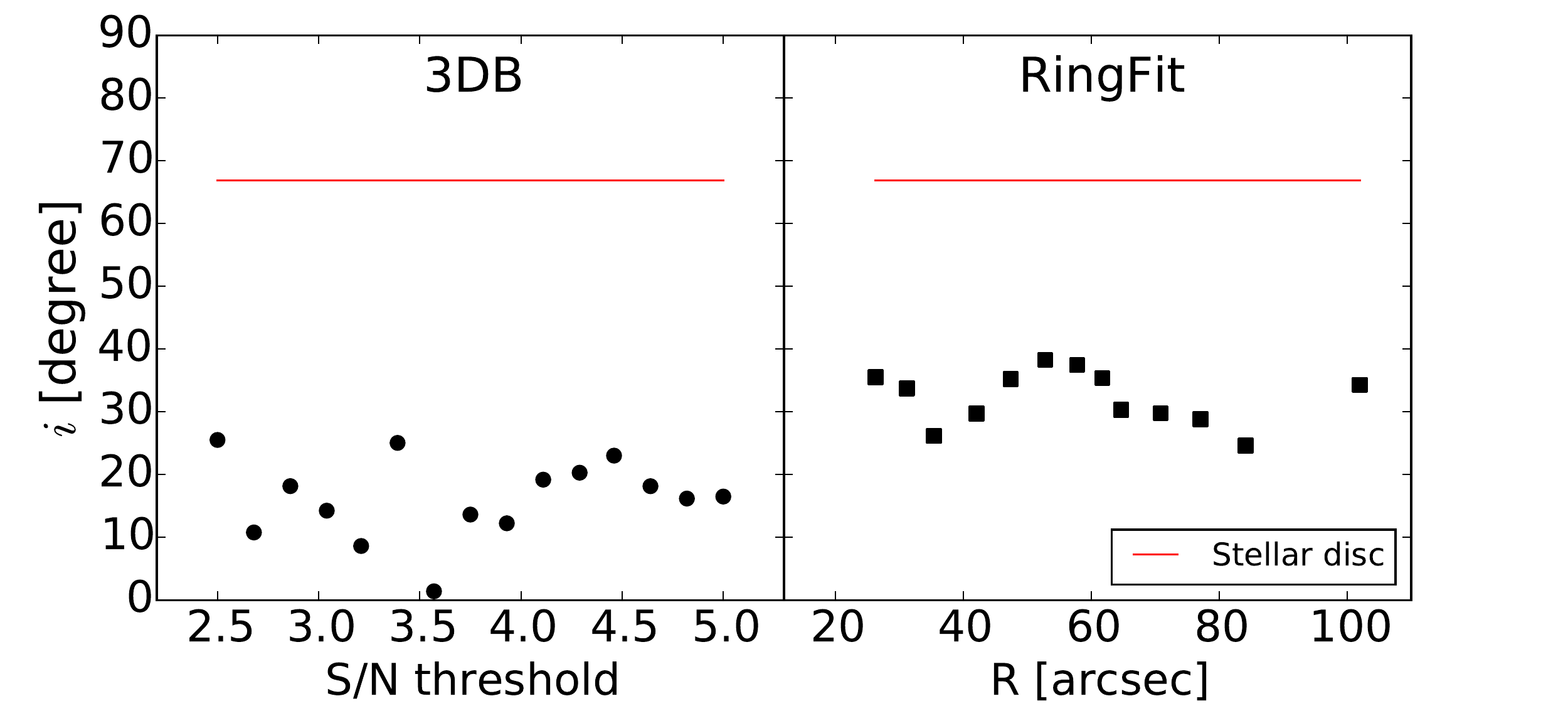}
 \caption{Same as Fig. \ref{fig:d52iguess} but for DDO 210.}
\label{fig:d210iguess}
\end{figure}

Figs. \ref{fig:d52iguess} and \ref{fig:d210iguess} show two examples of the estimate of $i$ with the three different methods. In the first case (DDO 52) the methods are in good agreement, while in the second case (DDO 210) the results are not compatible (notice that in this case the HI contours deviate strongly from elliptical shape, see Box C in  Fig. \ref{fig:DDO210}).
Usually, the optical disc has a more regular morphology with respect to the HI disc, therefore we give more weight to the geometrical parameters estimated from  \cite{HVBand}.
When the geometry of the stellar  and of the HI disc do not show a clear agreement (Fig. \ref{fig:d210iguess}) or the stellar disc presents peculiar features such as strong bars, we use the other two methods.
More details  can be found in the description of the individual galaxies in Sec. \ref{sec:gls}.

\section{Plot layout} \label{sec:plot_layout}
In this Appendix we report the caption that describes
Figs. 5-21.

\begin{itemize}

\item {\bf Box A}:
    Rotation velocity  estimated with 3DB (V$_\textrm{rot}$, blue line),
    asymmetric-drift correction (V$_\textrm{A}$, grey circles) and final corrected circular velocity (V$_\textrm{c}$, black circles).
\newline
\item {\bf Box B}:
    \begin{itemize}
    \item[] \textit{Top:} Inclination (left-hand panel) and position angle (right-hand panel) found with 3DB as functions of radius R. The red lines
    show the fit used to obtain the rotation velocity (see Sec. \ref{sec:tstage}), while the black dashed line indicates the median of the data.
    \item[] {\textit{Bottom:}} HI velocity dispersion estimated with 3DB (left-hand panel) and intrinsic surface density of the HI disc not corrected for Helium (right-hand panel) as function of radius R. In the left-hand panel the black dashed line indicates the median of the data.
    The empty circles (if present) represent the points excluded both from the calculation of the asymmetric-drift correction (see Sec. \ref{sec:asy}) and from the calculation of the median. 
    The excluded points belong to rings where the velocity dispersion found with 3DB is discrepant or peculiar with respect to the global trend (see Sec. \ref{sec:fnotes}).
    \end{itemize}
\item {\bf Box C}:
    \begin{itemize}
    \item[] \textit{Left-hand panel:} Stellar emission overlaid with the iso-density contours of the total HI emission. 
    The thick contour indicates  $\sigma_{3T}$, the  3-$\sigma$ pseudo noise of the total map (see Sec. \ref{sec:maprad} and Tab. \ref{tab:sample_obs}).
    \item[] \textit{Right-hand panel:}  Velocity field obtained as the 1th moment of the data cube.
    The thick contour highlights the systemic velocity (see Tab. \ref{tab:sample_res}). A physical scale is plotted on the bottom right corner of the panel, while the beam of the HI observation is shown in the bottom left corner.
    \end{itemize}
    The crosses and the ellipses show the assumed galactic centre and the last used ring respectively (see Tab. \ref{tab:sample_res}).
\newline
\item {\bf Box D}:
    PV diagram along the major (left-handed panel)  and the minor (right-handed panel)  axis of the HI disc.
    The black  and the red contours show  respectively the iso-density contours of the galaxy and the best-fit model  found with 3DB. The horizontal black dashed lines indicate the systemic velocity. 
\end{itemize} 

\section{An alternative scenario for DDO 216} \label{sec:d216b}

DDO 216 shows a peculiar kinematics:  a velocity gradient of about 10 km/s appears clearly in the velocity field  (right-hand panel in Box C in Fig. \ref{fig:DDO216})  and in the channel maps, but it could be entirely due
to a single `cloud' at a discrepant velocity in the approaching side of the galaxy \citep{d216gradient}.
The analysis of the channel maps partially supports this hypothesis: the receding and the approaching sides of the galaxy seem to have a slightly  distinct kinematics overlapping between $-186$ km/s and $-190$ km/s, while the V$_\text{sys}$ found with 3DB is around -180 km/s. At the same time, the total HI map does not show the clear presence of a separate component at the supposed position of the cloud (left-hand panel in Box C in Fig. \ref{fig:DDO216}). In conclusion we did not have sufficient elements to confirm or exclude the existence of a HI component other than the disc of DDO 216. We can summarises the two hypothesis as: 
\begin{itemize}
\item[-]{\textit{scenario1:}} The main velocity gradient is genuine and totally due to the rotation of the HI disc.
\item[-]{\textit{scenario2:}} The main velocity gradient is due to an HI sub-structure with a systemic velocity that differs of about 20 km/s with respect to the HI disc.
\end{itemize}
The scenario1 is analysed in Sec. \ref{sec:gls} while in this Appendix we describe the analysis of the scenario2.
First of all we masked the region of the datacube 
containing the emission of the supposed HI cloud. 
The mask has been  build by inspecting the channel maps using the task PYBLOT of GIPSY. Then we performed the kinematic fit of the  `clean' datacube with 3DB.
We set $i_\text{ini}$, PA$_\text{ini}$  and the galactic centre to the values obtained from the stellar disc \citep{HVBand}.
Contrarily  to the scenario1 (Sec. \ref{sec:gls}) we used the V$_\textrm{sys}$ estimated with 3DB, the two estimates differ by about 8 km/s, while the final best-fit PA (120$^\circ$) is tilted of about 10$^\circ$-15$^\circ$ with respect to the values found in the scenario1 (Box B in Fig. \ref{fig:DDO216}).

\begin{figure} 
 \centering 
 \includegraphics[width=1.0\columnwidth]{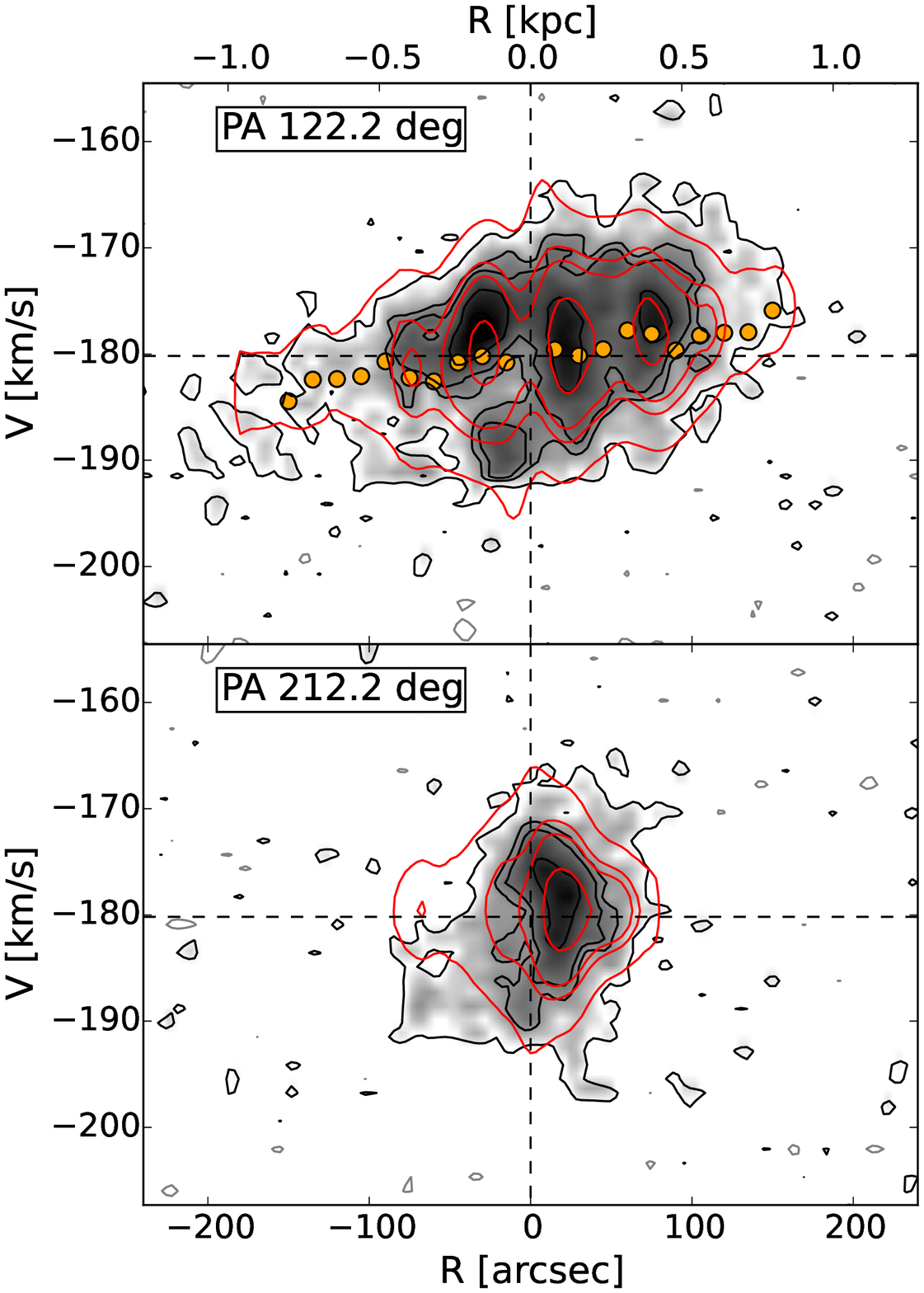}
 \caption{PV along the major axis (upper panel) and the minor axis (bottom panel) for the scenario2 of the galaxy DDO 216 (see text). The black contours indicate the datacube while the red contours show the best-fit model found with 3DB. Contours at $(2+6n)\sigma_\textrm{ch}$, where $\sigma_\textrm{ch}=0.91$ mJy bm$^{-1}$, the grey contours are at -$2 \sigma_\textrm{ch}$. The yellow points indicate the rotational velocity found with 3DB. }
\label{fig:d126pv}
\end{figure}

\begin{figure} 
 \centering 
 \includegraphics[width=1.0\columnwidth]{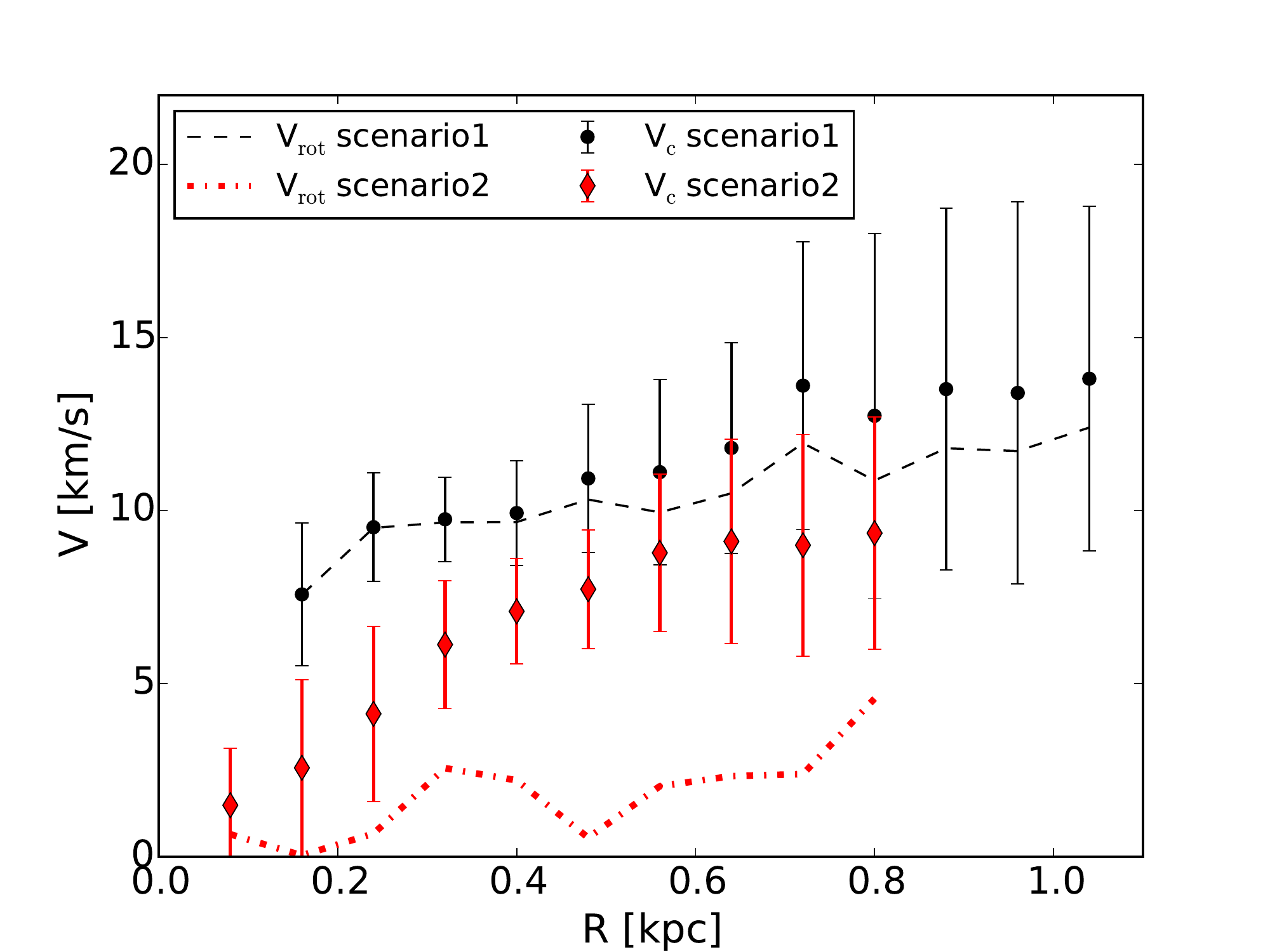}
 \caption{Rotational velocities (red dashed line) and circular velocities (red points) found for the scenario2 of DDO 216 (see text). The rotation curves of the scenario1 are also shown for comparison. }
\label{fig:d126rot}
\end{figure}

Without the `cloud' the velocity gradient almost disappears 
and the kinematics is dominated by the chaotic motion of the gas as it is clearly visible in the PVs in Fig. \ref{fig:d126pv}. As a consequence, the final rotation curve is 'shaped' by the asymmetric-drift correction.
Fig. \ref{fig:d126rot} compares the rotation curve found for the scenario1 and for the scenario2: the velocities are compatible within the errors,  but in scenario1 the V$_\text{c}$ is systematically higher by about 4 km/s and it extends to larger radii.

\cite{kirbypeg} found that the stellar component of DDO 216 is rotating in the same direction as the gas with a $\text{V}_\text{rot}$  of about 10 km/s, perfectly compatible with scenario1 (see Fig. \ref{fig:d126rot}). This velocity gradient is observed also in red giant stars, so the rotation is genuine  and  it is not caused by short-term hydrodynamical events (e.g. bubbles), which influence the gas kinematics but not the kinematics of the old stellar populations. However the centre of the rotation and $\text{V}_\text{sys}$ are quite different from what we found for scenario1 and they are instead compatible with scenario2. 
Further studies are needed to understand if the gas in this peculiar galaxy is rotating, which is crucial to accurately estimate its dynamical mass.

\section{Typical setup for $^\text{3D}$BAROLO} \label{sec:bsetup}
In this Appendix we report the keywords and their values for our typical parameter file passed to 3DB. For a detailed description of the keywords see the documentation of 3DB\footnote{{\tt http://editeodoro.github.io/Bbarolo/documentation/}} 
(`...' means that the values depends on the analysed galaxy).

\begin{itemize}
    \item[-]{FITSFILE ...;} name of the  datacube to fit.
    \item[-]{GALFIT TRUE;} enable the 3D fitting of the datacube.
    \item[-]{NRADII ...;} number of rings to be used in the fit.
    \item[-]{RADSEP ...;} separation between rings in arcsec.
    \item[-]{VROT 30;} initial guess for the rotation velocity in km/s.
    \item[-]{VDISP 8;} initial guess for the velocity dispersion in km/s.
    \item[-]{INC ...;} initial guess for $i$ in degree ($i_\text{ini}$, see Tab. \ref{tab:sample_res}).
    \item[-]{PA ...;} initial guess for the PA in degree (PA$_\text{ini}$, see Tab. \ref{tab:sample_res}).
    \item[-]{VSYS ...;} initial guess for the systemic velocity in km/s.
    \item[-]{XPOS ...;} initial guess for the X coordinate of the centre.
    \item[-]{YPOS ...;} initial guess for the Y coordinate of the centre.
    \item[-]{Z0 100/$f_c$;} initial guess for the scale height in arcsec, $f_c$ is the conversion factor from arcsec to pc and it is listed in Tab.\ref{tab:sample_obs} for each galaxy.
    \item[-]{LTYPE 1;} use a Gaussian profile for the HI vertical distribution. 
    \item[-]{FREE VROT VDISP INC PA;} parameters to fit with 3DB.
    \item[-]{FTYPE 2;} use the absolute differences between models and datacube as residuals.
    \item[-]{WFUNC 1;} use the cosine of the azimuthal angle as weigthing function for the fit (see Sec. \ref{sec:tstage}).
    \item[-]{CDENS 10;} this parameter sets (in a non-trivial way) the number of `clouds' to be used in the building of the ring models.
    \item[-]{NV 200;} number of sub-clouds in which each cloud (see above) is divided to populate the spectral axis of the ring models.
    \item[-]{TOL 1E-3;} tolerance condition to stop the fit.
    \item[-]{MASK Smooth;} create a mask for the fit smoothing the datacube.
    \item[-]{BLANKCUT 2.5;} include  in the mask (see above) only the pixels with  S/N$>2.5$.
    \item[-]{SIDE B;} fit the whole galaxy.
    \item[-]{TWOSTAGE TRUE;} enable the two-stage method described in Sec. \ref{sec:tstage}.
    \item[-]{BWEIGHT 1;} set the weight of the blank pixels (see \citealt{barolo} for details).
    \item[-]{DELTAINC 40;} $i$ can assume values only  between INC-DELTAINC and INC+DELTAINC.
    \item[-]{DELTAPA 80;} same as above, but for the PA.
    \item[-]{FLAGERRORS true;} enable the estimate of the errors on the fitted parameters.
    \item[-]{STARTRAD 1;} avoid the fit of the first ring placed by default at RADSEP/4.
    \item[-]{LINEAR 0.424;} instrumental resolution ($\sigma_\text{inst}$) in unity of channel width. In all our galaxies the instrumental FWHM is equal to the channel separation, so $\sigma_\text{inst}=1/2.355=0.424$.
\end{itemize} If the parameters XPOS,YPOS,VSYS,PA,INC are omitted, 3DB estimates them as described in Appendix \ref{sec:iguess}.

\bsp	
\label{lastpage}
\end{document}